\documentclass[preprint]{imsart} 
\RequirePackage[OT1]{fontenc}
\RequirePackage{amsthm, amsmath, amssymb, amsfonts, epsfig, subfigure, multirow,color}
\RequirePackage[authoryear]{natbib}
\RequirePackage[colorlinks=true,citecolor=blue,urlcolor=blue]{hyperref}
\startlocaldefs
\numberwithin{equation}{section}
\theoremstyle{plain}
\def\M{\mathbb M}

\def\R{\mathbb R}

\def\Z{\mathbb Z}
\def\P{\mathbb P}
\def\G{\mathbb G}
\def\argmin{\qopname\relax m{argmin}}
\def\argmax{\qopname\relax m{argmax}}
\def\F{\mathcal F}
\def\be{\begin{equation}}

\def\bes{\begin{equation} \begin{split}}
\def\ees{\end{split} \end{equation} }
\def\ee{\end{equation}}
\def\qed{\hfill$\square$}
\def\ed{\end{document}}

\newtheorem{prop}{Proposition}
\newtheorem{thm}{Theorem}
\newtheorem{cor}{Corollary}
\newtheorem{lem}{Lemma}
\newtheorem{rmk}{Remark}

\graphicspath{{plots_dsr_reg_th_estn/}}

\endlocaldefs

\begin{document}
\sloppy
\begin{frontmatter}
\title{M-estimation in multistage sampling procedures}
\runtitle{Multistage procedures}

\begin{aug}
\author{\fnms{Atul} \snm{Mallik}\thanksref{t1}\ead[label=e1]{atulm@umich.edu}},
\author{\fnms{Moulinath} \snm{Banerjee}\thanksref{t1}\ead[label=e2]{moulib@umich.edu}}
\and
\author{\fnms{George} \snm{Michailidis}\thanksref{t2}\ead[label=e3]{gmichail@umich.edu}}

\thankstext{t1}{Supported by NSF Grant DMS-1007751 and a Sokol Faculty Award, University 
of Michigan}
\thankstext{t2}{Supported by NSF Grants DMS-1161838 and DMS-1228164}
\runauthor{A. Mallik, M. Banerjee and G. Michailidis}


\address{Department of Statistics\\
University of Michigan\\
Ann Arbor, Michigan 48109 \\
\printead{e1}\\
\phantom{E-mail:\ }\printead*{e2}\\
\phantom{E-mail:\ }\printead*{e3}}
\end{aug}

\begin{abstract}
Multi-stage (designed) procedures, obtained by splitting the sampling budget suitably across stages, and designing the sampling at a particular stage based on information about the parameter obtained from previous stages, are often advantageous from the perspective of precise inference. We develop a generic framework for M-estimation in a multistage setting and apply empirical process techniques to develop limit theorems that describe the large sample behavior of the resulting M-estimates. Applications to change-point estimation, inverse isotonic regression, classification and mode estimation are provided: it is typically seen that the multistage procedure accentuates the efficiency of the M-estimates by accelerating the rate of convergence, relative to one-stage procedures. The step-by-step process induces dependence across stages and complicates the analysis in such problems, which we address through careful conditioning arguments.
\end{abstract}
\end{frontmatter}

\sloppy

\section{Introduction}\label{intro}

Multi-stage procedures, obtained by allocating the available sampling budget suitably across stages, and designing the sampling mechanism at a particular stage based on information about the parameter of interest obtained in previous stages, has been a subject of investigation in a number of recent
papers \citep{LBM09, TBM11, BGZ13}. Specifically, a two-stage procedure works as follows:
\begin{enumerate}
	\item In the first stage, utilize a fixed portion of the design budget to obtain an initial
estimate of the key parameter $d_0$, as well as nuisance parameters present in the model.
\item Sample the second stage design points in a shrinking neighborhood around the first stage estimator and use the earlier estimation approach (or a different one that leverages on the local behavior of the model in the vicinity of $d_0$) to obtain the final estimate of $d_0$ in this ``zoomed-in'' neighborhood. 
\end{enumerate}

Such two- (and in general multi-) stage procedures exhibit significant advantages in performance when estimating $d_0$ over their one stage
counterparts for a number of statistical problems. These advantages stem from {\em accelerating the convergence rate} of the multi-stage estimator
over the one-stage counterpart. Their drawback is that the application setting should allow one to generate values of the covariate $X$ at will anywhere
in the design space and obtain the corresponding response $Y$. Next, we provide a brief overview of related literature. \\
\noindent
(1) 
\cite{LBM09} considered the problem of estimating the \emph{change point} $d_0$ in a regression model $Y = f(X) + \epsilon$, where $f(x) = \alpha_0 1(x\leq d_0) + \beta_0 1(x> d_0)$, $\alpha_0 \neq \beta_0$. It was established that the two-stage estimate converges to $d_0$  at a rate much faster 
(almost $n$ times) than the estimate obtained from a one-stage approach. \\
\noindent
(2) In a non-parametric isotonic regression framework, where the response is related to the covariate by $Y = r(X) + \epsilon$ with $r$ being monotone, \cite{TBM11} achieve an acceleration up to the $\sqrt{n}$-rate of convergence (seen usually in parametric settings) for estimating thresholds $d_0$ of type $d_0 = r^{-1} (t_0)$ (for fixed known $t_0$), which represents
a marked improvement over the usual one-stage estimate which converges at the rate $n^{1/3}$. This involves using a local linear approximation for $r$ in a shrinking neighborhood of $d_0$, at stage two. While the $\sqrt{n}$-rate is attractive from a theoretical perspective, for functions which are markedly non-linear around $d_0$, this procedure performs poorly as illustrated in \cite{TANG13}, who alleviated this problem by another round of isotonic regression at the second stage. \\
\noindent
(3) \cite{BGZ13} considered the problem of estimating the location and size of the maximum of a multivariate regression function, where they  avoided the curse of dimensionality through a two-stage procedure.  

A significant technical complication that the multi-stage adaptive procedure introduces is that the second and higher stage data are no longer independent and identically distributed (i.i.d.), as those sampled in the first stage. This is due to the dependence of the design points on the first stage estimate of $d_0$. Moreover, in several cases, the second stage estimates are usually constructed by minimizing (or maximizing) a related empirical process sometimes over a random set based on the first stage estimates. 
Note that to establish the results on the rate of convergence of the multi-stage estimate of the parameter of interest, as well as derive its
limiting distribution, the above mentioned papers used the specific structure of the problem under consideration and a variety of technical tools starting from first principles. This begs the question whether for statistical models exhibiting similarities to those discussed above, a {\em unified approach} within the context of M-estimation can be established for obtaining the rate and the limiting distribution of the multistage estimate. 

We address this issue rigorously in this paper for two-stage procedures. To accomplish this task, we extend empirical process results originally developed for the i.i.d. setting to situations with dependence of the above nature. In particular, we present results for deriving the rate of convergence and deducing the limit distribution of estimators obtained in general two-stage problems (see Section \ref{sec:genres}); to this end, a process convergence result in a two-stage sampling context is established. Our general results, which are also expected to be of independent interest, are illustrated on: (i) a variant of the change-point problem (Section \ref{sec:cp}), (ii) the inverse isotonic regression, under a fully non-parametric scheme studied empirically in \cite{TANG13} (Section 2.4), (iii) a classification problem (Section \ref{sec:classy}) and 
(iv) mode estimation for regression (Section \ref{sec:mode}). A key insight gleaned from the general theory and the illustrative examples is that acceleration of the convergence rate occurs when the parameter of interest corresponds to a ``local" feature of the model (e.g. the change-point in a regression curve), but also depends on the statistical criterion used.

\section{Problem formulation and general results} \label{sec:genres}
A typical two-stage procedure involves estimating certain parameters, say a vector $\theta_n$, from the first stage sample. Let $\hat{\theta}_n$ denote this first stage estimate. Based on $\hat{\theta}_n$, a suitable sampling design is chosen to obtain the second stage estimate of the parameter of interest $d_0$ by minimizing (or maximizing) a random criterion function $\M_n(d, \hat{\theta}_n)$ over  domain $\mathcal{D}_{\hat{\theta}_n} \subset \mathcal{D}$, i.e.,
\be
\hat{d}_n = \argmin_{d \in \mathcal{D}_{\hat{\theta}_n} }  \M_n(d, \hat{\theta}_n).
\label{eq:destim}
\ee
We denote the domain of optimization for a generic $\theta$ by $\mathcal{D}_{\theta}$. 
We will impose more structure on $\M_n$ as and when needed. We start with a general theorem about deducing the rate of convergence of $\hat{d}_n$ arising from such criterion. In what follows, $M_n$ is typically a population equivalent of the criterion function $\M_n$, e.g., $M_n(d, \theta_n) = E \left[\M_n (d, \theta_n) \right]$, which is at its minimum at the parameter of interest $d_0$ or at a quantity $d_n$ asymptotically close to $d_0$.  
\begin{thm} \label{th:ratethmc6}
Let $\{ \M_n(d,  \theta ),\ n\geq 1 \}$ be stochastic processes and $\{ M_n( d, \theta),\ n\geq 1 \}$ be deterministic functions, indexed by $d \in \mathcal{D}$ and $\theta \in \Theta$. Let $d_n \in \mathcal{D}$, $\theta_n \in \Theta$  and $d \mapsto \rho_n(d, d_n)$ be a measurable map from $\mathcal{D}$ to $[0,\infty)$.
Let $\hat{d}_n$ be a (measurable) point of minimum of $\M_n(d, \hat{\theta}_n)$ over $d \in \mathcal{D}_{\hat{\theta}_n}\subset \mathcal{D}$, where $\hat{\theta}_n$ is a random map independent of the process $\M_n(d, {\theta})$. For each $\tau>0$ and some $\kappa_n>0$ (not depending on $\tau$), suppose that the following hold:
\begin{itemize}
	\item[(a)] There exists a sequence of sets $\Theta_n^\tau$ in $\Theta$ such that $P[ \hat{\theta}_n \notin \Theta_n^\tau] < \tau$.
	\item[ (b) ] There exist constants $c_\tau  > 0$, $N_{\tau} \in \mathbb{N}$ such that for all $\theta \in \Theta_n^\tau$, $d \in \mathcal{D}_\theta$ with  $\rho_n(d, d_n) <\kappa_n$, and $n > N_{\tau}$,
	\be
M_n( d, \theta) - M_n( d_n, \theta)  \geq c_\tau \rho_n^2(d, d_n).
\label{eq:cond00}
\ee
Also, for any $\delta \in (0, \kappa_n)$ and $n > N_{\tau}$,
{ \begin{eqnarray}
\lefteqn{ \sup_{\theta \in \Theta_n^\tau} E^* \sup_{\substack{\rho_n(d, d_n) < \delta, \\ d \in \mathcal{D}_\theta}} \left| (\M_n (d, \theta) -M_n(d, \theta))  - (\M_n(d_n, \theta)  -M_n(d_n, \theta) ) \right| } \hspace{3.5in} \nonumber \\
& \leq & C_\tau \frac{\phi_n(\delta)}{\sqrt{n}},
\label{eq:cond01}
 \end{eqnarray}}
for  a constant  $C_\tau >0$ and functions $\phi_n$ (not depending on $\tau$) such that $\delta \mapsto \phi_n(\delta)/\delta^\alpha$ is decreasing for some $\alpha < 2$.
\end{itemize}
Suppose that $r_n $ satisfies $
r^2_n\ \phi_n\left(\frac{1}{r_n}\right) \lesssim \sqrt{n}, \nonumber$ 
and $P\left(\rho_n(\hat{d}_n, d_n)\geq  \kappa_n \right)$ converges in probability to zero, then $r_n\ \rho_n(\hat{d}_n, d_n) = O_p(1)$.

Further, if the assumptions in part (b) of the above theorem hold for all sequences $\kappa_n >0$ in the sense that there exist constants $c_\tau  > 0$, $C_\tau  > 0$, $N_{\tau} \in \mathbb{N}$ such that for all $\theta \in \Theta_n^\tau$, $d \in \mathcal{D}_\theta$, $\delta>0$ and $n > N_{\tau}$, \eqref{eq:cond00} and \eqref{eq:cond01} hold,
 then justifying the convergence of $P\left(\rho_n(\hat{d}_n, d_n) \geq  \kappa_n\right)$ to zero is not necessary.
\end{thm}
The proof uses shelling arguments and is given in Section \ref{pf:ratethm} of the Appendix. The shelling arguments need substantially more careful treatment than those employed in i.i.d. scenarios since the $\mathbb{M}_n$ processes depend on the second stage data which are correlated through their dependence on the first stage estimate. 
\newline
An intermediate step to applying the above result involves justifying the convergence of $P\left(\rho_n(\hat{d}_n, d_n) \geq  \kappa_n\right)$ to zero. As mentioned in the result, if the assumptions in part (b) of the above theorem hold for all sequences $\kappa_n >0$, then justifying this condition is {\it not} necessary. This is the case with {\it most } of the examples that we study in this paper. The following result is used otherwise. 
\begin{lem}\label{lm:const}
Let $\M_n$, $M_n$ and $\rho_n$  be as defined in Theorem \ref{th:ratethmc6}. For any fixed $\tau>0$, let 
\[
c^\tau_n(\kappa_n) =\inf_{\theta \in \Theta_n^{\tau}}\, \inf_{\rho_n(d, d_n) \geq  \kappa_n, d \in \mathcal{D}_\theta} \left\{M_n(d, \theta) - M_n(d_n, \theta) \right\}\,.
\]
Suppose that 
\be \label{eq:minicond}
\sup_{\theta \in \Theta_n^\tau} P\left( 2\sup_{\substack{d \in \mathcal{D}_\theta}} \left| \M_n({d}, \theta) - M_n(d, \theta) \right| \geq c^\tau_n(\kappa_n) \right) 
{\rightarrow} \;0.
\ee
Then, $P\left(\rho_n(\hat{d}_n, d_n) \geq  \kappa_n\right)$ converges to zero .
\end{lem}
Condition \eqref{eq:minicond} requires $c^\tau_n(\kappa_n) $ to be positive (eventually) which ensures that $d_n$ is the unique minimizer of $M_n(d, \theta)$ over the set $d \in \mathcal{D}_\theta$. The proof is given in Section \ref{pf:const} of the Supplement.

The conclusion of Theorem \ref{th:ratethmc6}, $r_n\ \rho_n(\hat{d}_n, d_n) = O_p(1)$,  typically leads to a result of the form $s_n (\hat{d}_n - d_n) = O_p(1)$, $s_n \rightarrow \infty$. 
Once such a result has been established,  the next step is to study the limiting behavior of the local process
$$ Z_n(h, \hat{\theta}_n) = v_n \left[\M_n\left(d_n + \frac{h}{s_n}, \hat{\theta}_n\right) - \M_n\left(d_n ,\hat{\theta}_n\right)\right]$$
for a properly chosen $v_n$. Note that  
$$s_n (\hat{d}_n - d_n)  = \argmin_{h: d_n + h/s_n \in \mathcal{D}_{\hat{\theta}_n}} Z_n(h, \hat{\theta}_n).$$ Note that $Z_n$ can be defined in such a manner so that the right hand side is the minimizer of $Z_n$ over the entire domain. To see this, let $\mathcal{D}_{\hat{\theta}_n} = [a_n(\hat{\theta}_n), b_n(\hat{\theta}_n)]$, say (in one dimension). 
If we extend the definition of $Z_n$ to the entire line by defining 
\begin{equation} \label{zextn}
Z_n(h,\hat{\theta}_n) =\left\{ 
  \begin{array}{l l}
     Z_n(s_n(b_n(\hat{\theta}_n) - d_n)) & \quad \mbox{ for $h > s_n(b_n(\hat{\theta}_n) - d_n)$ and }\\
     Z_n(s_n(a_n(\hat{\theta}_n) - d_n))  & \quad  \mbox{ for $h < s_n(a_n(\hat{\theta}_n) - d_n)$,} 
  \end{array} \right.
\end{equation}
then, clearly: 
\[ s_n(\hat{d}_n - d_n) = \argmin_{\mathbb{R}}\,Z_n(h,\hat{\theta}_n) \,.\] 
In $p$ dimensions, define $Z_n$ outside of the actual domain, the translated $\hat{D}_{\hat{\theta}_n}$, to be the supremum of the process $Z_n$ on its actual domain. Then the infimum of $Z_n$ over the entire space is also the infimum over the actual domain. 
Such an extension then allows us to apply the argmin continuous mapping theorem \cite[Theorem 2.7]{KP90} to arrive at the limiting distribution of $s_n (\hat{d}_n - d_n) $.  


In our examples and numerous others, $Z_n$ can be expressed as an empirical process acting on a class of functions changing with $n$, indexed by the parameter $h$ over which the argmax/argmin functional is applied and by the parameter $\theta$ which gets estimated from the first stage data, e.g.,
\be \label{eq:mnform}
Z_n(h, \theta)  = \frac{1}{\sqrt{n}} \sum_{i=1}^n f_{n, h,\theta}(V_i) = \G_n f_{n,h, \theta} + \zeta_n(h, \theta). 
\ee
Here, $V_i \sim P$ are i.i.d. random vectors, $\G_n = \sqrt{n}(\P_n - P)$ and $ \zeta_n(h, \theta) = \sqrt{n}P f_{n,h, \theta} $ with $\P_n$ denoting the empirical measure induced by $V_i$s.  The parameter $\theta$ could be multi-dimensional and would account for the nuisance/design parameters which are estimated from the first stage sample. The term $\sqrt{n}P f_{n, h,\theta}$ typically contributes to the drift of the limiting process.  
We first provide sufficient conditions for tightness of the centered $Z_n (h, \hat{\theta}_n)$ and then deal with its limit distribution. 
\begin{thm} \label{th:tight}
Let $\hat{\theta}_n$ be a random variable taking values in $\Theta$ which is independent of the process $Z_n$ defined in \eqref{eq:mnform}. As in Theorem \ref{th:ratethmc6}, let there exist a (non-random) set $\Theta_n^\tau \subset \Theta$ such that $P[\hat{\theta}_n \notin \Theta_n^\tau] < \tau$, for any fixed $\tau>0$. For each $\theta \in \Theta$, let $\mathcal{F}_{n,\theta}  = \{ f_{n,h, \theta}: h \in \mathcal{H} \}$ with measurable envelopes $F_{n,\theta}$. Let $\mathcal{H}$ be totally bounded with respect to a semimetric $\tilde{\rho}$. Assume that for each $\tau, \eta >0$ and every $\delta_n \rightarrow 0$, 
\begin{eqnarray}
\sup_{\theta \in \Theta_n^\tau} P F_{n, \theta}^2& = & O(1), \label{eq:cond1}\\
\sup_{\theta \in \Theta_n^\tau} P F^2_{n, \theta} 1\left[F_{n, \theta} > \eta \sqrt{n}\right]  & \rightarrow & 0  \label{eq:cond2} \\
\sup_{\substack{\theta \in \Theta_n^\tau \\ \tilde{\rho} (h_1,h_2) < \delta_n }} P (f_{n,h_1, \theta} - f_{n,h_2, \theta})^2 & \rightarrow & 0  \mbox{ and } \label{eq:cond3} \\
	\sup_{\substack{\theta \in \Theta_n^\tau \\ \tilde{\rho} (h_1,h_2) < \delta_n }} | \zeta_n(h_1,\theta) - \zeta_n(h_2,\theta)|  & \rightarrow &  0. \label{eq:cond4}
\end{eqnarray}
Assume that, for $\delta>0$, $\mathcal{F}_{n, \delta} = \{ f_{n,h_1,\hat{\theta}}- f_{n,h_2,\hat{\theta}}: \tilde{\rho}(h_1,h_2) < \delta\}$ is {\it suitably measurable} (explained below), for each $\theta \in \Theta_n^\tau$, $\mathcal{F}^2_{n,\theta,\delta} = \{ (f_{n,h_1,{\theta}}- f_{n,h_2,{\theta}})^2: \tilde{\rho}(h_1,h_2) < \delta\} $ is $P$-measurable,  and 
\be
\sup_{\theta \in \Theta_{n}^\tau} \int_0^{\infty} \sup_{Q}\sqrt{\log N \left(u \|F_{n,\theta} \|_{L_2(Q)}, \mathcal{F}_{n,\theta}, L_2(Q)\right)} du = O(1) \label{eq:cond5}
\ee
or \be
\sup_{\theta \in \Theta_{n}^\tau} \int_0^{\infty} \sqrt{\log N_{[\;]} \left(u \|F_{n,\theta} \|_{L_2(P)}, \mathcal{F}_{n,\theta}, L_2(P)\right)}du = O(1)
\ee
Then, the sequence $\{Z_n(h, \hat{\theta}_n): h \in \mathcal{H}   \} $ is asymptotically tight in $l^\infty(\mathcal{H})$. Here, $N_{[\;]}()$ and $N()$ denote the bracketing and covering numbers respectively and the supremum in \eqref{eq:cond5} is taken over all discrete probability measures $Q$. 
\end{thm}
The measurability required for the class $\mathcal{F}_{n, \delta}$ is in the following sense.
For any vector $\{ e_1, \ldots, e_n \} \in \{ -1, 1\}^n$, the map
\be
(V_1, V_2, \ldots, V_n, \hat{\theta}, e_1,\ldots,e_n) \mapsto \sup_{g_{n,\hat{\theta}} \in \mathcal{F}_{n, \delta}} \left| \frac{1}{\sqrt{n}} \sum_{i=1}^n e_i g_{n,\hat{\theta}}(V_i) \right| 
\label{eq:condmes}
\ee 
is assumed to be jointly measurable. This is very much in the spirit of the $P$-measurability assumption made for Donsker results involving covering numbers (e.g., \citet[Theorem 2.5.2]{VW96}) and can be justified readily in many applications. 
We prove the above result assuming \eqref{eq:cond5}. The broad brushstrokes of the proof rely on symmetrization by Rademacher random variables and the resulting sub-Gaussianity of the symmetrized processes (conditional on the data), followed by chaining arguments, and control of the resulting covering entropy bounds. While this general approach arises in the proofs of standard Donsker theorems under bounded  uniform entropy integral 
conditions, the arguments are considerably more delicate in this case, since the random $\hat{\theta}_n$ sits in the second co-ordinate of the parameters indexing the empirical process. 

The form of the limit process, which may depend on the weak limit of the first stage estimates, can be derived using the following lemma.
\begin{lem}\label{lm:fidiconvgen}
For a generic $\theta$, let $\Delta_\theta = n^\nu({\theta} - \theta_n)$. Consider the setup of Theorem \ref{th:tight}. Additionally, assume that
\begin{enumerate}
	\item  ${\Delta}_{\hat{\theta}_n} = n^\nu(\hat{\theta}_n - \theta_n)$ converges in distribution to a random vector $\xi$.
	\item For any $\tau > 0$, the covariance function 
\be
\begin{split}
	 C_n (h_1, h_2, \Delta_\theta) & = P f_{n,h_1, \theta_n + n^{-\nu} \Delta_\theta}f_{n,h_2,\theta_n + n^{-\nu} \Delta_\theta} \\
	 & - P f_{n,h_1, \theta_n + n^{-\nu} \Delta_\theta} Pf_{n,h_2, \theta_n + n^{-\nu} \Delta_\theta} \nonumber
	 \end{split}
	 \ee 
	 converges pointwise to $C(h_1, h_2, \Delta_\theta)$ on $\mathcal{H} \times \mathcal{H}$, uniformly in $\Delta_\theta$, $\theta \in \Theta_n^\tau$.
	\item For any $\tau > 0$, the functions $\zeta_n(h,\theta_n + n^{-\nu} \Delta_\theta)  $ converges pointwise to a function $\zeta( h, \Delta_\theta)$ on $\mathcal{H}$, uniformly in $\Delta_\theta$, $\theta \in \Theta_n^\tau$. 
	\item  The limiting functions $C(h_1, h_2, \Delta_\theta)$ and $\zeta(h, \Delta_\theta)$ are continuous in $\Delta_\theta$. 
\end{enumerate}
Let $Z(h, \xi)$ be a stochastic process constructed in the following manner. 
For a particular realization $\xi_0$ of $\xi $, generate a Gaussian process $Z(h, \xi_0)$ (independent of $\xi$)  with drift $\zeta(\cdot, \xi_0)$ and covariance kernel $C(\cdot, \cdot, \xi_0)$. Then, the process $Z_n(\cdot, \hat{\theta}_n)$ converges weakly $Z(\cdot, \xi)$ in $\ell^\infty(\mathcal{H})$.
\end{lem}
The proof is given in Section \ref{pf:fidiconv} of the Supplement. For notational ease, we assumed each element of the vector $\hat{\theta}_n$ converges at the same rate  ($n^\eta$). The extension to the general situation where different elements of $\hat{\theta}_n$ have different rates of convergence is not difficult.

In most of our examples, the second stage limit process does not depend on the behavior of the first stage estimate. This happens when the limits of $C_n$ and $\zeta_n$ in the above lemma are free of the third argument $\Delta_{\theta}$, in which case the following result holds. 
\begin{cor}\label{lm:fidiconv}
Consider the setup of Theorem \ref{th:tight}. Additionally, assume that for any $\tau > 0$,
\begin{enumerate}
	\item  The covariance function 
\be
	 C_n (h_1, h_2, \theta)  = P f_{n,h_1, \theta}f_{n,h_2,\theta}  - P f_{n,h_1, \theta} Pf_{n,h_2, \theta} \nonumber
	 \ee 
	 converges pointwise to $C(h_1, h_2)$ on $\mathcal{H} \times \mathcal{H}$, uniformly in $\theta$, $\theta\in \Theta_n^\tau$.
	\item  The functions $\zeta_n(h,\theta)  $ converges pointwise to a function $\zeta( h)$ on $\mathcal{H}$, uniformly in $\theta$, $\theta \in \Theta_n^\tau$. 
\end{enumerate}
Let $Z(h)$ be a Gaussian process with drift $\zeta(\cdot)$ and covariance kernel $C(\cdot, \cdot)$. Then, the process $Z_n(\cdot, \hat{\theta}_n)$ converges weakly to $Z(\cdot)$ in $\ell^\infty(\mathcal{H})$.
\end{cor}
\begin{rmk}
The asymptotic dependence of the second stage processes on the limit of the first stage process, alluded to above, does appear in connection with certain curious aspects of the mode estimation problem considered in Section 6. See Theorem \ref{th:twostagenf} and its proof. 
\end{rmk}
In our applications, the process $Z_n(h,\hat{\theta}_n)$ is defined for $h$ in a Euclidean space, say $\tilde{\mathcal{H}} = \mathbb{R}^p$ and Theorem \ref{th:ratethmc6} is used to show that $\hat{h}_n :=s_n(\hat{d}_n - d_n)$, which assumes values in $\tilde{\mathcal{H}}$, is $O_p(1)$. The process $Z_n$ is viewed as living in $\mathcal{B}_{loc}(\mathbb{R}^p) =\{ f : \R^p \mapsto \R: f $ is bounded on $[-T, T]^p$ for any $T >0 \}$, the space of locally bounded functions on $\mathbb{R}^p$. 

To deduce the limit distribution of $\hat{h}_n$, we first show that for a process $Z(h, \xi)$ in 
$C_{min}(\mathbb{R}^p) = \{ f\in \mathcal{B}_{loc}(\mathbb{R}^p): f $ possesses a unique minimum and $f(x) \rightarrow \infty$ as $\|x\| \rightarrow \infty \}$, the process $Z_n(h,\hat{\theta}_n)$ converges to $Z(h, \xi)$ in $\mathcal{B}_{loc}(\mathbb{R}^p)$. This is accomplished by showing that on every $[-T,T]^p$, $Z_n(h,\hat{\theta}_n)$ 
converges to $Z(h, \xi)$ on $\ell^{\infty}([-T,T]^p)$, using Theorem \ref{th:tight} and Lemma \ref{lm:fidiconvgen}. 
An application of the argmin continuous mapping theorem (Theorem 2.7) of \citet{KP90} now yields the desired result, i.e., $\hat{h}_n \stackrel{d}{\rightarrow} \argmin_{h \in \R^p} Z(h, \xi).$

Next, based on our discussion above, we provide a road-map for establishing key results in multi-stage problems.
\begin{enumerate}
\item[I]{\it Rate of convergence.}
	\item With $\hat{\theta}_n$ denoting the first stage estimate, identify the second stage criterion as a bivariate function $\M_n(d, \hat{\theta}_n)$ and its population equivalent 	$M_n(d,\hat{\theta}_n)$. A useful choice for $M_n$ is $M_n( d, \theta) = E\left[\M_n(d, \theta)\right]$. The non-random process $M_n$ is at its minimum at $d_n$ which either equals the parameter of interest $d_0$ or is asymptotically close to it.
	\item Arrive at $\rho_n(d, d_n)$ using \eqref{eq:cond00} which typically involves a second order Taylor expansion when $M_n$ is smooth (Section 3 deals with a non-smooth case). The distance $\rho_n$ is typically some function of the Euclidean metric. 
	\item Justify the convergence $P\left(\rho_n(\hat{d}_n, d_n) \geq \kappa_n\right)$ to zero using Lemma \ref{lm:const}, if needed and derive a bound on the modulus of continuity as in \eqref{eq:cond01}.  This typically requires VC or bracketing arguments such as Theorem 2.14.1 of \citet{VW96}. With suitably selected $K_\tau$, $\Theta_n^\tau$ can be chosen to be shrinking sets of type $[\theta_n - K_\tau/n^{\nu},\theta_n + K_\tau/n^{\nu}]$, when a result of the type $n^\nu(\hat{\theta}_n - \theta_n) = O_p(1)$ holds. Such choices typically yield efficient bounds for \eqref{eq:cond01}.
		\item Derive the rate of convergence using Theorem \ref{th:ratethmc6}.

\item[II]{\it Limit Distribution.}
		\item Express the local process $Z_n$ as an empirical process acting on a class of functions and a drift term \eqref{eq:mnform}.
		\item Use Theorem \ref{th:tight} and Lemma \ref{lm:fidiconvgen} or Corollary \ref{lm:fidiconv} to derive the limit process $Z$ and apply argmin continuous mapping to derive the limiting distribution of $\hat{d}_n$.
\end{enumerate}

\begin{rmk}
Note that our results are also relevant to situations where certain extra/nuisance parameters are estimated from separate data and argmax/argmin functionals of the empirical process acting on functions involving these estimated parameters are considered. We note here that \cite{VW07} considered similar problems where they provided sufficient conditions for replacing such estimated parameters by their true values, in the sense that $\sup_{d \in \mathcal{D}} \left|\G_n(f_{d, \hat{\theta}} - f_{d, {\theta}_0} )\right|$ converges in probability to zero. Here, $\G_n =  \sqrt{n}(\P_n - P)$, with $\P_n$ denoting the empirical measure, $f_{d,\theta}$ are measurable functions indexed by $(d, \theta) \in \mathcal{D} \times \Theta$ and $\hat{\theta}$ denotes a suitable estimate of the nuisance parameter $\theta_0$. We show that while a result of the above form does not generally hold for our examples, (see Proposition \ref{lm:vwcont}), the final limit distribution can still have a form with estimated nuisance parameters replaced by their true values.
\end{rmk}

In the following sections, we illustrate the above results. Specifically, in Section
\ref{sec:cp} we study a variant of the change-point problem in a regression function, presented in \cite{LBM09}.
While in that paper the signal at the change-point $d_0$ was assumed to be constant, in this study
it is assumed to decrease as a function of the sample size $n$. The change from a constant to a decreasing signal-to-noise ratio has telling consequences for the asymptotic behavior of the least squares estimate of the change-point as will be seen shortly, since the limiting process 
changes from Poisson in the former to Gaussian in the latter. For details, see Section \ref{sec:cp} and also the discussion in Section \ref{disco}. 
Moreover, this model represents a canonical
example for illustrating the results and the techniques established above. Our second illustration, presented in Section \ref{sec:inviso}, 
rigorously establishes asymptotic results for 
the two-stage isotonic regression estimator empirically studied in \cite{TANG13}.
The third example, presented in Section \ref{sec:classy}, 
examines a flexible classifier, where the adaptive sampling design shares strong similarities with {\em active learning} procedures.
Our final example in Section \ref{sec:mode} addresses the problem of mode estimation in a fully nonparametric fashion,
unlike the parametric second-stage procedure employed in \cite{BGZ13}.


\section{Change-point model with fainting signal}\label{sec:cp}
We consider a change-point model of the form $Y = m_n(X) + \epsilon$, 
where $$m_n(x) = \alpha_n 1[x \leq d_0] + \beta_n 1[x > d_0] $$ for an unknown $d_0 \in (0,1)$ and $\beta_n - \alpha_n = c_0 n^{-\xi}$, $c_0 > 0$ and $\xi < 1/2$. The  errors $\epsilon$ are independent of $X$ and have mean 0 and variance $\sigma^2$. In contrast with the change-point model considered in \cite{LBM09}, the signal in the model $\beta_n -\alpha_n$ decreases with $n$. A similar model with decreasing signal was studied in \cite{MlrSng97}.  We assume that the experimenter has the freedom to choose the design points to sample and budget (of size) $n$ at their disposal. We apply the following two-stage approach.
\begin{enumerate}
	\item At stage one, sample $n_1 = p n$ covariate values, ($p \in (0,1)$),  from a uniform design on $ \mathcal{D} = [0, 1]$ and, from the obtained data, $\{(Y_i^{(1)}, X_i^{(1)})\}_{i=1}^{n_1}$, estimate $\alpha_n,$ $\beta_n$ and  $d_0$ by
\begin{eqnarray}
\lefteqn{\hat{\theta}_{n_1}  = \left(\hat{\alpha}, \hat{\beta}, \hat{d}_1\right)} \nonumber\\
& = & \argmin_{\alpha, \beta, d} \sum_{i=1}^{n_1}\left[( Y_i^{(1)} - \alpha)^2 1\left[X_i^{(1)} \leq d\right] + (Y_i^{(1)} - \beta)^2 1\left[X_i^{(1)} > d\right]\right]. \nonumber
\end{eqnarray}
These are simply the least squares estimates. 
	\item For $K>0$ and $\gamma >0$, sample the remaining $n_2 = (1-p)n $ covariate-response pairs $\{\left( Y_i^{(2)}, X_i^{(2)} \right)\}_{i=1}^{n_2}$, where
	$$Y_i^{(2)} = \alpha_n 1[X^{(2)}_i \leq d_0] + \beta_n 1[X^{(2)}_i > d_0] + \epsilon_i $$ and 
	$X_i^{(2)}$'s are sampled uniformly from the interval $\mathcal{D}_{\hat{\theta}_{n_1}} = [\hat{d}_1 - K {n_1}^{-\gamma}, 
\hat{d}_1 + K {n_1}^{-\gamma}]$. The $X_i^{(2)}$'s are viewed as arising from $n$ i.i.d. Uniform$[-1,1]$ random variables $\{U_i\}_{i=1}^{n_2}$:  specifically, $X_i^{(2)} := \hat{d}_1 + U_i\,K\,n_1^{-\gamma}$, with the $\{U_i\}_{i=1}^{n_2}$ being independent of the i.i.d. sequence of errors 
$\{\epsilon_i\}_{i=1}^{n_2}$, and both $U$'s and $\epsilon$'s are independent of the first stage data. 
Obtain an updated estimate of $d_0$ by
\be \label{eq:argcp}
\hat{d}_2 = \argmin_{d \in \mathcal{D}_{\hat{\theta}_n}} \sum_{i=1}^{n_2} \left[(Y_i^{(2)} - \hat{\alpha})^2 1\left[X_i^{(2)} \leq d\right] + (Y_i^{(2)} - \hat{\beta})^2 1\left[X_i^{(2)} > d\right]\right].
\ee
\end{enumerate}
Here, $\gamma$ is chosen such that $P\left(d_0 \in [\hat{d}_1 - K {n_1}^{-\gamma}, \hat{d}_1+ K {n_1}^{-\gamma}]\right)$ converges to 1. 
Intuitively, this condition compels the second stage design interval to contain $d_0$ with high probability. This is needed as the objective function relies on the dichotomous behavior of the regression function on either side of $d_0$ for estimating the change-point. If the second stage interval does not include $d_0$ (with high probability), the stretch of the regression function, $m_n$,  observed (with noise) is simply flat, thus failing to provide information about $d_0$.

In \cite{BB76} and \cite{Btchrya87}, similar models were studied in a one-stage fixed design setting. By a minor extension of their results, it can be shown that ${n_1}^\nu (\hat{d}_1 - d_0) = O_p(1)$ for $\nu = 1 - 2 \xi$,  $\sqrt{n_1}(\hat{\alpha}- \alpha_n) = O_p(1)$ and  $\sqrt{n_1}(\hat{\beta}- \beta_n) = O_p(1)$. Hence, any choice of $\gamma < \nu$ suffices. 

For simplicity, we assume that the experimenter works with a uniform random design at both stages.
An extension to designs with absolutely continuous positive densities supported on an interval is straightforward. 

The expression in \eqref{eq:argcp} can be simplified to yield
\be \label{eq:dhatdef}
\hat{d}_2 = \argmin_{d \in \mathcal{D}_{\hat{\theta}_{n_1}}} {\M}_{n_2}(d, \hat{\theta}_{n_1}) 
\ee
where for $\theta = (\alpha, \beta, \mu) \in \R^3$, 
\begin{eqnarray*}
{\M}_{n_2}(d, {\theta}) & = &\frac{\mbox{sgn}(\beta- \alpha)}{n_2}\sum_{i=1}^{n_2}\left(Y_i^{(2)} - \frac{\alpha + \beta}{2}\right) \left(1\left[X_i^{(2)} \leq d\right] - 1\left[X_i^{(2)} \leq d_0\right]\right)  \nonumber 
\end{eqnarray*}
with $X_i^{(2)} \sim \mbox{Uniform}[\mu- K {n_1}^{-\gamma}, \mu + K {n_1}^{-\gamma}]$, $\hat{\theta}_{n_1} = (\hat{\alpha}, \hat{\beta}, \hat{d}_1)$ and sgn denoting the sign function. 
We take $M_{n_2}(d, \theta) = E \left[\M_{n_2}(d, \theta)\right]$ to apply Theorem \ref{th:ratethmc6}, which yields the following result on the rate of convergence of $\hat{d}_2$.
\begin{thm}\label{th:ratecp}
For $\hat{d}_2$ defined in \eqref{eq:dhatdef} and $\eta =1 + \gamma - 2 \xi$
$$ n^\eta (\hat{d}_2 - d_0)  = O_p(1). $$
\end{thm}
The proof, which is an application of Theorem \ref{th:ratethmc6}, illustrates the typical challenges involved in verifying its conditions and is given in Section \ref{pf:ratecp}.  
\newline
\newline
To deduce the limit distribution of $\hat{d}_2$, consider the process
{\small \be \label{eq:localcp}
Z_{n_2}(h, \theta) = \frac{1}{n_2^\xi} \sum_{i=1}^{n_2} \left(Y_i^{(2)} - \frac{\alpha + \beta}{2}\right) \left(1\left[X_i^{(2)} \leq  d_0 + hn^{-\eta}\right] - 1\left[X_i^{(2)} \leq  d_0 \right]\right) 
\ee}
with   $X_i^{(2)} \sim \mbox{Uniform}[\mu- K {n_1}^{-\gamma}, \mu + K {n_1}^{-\gamma}]$.
Note that $n^\eta (\hat{d}_2 - d_0) = \argmin_h Z_{n_2}(h, \hat{\theta}).$ Letting $V = (U,\epsilon)$ denote a generic $(U_i, \epsilon_i)$, it is convenient to write $Z_{n_2}$ as 
\be
Z_{n_2}(h, \theta) = \G_{n_2} f_{n_2, h, \theta}(V) + \zeta_{n_2}(h,\theta),
\ee
where  $\zeta_{n_2}(h,\theta) = \sqrt{n_2} P f_{n_2, h, \theta}(V)$  and 
{\begin{eqnarray*}
f_{n_2, h, \theta} (V) & =&  n_2^{1/2- \xi} \left(m_n(\mu + U K n_1^{-\gamma}) + \epsilon - \frac{\alpha+ \beta}{2}\right) \times\\
& & \left(1\left[\mu + U K n_1^{-\gamma} \leq  d_0 + hn^{-\eta}\right] - 1\left[\mu + U K n_1^{-\gamma} \leq  d_0 \right]\right) . \nonumber
\end{eqnarray*}}
%
%
This is precisely the form of the local process needed for Theorem \ref{th:tight}.  We next use it to deduce the weak limit of the process $Z_{n_2}(h, \hat{\theta})$.
\begin{thm} \label{th:limproccp}
Let $B$ be a standard Brownian motion on $\R$ and
$$Z(h) = \sqrt{\frac{(1-p)^{1-2\xi} p^\gamma}{2K}}\sigma B(h) + \frac{(1-p)^{1-\xi} p^\gamma}{2K} \frac{c_0}{2} |h|.$$
Then, the sequence of stochastic process $Z_{n_2}(h),$ $h \in \R$ are asymptotically tight and converge weakly to the process $Z(h)$. 
\end{thm}
The proof, which uses Theorem \ref{th:tight} and Lemma \ref{lm:fidiconv}, is provided in Section \ref{pf:limproccp}.
\newline
\newline
{\it Comparison with results from \cite{VW07}.}
As mentioned earlier, \cite{VW07} derived sufficient conditions to prove results of the form 
$\sup_{d \in \mathcal{D}} \left|\G_n(f_{d, \hat{\theta}} - f_{d, {\theta}_0} )\right| \stackrel{p}{\rightarrow} 0$, where $\{ f_{d, \theta}: d \in \mathcal{D}, \theta \in \Theta\}$ is a suitable class of measurable functions and $\hat{\theta}$ is a consistent estimate of ${\theta}_0$. If such a result were to hold in the above model, the derivation of the limit process would boil down to working with the process $\{ \G_n f_{d, {\theta}_0}: d \in \mathcal{D} \} $, which is much simpler to work with. 
However, we show below that for $h\neq 0$,
\be \label{eq:notconv}
T_{n_2} :=  (Z_{n_2}(h, \alpha_n, \beta_n, \hat{d}_1) - Z_{n_2}(h,{\alpha}_n, {\beta}_n, {d}_0)) 
\ee
{\it does not converge in probability to zero,} let alone the supremum of the above over $h$ in compact sets and hence, the results in \cite{VW07} do not apply. Similar phenomena can be shown to hold for the examples we consider in later sections. 
\begin{prop} \label{lm:vwcont}
Let $\pi_0^2 : = \sigma^2 p^\gamma (1-p)^{1-2\xi} {|h|}/{K}$ and $T_{n_2}$ be as defined in \eqref{eq:notconv}.
Then, for $h \neq 0$, $T_{n_2}$ converges to a normal distribution with mean 0 and variance $\pi_0^2$.
\end{prop}
The proof is given in Section \ref{pf:vwcont} of the Supplement.  
We now provide the limiting distribution of  $\hat{d}_2$.
\begin{thm}\label{limitcp}
The process $Z$ possesses a unique tight argmin almost surely and  for $\lambda_0   = (8 K \sigma^2 )/ (c_0^2  (1-p) p^\gamma)$,
$$n^\eta(\hat{d}_2 - d_0) \stackrel{d}{\rightarrow} \argmin_h Z(h) \stackrel{d}{=} \lambda_0 \argmin_{v} \left[ B(v) + |v| \right].$$
\end{thm}
\begin{rmk}
We considered a uniform random design for sampling at both stages. The results extend readily to other suitable designs. For example, if the second stage design points are sampled as $X_i^{(2)} = \hat{d}_1 + V_i K n_1^{-\gamma}$, where $V_i$'s are i.i.d. realizations from a distribution with a (general) positive continuous density $\psi$ supported on $[-1,1]$, it can be shown that $\hat{d}_2$ attains the same rate of convergence. The limit distribution has the same form as above with $\lambda_0$ replaced by $\lambda_0/(2\,\psi(0))$. 
\end{rmk}
The proof is given in Section \ref{pf:limitcp}. 
\newline
\newline
{\it Optimal allocation.} The interval from which the covariates are sampled at the second stage is chosen such that the change-point $d_0$ would be contained in the prescribed interval with high probability, i.e., we pick $K$ and $\gamma$ such that 
$P\left(d_0 \in [\hat{d}_1 - K {n_1}^{-\gamma}, \hat{d}_1+ K {n_1}^{-\gamma}]\right)$ converges to 1. But, \emph{in practice} for a fixed $n$, a suitable choice would be 
$$ K {n_1}^{-\gamma}   \approx \frac{C_{\tau/2}}{n_1^{1- 2\xi}}$$
for a small $\tau$, with $C_{\tau/2}$ being the $(1- \tau/2)$th quantile of the limiting distribution of $n_1^{1- 2\xi}(\hat{d}_1 - d_0)$ which is symmetric around zero. 
As $\argmin_{v} \left[ B(v) + |v| \right]$ is a symmetric random variable, the variance of $(\hat{d}_2 - d_0)$ would then be (approximately) smallest when 
\begin{eqnarray*}
\frac{\lambda_0}{n^{\eta}} & =& \frac{8K \sigma^2  }{ c_0^2 (1-p) p^\gamma n^\eta}  = \frac{8\sigma^2  C_{\tau/2} }{ c_0^2 (1-p) p^\gamma n^\eta n_1^{1 - \gamma - 2\xi}}\\
& = &\frac{8\sigma^2  C_{\tau/2} }{ c_0^2 (1-p) p^{1- 2 \xi} n^{2(1- 2 \xi)}}
\end{eqnarray*}
is at its minimum. This yields the optimal choice of $p$ to be $p_{opt} = (1- 2\xi)/(2(1 - \xi))$.

\section{Inverse isotonic regression}\label{sec:inviso}
In this section, we consider the problem of estimating the \emph{inverse} of a monotone regression function at a pre-specified point $t_0$ using multi-stage procedures. Responses $(Y,X)$ are  obtained from a model of the form $Y = r(X) + \epsilon$, where $r$ is a monotone function on [0,1] and the experimenter has the freedom to choose the design points. It is of interest to estimate the threshold $d_0 = r^{-1}(t_0)$ for some $t_0$ in the interior of the range of $r$ with $r'(d_0)>0$. 
\newline
The estimation procedure is summarized below: First, sample $n_1 = p \times n$ covariate values uniformly from $[0, 1]$ and obtain the corresponding responses. From the data, $\{(Y_i^{(1)}, X_i^{(1)})\}_{i=1}^{n_1}$, obtain the isotonic regression estimate $\hat{r}_{n_1}$ of $r$ (see \citet[Chapter 1]{RWD88}) and, subsequently, an estimate $\hat{d}_1 = \hat{r}^{-1}_{n_1} (t_0)$ of $d_0$.  Sample the remaining $n_2 = (1-p)n$ covariate-response pairs $\{ (Y_i^{(2)}, X_i^{(2)}) \}_{i=1}^{n_2}$, in the same way as in Step 2 of the two-stage approach in Section \ref{sec:cp}, but now $\gamma < 1/3$ and $Y_i^{(2)} = r(X_i^{(2)})+ \epsilon_i^{(2)}$.
Obtain an updated estimate $\hat{d}_2 = \hat{r}_{n_2}^{-1} (t_0) $ of $d_0$, $\hat{r}_{n_2}$ being the isotonic regression estimate based on $\{ Y_i^{(2)}, X_i^{(2)} \}_{i\leq n_2}$, and $\hat{r}_{n_2}^{-1} $ the right continuous inverse of $\hat{r}_{n_2}$. 
\newline
In this study, we rigorously establish the limiting properties of $\hat{d}_2$.
The parameter $\gamma$ is chosen such that $P\left(d_0 \in [\hat{d}_1 - K n_1^{-\gamma}, \hat{d}_1+ K n_1^{-\gamma}]\right)$ converges to 1. As $n_1^{1/3} (\hat{d}_1 - d_0) = O_p(1)$ (see, for example, \citet[Theorem 2.1]{TBM11}), any choice of $\gamma < 1/3$ suffices. 

The \emph{switching relationship} \citep{Gr85,Gr89} is useful in studying the limiting behavior of $\hat{r}_{n_2}$ through M-estimation theory. It simply relates the estimator $\hat{r}_{n_2}$ to the minima of a tractable process as follows. Let
$$ V^0(x) = \frac{1}{n_2} \sum_{i=1}^{n_2} Y^{(2)}_i 1\left[X^{(2)}_i \leq x\right] \mbox{ and } G^{0}(x) = \frac{1}{n_2} \sum_{i=1}^{n_2} 1\left[X^{(2)}_i \leq x\right]. $$
For  $\hat{\theta}_{n_1} = \hat{d}_1$ and any $d \in [\hat{\theta}_{n_1} - K n_1^{-\gamma}, \hat{\theta}_{n_1} + K n_1^{-\gamma}]$, the following (switching) relation holds with probability one:
\be \label{eq:switch}
\hat{r}_{n_2}(d) \leq t \,  \Leftrightarrow \, \argmin_{x \in [\hat{\theta}_{n_1} - K n_1^{-\gamma}, \hat{\theta}_{n_1} + K n_1^{-\gamma}]} \{V^0(x) - t G^0(x) \} \geq X^{(2)}_{(d)},
\ee
where $X^{(2)}_{(d)}$ is the last covariate value $X_i^{(2)}$ to the left of $d$ and the argmin denotes the smallest minimizer (if there are several).  As $\hat{r}_{n_2}^{-1}$ is the right continuous inverse
 of $\hat{r}_{n_2}$, $ \hat{r}_{n_2}(d) \leq t  \Leftrightarrow d \leq   \hat{r}_{n_2}^{-1}(t)$ and hence, using \eqref{eq:switch} at $t = t_0 = r(d_0)$, we get
\be \label{eq:switch2}
\hat{d}_2 = \hat{r}_{n_2}^{-1}(t_0) \geq d \,  \Leftrightarrow \, \argmin_{x \in [\hat{\theta}_{n_1} - K n_1^{-\gamma}, \hat{\theta}_{n_1} + K n_1^{-\gamma}]} \{V^0(x) - r(d_0) G^0(x) \} \geq X^{(2)}_{(d)}. 
\ee
Let
$$ \hat{x}=\argmin_{x \in [\hat{\theta}_{n_1} - K n_1^{-\gamma}, \hat{\theta}_{n_1} +  K n_1^{-\gamma}]} \{V^0(x) - r(d_0) G^0(x) \}.$$
Note that both $\hat{x}$ and $\hat{d}_2$ are order statistics of $X$ (since $\hat{r}_{n_2}(\cdot)$ and $V^0(\cdot) - r(d_0) G^0(\cdot)$ are piecewise constant functions). In fact, it can be shown using \eqref{eq:switch2} twice (once at $d= \hat{d}_2$ and the second time with $d$ being the order statistic to the immediate right of $\hat{d}_2$) that they are consecutive order statistics with probability one. Hence, 
\be \label{eq:switchrel}
\hat{d}_2 =\hat{x}+ O_p\left((2Kn_1^{-\gamma})\frac{\log n_2}{n_2}\right) = \hat{x}+ O_p\left(\frac{\log n}{n^{1+\gamma}}\right).
\ee
The $O_p$ term in the above display corresponds to the order of the maximum of the differences between consecutive order statistics (from $n_2$ realizations from a uniform distribution on an interval of length $2K n_1^{-\gamma}$). 
We will later show that $ n^{(1+\gamma)/3} (\hat{x} - d_0) = O_p(1)$. As $n^{(1+\gamma)/3} = o(n^{1+\gamma}/ \log n)$, it suffices to study the limiting behavior of $\hat{x}$ to arrive at the asymptotic distribution of $\hat{d}_2$.
To this end, we start with an investigation of a version of the process $\{V^0(x) - r(d_0) G^0(x) \}$ at the resolution of the second stage ``zoomed-in'' neighborhood, given by
$$\mathbb{V}_{n_2}(u) = \P_{n_2} (Y^{(2)} - r(d_0))1\left[X^{(2)} \leq d_0 + u n_2^{-\gamma}\right].$$
For $\mathcal{D}_{\hat{\theta}_{n_1}} = \left[n_2^{\gamma}(\hat{\theta}_{n_1} - K n_1^{-\gamma}), n_2^{\gamma}(\hat{\theta}_{n_1} + K n_1^{-\gamma})\right]$,
\be
\hat{u} := n_2^\gamma(\hat{x} - d_0) = \argmin_{u \in \mathcal{D}_{\hat{\theta}_{n_1}}} \mathbb{V}_{n_2}(u). \nonumber
\ee
Further, let $U \sim \mbox{Uniform}[-1,1]$ and $V = (U,\epsilon)$. Note that $X^{(2)} = \hat{\theta}_{n_1} + U K n_1^{-\gamma}$ and $Y^{(2)} = r(\hat{\theta}_{n_1} + U K n_1^{-\gamma}) + \epsilon$. Let
%
\begin{eqnarray*}
g_{n_2,u,\theta} (V) 
& = & n_2^\gamma\left(r(\theta + U K n_1^{-\gamma} ) + \epsilon - r(d_0) \right) \times\\
& &\left(1\left[\theta + U Kn_1^{-\gamma} \leq d_0 + u n_2^{-\gamma }\right] - 1\left[\theta + U K n_1^{-\gamma} \leq d_0 \right]\right). \nonumber
\end{eqnarray*}
Also, let
\be
\M_{n_2}\left(u, \theta \right) = \P_{n_2} \left[g_{n_2,u, \theta}(V)\right].\nonumber
\ee 
Then, $\hat{u} = \argmin_{u \in \mathcal{D}_{\hat{\theta}_{n_1}}} \M_{n_2}\left(u, \hat{\theta}_{n_1}\right)$. 
Let $M_{n_2}(u, \theta) = P {g}_{n_2,u,\theta}$ which, by monotonicity of $r$, is non-negative. Also, let $\theta_0 = d_0$ and $\Theta_{n_1}^\tau = \{ \theta: |\theta -\theta_0| \leq K_\tau n_1^{-1/3} \}$ where $ K_\tau $ is chosen such that $ P\left(\hat{\theta}_{n_1} \in \Theta_{n_1}^\tau \right) > 1- \tau$ for $\tau>0$. As $\gamma < 1/3$, $0$  is contained in all the intervals $\mathcal{D}_\theta$, $\theta \in \Theta_{n_1}^\tau$ (equivalently, $d_0 \in [\theta - K n_1^{-\gamma}, \theta + K n_1^{-\gamma}]$), eventually.  Note that $M_{n_2}(0,\theta) = 0$. Hence, $0$ is a minimizer of $M_{n_2}(\cdot, \theta)$ over $\mathcal{D}_\theta$ for each $\theta \in \Theta_n^\tau$. The process $M_{n_2}$ is a population equivalent of $\M_{n_2}$ and hence, $\hat{u}$ estimates 0. We have the following result for the rate of convergence of $\hat{u}$.
\begin{thm}\label{th:rateu}
Assume that $r$ is continuously differentiable in a neighborhood of $d_0$ with $r'(d_0) \neq 0$. Then, for $\alpha = (1 - 2\gamma)/3$, $n_2^\alpha\hat{u} = O_p(1)$.
\end{thm}
The proof, which relies on Theorem \ref{th:ratethmc6} is given in Section \ref{pf:rateu} of the Supplement. 
Next, we derive the limiting distribution of $\hat{d}_2$ by studying the limiting behavior of $\hat{w} = n_2^{\alpha}\hat{u} = n_2^{(1+\gamma)/3}\,(\hat{x} - d_0)$.  
Let $f_{n_2,w,\theta} = {n_2}^{1/6 - 4\gamma/3}g_{n_2,w n_2^{-\alpha},\theta}$, $\zeta_{n_2}(w, \theta) = \sqrt{n_2} P f_{n_2,w,\theta}$ and 
\be
Z_{n_2}(w, {\theta}) = \G_{n_{2}} f_{n_2,w,\theta} + \zeta_{n_2}(w, \theta). \nonumber  
\ee
Then, $n_2^\alpha\hat{u} = \hat{w} = \argmin_{w: n_2^{-\alpha}w \in \mathcal{D}_{\hat{\theta}_{n_1}}} Z_{n_2}(w, \hat{\theta}_{n_1})$.
We have the following result for the weak convergence of $Z_{n_2}$.
\begin{thm}\label{th:limprociso}
Let $B$ be a standard Brownian motion on $\R$ and 
$$Z(w) = \sigma \sqrt{\frac{p^\gamma}{2K(1-p)^{\gamma}}}B(w) +  \left(\frac{p}{1-p}\right)^\gamma \frac{r'(d_0)}{4K} w^2 . $$
The processes $Z_{n_2}(w, \hat{\theta}_{n_1})$ are asymptotically tight and converge weakly to $Z$.
Further, 
\be
n^{(1+\gamma)/3} (\hat{d}_2 - d_0) \stackrel{d}{\rightarrow}\left(\frac{8 \sigma^2 K}{(r'(d_0))^2 p^\gamma (1-p)}\right)^{1/3}\argmin_w\{ B(w) + w^2 \}. \nonumber
\ee 
\end{thm}
The proof is given in Section \ref{pf:limprociso} of the Supplement where the first part of the theorem is established by an application of Theorem 2 and Corollary 1. Next, an application of an argmin continuous mapping theorem \cite[Theorem 2.7]{KP90} shows the limit distribution of $n_2^{(1+\gamma)/3}\,(\hat{x}_2 - d_0)$ to be that of the unique minimizer of $Z(h)$, which, along with \eqref{eq:switchrel} and rescaling arguments gives us the final result. 
\newline
Again, similar to the change-point problem, extensions of the above result to non-uniform random designs are possible as well. Also, the proportion $p$ can be optimally chosen (to be $1/4$) to minimize the limiting variance of the second stage estimate. More details on this and related implementation issues can be found in \citet[Section 2.4]{TANG13}.

\section{A classification problem} \label{sec:classy}
In this section, we study a non-parametric classification problem where we show that a multi-stage procedure yields a better classifier in the sense of approaching the misclassification rate of the Bayes classifier. 

Consider a model $ Y \sim Ber (r(X))$, where $r(x) =P\left(Y=1 \mid X =x\right)$ is a function on $[0,1]$ and the experimenter has freedom to choose the design distribution (distribution of $X$). Interest centers on using the training data $\{Y_i, X_i \}_{i=1}^n$ (obtained from a designed setting) to develop a classifier that predicts $Y$ at a given realization $X =x$. 
A classifier $f$ in this case is, simply, a function from $[0,1]$ to $\{0,1\}$ which provides a decision rule; assign $x$ to the class $f(x)$.  The misclassification rate or the risk $f$ with respect to \emph{test data}, $(\tilde{Y}, \tilde{X})$ is given by
$$ \mathcal{R}(f) = \tilde{P} \left[ \tilde{Y}\neq f(\tilde{X})\right]\,,$$
where $\tilde{P}$, the distribution of the test data, can have an arbitrary marginal distribution for $\tilde{X}$, but the conditional of $\tilde{Y}$ given $\tilde{X}$ has to match that in the training data. 
As $\mathcal{R}(f) = E \left[P \left[Y \neq f(X) \mid X\right]\right]$ which equals $$E\left[1\left[f(X) = 0\right]r(X) +  1\left[f(X) =1\right](1-r(X))\right],$$ it is readily shown that $\mathcal{R}(f)$ is at its minimum for the Bayes classifier $f^*(x) = 1\left[r(x) \geq 1/2\right] $, which, of course, is unavailable as $r(\cdot)$ is unknown.
It is typical to evaluate the performance of a classifier $f$ (which is typically based on the training data and therefore random) by comparing its risk to that of the Bayes classifier which is the best performing decision rule in terms of $\mathcal{R}(\cdot)$.

We study the above model under the shape-constraint that $r(\cdot)$ is monotone. This is a natural constraint to impose as many popular parametric classification models, such as the logit and the probit involve a non-decreasing $r(\cdot)$. 
In this setting, $r^{-1}(1/2)$ can be estimated in an efficient manner through the  multi-stage procedure spelled out in Section \ref{sec:inviso}.
Note that the multi-stage procedure shares similarities to active learning procedures \cite{Cohn94}, especially those based on adaptive sampling strategies
\cite{Iyengar00}. Let $\hat{d}_2 = \hat{r}_{n_2}^{-1}(1/2)$ denote the second stage estimate. In contrast to Section \ref{sec:inviso}, we now have a binary regression model with the underlying regression function being monotone.  The asymptotic results for $\hat{d}_2$ in this model parallel those for a heteroscedastic isotonic regression model (since Var($Y\mid X) = r(x) (1- r(x)$)) and can be established by using very similar techniques to those needed for the previous section. Specifically, it can be shown that
\be \label{limiisoh}
n^{(1+\gamma)/3} (\hat{d}_2 - d_0) \stackrel{d}{\rightarrow} \left(\frac{8 K r(d_0)(1- r(d_0))}{(r'(d_0))^2 p^\gamma (1-p)}\right)^{1/3}\argmin_w\{ B(w) + w^2 \},
\ee
where $d_0 = r^{-1}(1/2)$. Here, the variance $\sigma^2$ in Theorem \ref{th:limprociso} gets replaced by Var($Y\mid X = d_0) =r(d_0)(1- r(d_0))$. 

Now, the approximation to the Bayes classifier can be constructed as 
$$\hat{f}(x) = 1\left[\hat{r}_{n_2} (x) \geq 1/2\right] = 1\left[x \geq \hat{d}_2\right]. $$
We compare the limiting risk of this classifier to that for the Bayes rule $f^*$ for a fixed \emph{test data} covariate distribution, which we take to be  the uniform distribution on $[0,1]$. This is the content of the following theorem, where $\mathcal{R}(\hat{f})$ is interpreted as $\mathcal{R}({f})$ computed at $f = \hat{f}$.  
\begin{thm}\label{th:risk}
Assume that $r$ is continuously differentiable in a neighborhood of $d_0$  with $r'(d_0) \neq 0$. Then, 
$$n^{2(1+\gamma)/3}(\mathcal{R}(\hat{f}) - \mathcal{R}(f^*)) \stackrel{d}{\rightarrow }  \left(\frac{8 K r(d_0)(1- r(d_0))}{\sqrt{r'(d_0)} p^\gamma (1-p)}\right)^{2/3}\left[\argmin_w \{ B(w) + w^2 \}\right]^2.$$
\end{thm} 
This is a significant improvement over the corresponding single stage procedure, 
whose risk approaches the Bayes risk at the rate $n^{2/3}$, even in the presence of `oracle-type' information which allows the sampling to be finessed. To elaborate: consider a \emph{single stage} version of this problem with $n$ being the total budget for the training data. The goal is, of course, to estimate $d_0 = f^{-1}(1/2)$, in order to get the estimated Bayes' classifier. Suppose, `oracle type' information is available to the experimenter in the form of a density $g$ on $[0,1]$ that is peaked around the true $d_0$ and can therefore be used to sample more heavily around the parameter of interest. Thus, $X_1, \ldots, X_n$ are sampled from the density $g$ and conditional on the $X_i$'s, the $Y_i$'s are independent Bernoulli($r(X_i)$) random variables. If $\tilde{d}$ is the inverse isotonic estimate of $d_0$, by calculations similar to \citet[Theorem 2.1]{TBM11}, it can be shown that:
\[ n^{1/3}\,(\tilde{d} - d_0) \rightarrow_d \,\left(\frac{4 K r(d_0)(1- r(d_0))}{(r'(d_0))^2 g(d_0)}\right)^{1/3}\argmin_w\{ B(w) + w^2 \} \,.\] 
The limit behavior of the Bayes' risk of the corresponding classifier: $\tilde{f}(x) = 1(x \geq \tilde{d})$, with respect to the Uniform$[0,1]$ test-data distribution is given by the following theorem.
\begin{thm}
\label{th:risk1}
Under the same conditions as in Theorem \ref{th:risk}  
$$n^{2/3}(\mathcal{R}(\tilde{f}) - \mathcal{R}(f^*)) \stackrel{d}{\rightarrow }  \left(\frac{4 r(d_0)(1- r(d_0))}{\sqrt{r'(d_0))}g(d_0) }\right)^{2/3}\left[\argmin_w\{ B(w) + w^2 \}\right]^2.$$
\end{thm} 
So, for large values of $g(d_0)$, the excess risk of the estimated classifier over the Bayes' classifier will be small. However, a comparison of the two theorems in this section shows that the two-stage procedure, even in the absence of `oracle type' information, produces a classifier that 
eventually beats the one-stage classifier equipped with the `handicap' $g$.  
The proof of Theorem \ref{th:risk} is given in Section \ref{pf:risk} of the Supplement, while that of Theorem \ref{th:risk1} follows along the same lines starting from the limit 
distribution of $\tilde{d}_1$ and thus is omitted.

\begin{rmk}\label{smooth-estimates}
The above procedure illustrates rate acceleration based on a monotone model using the classical isotonic regression estimate. If one is willing to make
additional smoothness assumptions on $r$, a similar acceleration phenomenon would be observed with smoothed monotone estimates, the difference being that a faster rate would 
be achieved at stage two, given that the corresponding estimator at stage one would converge faster than $n_1^{1/3}$.
There is reason to believe that an analogous result would hold in non-parametric classification problems involving multiple covariates, although such an
investigation is outside the scope of the current paper.
\end{rmk}



\section{A mode estimation problem}\label{sec:mode}
Consider a model of the form $Y = m(X) + \epsilon$ in a design setting where $m(x) = \tilde{m}(||x- d_0||)$ with $\tilde{m}:  [0,\infty) \mapsto \R$ being a monotone decreasing function. Consequently, the regression function $m$ is unimodal and symmetric around $d_0$. Interest centers on estimating the point of maximum $d_0$ which can be thought of as a target or a source emanating signal isotropically in all directions. This is a canonical problem that has received a lot of attention in the statistics literature (see discussion in \cite{BGZ13}), but also has interesting applications in target detection problems using wireless sensor technology; see \cite{Katenka08}. In the latter case, one is interested in estimating the location of a target $d_0$ from noisy signals $Y_i=\tilde{m}(||X_i-d_0||)+\epsilon_i$, obtained from sensors at locations $X_i$. In many practical settings, in order for the sensors to save on battery and minimize communications, only a fraction of the available sensors is turned on and if a target is detected additional sensors are switched on to improve its localization. In this section we study this problem under multistage sampling and for simplicity restrict to a one-dimensional covariate (but see the discussion at the end of Section \ref{disco} for multivariate regressors). 

We assume that $\tilde{m}'(0) <0$, which corresponds to a cusp-like assumption on the signal.
We propose the following two-stage, computationally simple approach, which is adapted from the shorth procedure (see, for example, \citet[Section 6]{KP90}) originally developed to find the mode of a symmetric density. 
\begin{enumerate}
	\item At stage one, sample $n_1 = p n$ ($p \in (0,1)$) covariate values  uniformly from $[0, 1]$ and, from the obtained data, $(Y_i^{(1)}, X_i^{(1)})_{i=1}^{n_1}$, estimate $d_0$ by $\hat{d}_1 = \argmax_{d \in (b,1-b)} \M_{n_1}(d)$, where 
\be
\M_{n_1}(d) = \P_{n_1} Y^{(1)} 1\left[|X^{(1)}- d| \leq b\right],
\ee
where the bin-width $b>0$ is sufficiently small so that $[d_0-b, d_0+b] \subset (0,1)$.
Note that the estimate is easy to compute as the search for the maximum of $\M_{n_1}$ is restricted to points  $d$ such that either $d-b$ or $d+b$ is a design point.
	\item For $K>b>0$ and $\gamma >0$, sample the remaining $n_2 = (1-p)n $ covariate-response pairs $\{ Y_i^{(2)}, X_i^{(2)} \}$, where
	$$Y_i^{(2)} = m(X_i^{(2)})  + \epsilon_i^{(2)}, \ \ \  X_i^{2} \sim \mbox{Uniform}[\hat{d}_1 - K {n_1}^{-\gamma}, \hat{d}_1 + K {n_1}^{-\gamma}].$$
Obtain an updated estimate of $d_0$ by 
$$\hat{d}_2 = \argmax_{d \in \mathcal{D}_{\hat{\theta}_{n_1}}} \M_{n_2}(d), \mbox{ where}$$ 
\be
\M_{n_2}(d) = \P_{n_2} Y^{(2)} 1\left[|X^{(2)}- d| \leq b n_1^{-\gamma}\right], 
\ee
\end{enumerate}
 $ \hat{\theta}_{n_1} = \hat{d}_1$ and $\mathcal{D}_{\hat{\theta}_{n_1}} = [ \hat{\theta}_{n_1} - (K-b)n_1^{-\gamma}, \hat{\theta}_{n_1} + (K-b)n_1^{-\gamma} ]$. 
Here, $\gamma$ is chosen such that $P\left(d_0 \in [\hat{d}_1 - (K-b) {n_1}^{-\gamma}, \hat{d}_1+ (K-b) {n_1}^{-\gamma}]\right)$ converges to 1. It will be shown that ${n_1}^{1/3} (\hat{d}_1 - d_0) = O_p(1)$. Hence, any choice of $\gamma < 1/3$ suffices.

The limiting behavior of the one-stage estimate, which corresponds to the case $n_1 = n$, is derived next.
\begin{thm}\label{th:fstage}
We have ${n_1}^{1/3} (\hat{d}_1 - d_0) = O_p(1)$ and 
\be
{n_1}^{1/3} (\hat{d}_1 - d_0)  \stackrel{d}{\Rightarrow} \mathcal{Z}:= \left(\frac{a}{c}\right)^{2/3}\argmax \left\{ B(h) -  h^2\right\}
\ee
where $ a = \sqrt{2(m^2(d_0 + b) + \sigma^2)}$ and $c = -m'(d_0 + b) >0$.
\end{thm}
The proof follows from applications of standard empirical process results and is outlined in Section \ref{pf:fstge} of the Supplement.
\begin{rmk}
\label{one-stage-mode}
We note that the one-stage result does not require the assumption that $\tilde{m}'(0) < 0$ and is valid for both smooth and non-smooth signals at 0. The criticality of that assumption for obtaining gains out of a two-stage procedure will be clear from the following theorem. 
\end{rmk}
For the second stage estimate,
employing the general results from Section \ref{sec:genres}, we establish the following in Section \ref{pf:twostage} of the Supplement.
\begin{thm}\label{th:twostage}
We have ${n_2}^{(1+\gamma)/3} (\hat{d}_2 - d_0) = O_p(1)$ and 
\be
{n}^{(1+\gamma)/3} (\hat{d}_2 - d_0)  \stackrel{d}{\rightarrow} \left(\frac{4K(m^2(d_0)+\sigma^2)}{({m}'(d_0+))^2 p^\gamma(1-p)}\right)^{1/3}\argmax \left\{ B(h) -  h^2\right\}
\ee
\end{thm}
\begin{rmk}
\label{cusp-assumption}
It follows from the above result that small magnitudes of $m'(d_0+)$ lead to higher variability in the second stage estimate and suggests that for smooth functions, when $m'(d_0) = 0$, the limiting variance of $n^{(1+\gamma)/3}(\hat{d}_2 - d_0)$ blows up to infinity. That this is indeed the case will be seen shortly, as the actual rate of convergence of the two-stage estimator obtained via the above procedure is \emph{slower} for smooth $m$. 
\end{rmk}

\begin{rmk}\label{symmet}
It is worthwhile to point out that the symmetry of the function $m$ around $d_0$ is also crucial. If $m$ were not symmetric, our estimate from stage one, which reports the center of the bin (with width $2b$) having the maximum average response as the estimate of $d_0$, need not be consistent. For example, when $ m(x) = \exp(-a_1 |x- d_0|)$ for $x \leq d_0$, and $m(x) = \exp(-a_2|x-d_0|)$ for $x> d_0$, ($a_1 \neq a_2$) it can be shown that the expected criterion function,  $E\left[\M_{n_1} (d)\right]$ is minimized at $d^* = d_0 + (a_1 - a_2) b / (a_1 + a_2) \neq d_0$ and that $\hat{d}_1$ is a consistent estimate of $d^*$.  
\end{rmk}

\begin{rmk}\label{nonunif}
It is critical here to work with a uniform design for this problem. The uniform design at each stage ensures that the population criterion function is maximized at the true parameter $d_0$. With a non-uniform design at stage one, $\hat{d}_1$ will generally not be consistent for $d_0$. Further, if a non-uniform random design (symmetric about $\hat{d}_1$) is used at stage two (with a uniform design at stage one), $\hat{d}_2$ cannot be expected to converge at a rate faster than $n^{1/3}$ as it effectively ends up estimating an intermediate point between $d_0$ and $\hat{d}_1$. See Remark \ref{nonuniftech} for more (technical) details. 
\end{rmk}
\begin{rmk}\label{robmonro}
Root finding algorithms \citep{RM51} and their extensions \citep{KW52} provide a classical approach for locating the maximum of a regression function in an experimental design setting. However, due to the non-smooth nature of our problem ($m$ not being differentiable at $d_0$), $d_0$ is no longer the solution to the equation $m'(d) =0$, and therefore, these algorithms do not apply.
\end{rmk}

As was the case with the change-point and inverse isotonic regression problem, an optimal choice for the proportion $p$ exists that minimizes the limiting variance of the second stage estimate. As before, $K$ and $\gamma$ are chosen in practice such that 
$K {n_1}^{-\gamma}   \approx {C_{\tau/2}}/{n_1^{1/3}},$
where $C_{\tau/2}$ is the $(1- \tau/2)$'th quantile of the limiting distribution of $n_1^{1/3}(\hat{d}_1 - d_0)$.
The variance of $(\hat{d}_2 - d_0)$ would be (approximately) at its minimum when 
\begin{eqnarray*}
\frac{1}{n^{(1+\gamma)/3}}\left(\frac{4K(m^2(d_0)+\sigma^2)}{({m}'(d_0+))^2 p^\gamma(1-p)}\right)^{1/3} \approx \frac{1}{n^{4/9}}\left(\frac{4C_{\tau/2}(m^2(d_0)+\sigma^2)}{({m}'(d_0+))^2 p^{1/3}(1-p)}\right)^{1/3} 
\end{eqnarray*}
is at its minimum. Equivalently, $p^{1/3}(1-p)$ needs to be at its maximum. This yields the optimal choice of $p$ to be $p_{opt} \approx 0.25.$
\newline
\newline
{\it The case of a smooth $m$.} Next, we address the situation where $m$ is smooth, i.e., $m'(d_0)$ exists and equals zero. In this setting, the above approach is not useful. In contrast to the rate acceleration observed for non-smooth (at 0) $m$ case, here the rate actually \emph{decelerates}: it can actually be shown that the second stage estimate converges at a slower rate ($n^{(1-\gamma)/3}$) than the first stage estimate (see Remark \ref{derzero} in the Supplement). This is due to the fact that the function $m$ appears almost flat in the (second stage) zoomed-in neighborhood and our criterion that simply relies on finding the bin with maximum average response is not able to distinguish $d_0$ well from other local points in the zoomed-in neighborhood. However, if one were to use a
a symmetric (non-uniform) design centered at the first stage estimate for the second stage of sampling, 
an $n^{1/3}$-rate of convergence can be maintained for the second stage estimate (see Remark \ref{gexpln} in the Supplement for a technical explanation). 
\newline
More formally, let $W_i$'s, $1\leq i \leq n_2$, be i.i.d. realizations from density $g$, which is symmetric around 0. We assume $g$ to be Lipschitz of order 1, supported on [-1,1], with $g'(x) \neq 0$ on $(-1,1,) \backslash \{0\}$.  The second stage design points are now taken to be $X_i^{(2)} = \hat{d}_1
+ W_i K  n_1^{-\gamma}$, $1\leq i \leq n_2$. The rest of the procedure remains the same (as described at the beginning of this section) for constructing the second stage estimate $\hat{d}_2$. The following result can then be deduced.  
\begin{thm}\label{th:twostagenf}
Assume that the design density $g$ is Lipschitz of order 1. Then ${n_2}^{1/3} (\hat{d}_2 - d_0) = O_p(1)$ and 
\be
{n_2}^{1/3} (\hat{d}_2 - d_0)  \Rightarrow \left(\frac{1-p}{p}\right)^{1/3} \mathcal{Z} 
\ee
Consequently, ${n}^{1/3} (\hat{d}_2 - d_0)  \Rightarrow p^{-1/3}\mathcal{Z} $.
\end{thm}
A sketch of the proof is given in Section \ref{pf:twostagenf} of the Supplement. In particular, it is interesting to note that the 
asymptotic randomness in $\hat{d}_2$ comes from the first stage, unlike the other examples examined.
The form of the limit distribution shows that a larger $p$ yields a smaller limiting variance, and that the precision of the estimate is greatest when $p = 1$, i.e. a one-stage procedure, which tallies with the result in Theorem \ref{th:fstage}. 

We end this section by pointing out the contrasts between the mode estimation problem and the change-point/ isotonic regression problems. In the latter problems, the design density at $d_0$ appears as a variance reducing factor in the limit distribution of the first stage estimator itself; see, for example, \citet[Theorem 2.1]{TBM11} for the result on the isotonic regression problem with general sampling designs. A two-stage procedure is formulated to leverage on this phenomenon by sampling more points close to $\hat{d}_1$, the first stage estimate of $d_0$. A second stage design peaking at $\hat{d}_1$ (instead of a flat design) then leads to further gains (see Remark \ref{limitcp}). In contrast with these problems, the mode estimation procedure need not be consistent  at the first stage when the covariates are sampled from a non-flat design (see Remarks \ref{nonunif} and \ref{nonuniftech}). The interaction with the sampling design is much more complex than the design density simply appearing as a variance reducing factor. Hence, moving to a two-stage procedure and the use of non-flat densities do not \emph{necessarily} buy us gains, as demonstrated by the theorems in this section. 

%
%
%
%
%
%
 
There are some other multistage methods applicable to this smooth $m$ setting as well. Once could conceive fitting a quadratic curve (which is the local nature of the regression function $m$, as $m''(d_0) \neq 0$) to the data obtained from the second stage, akin to the ideas in \cite{BGZ13} and 
\cite{Hg41}.  The Kiefer-Wolfowitz procedure \citep{KW52} previously mentioned, that involves sampling 2 points at each of the $n/2$ stages, can be used to estimate the location of the maximum as well, since $m'(d_0) = 0$.

\section{Conclusions}\label{disco} 

\noindent
{\it Poisson limits.} In this paper we have considered the situation where the limit distribution of the second stage estimate is governed by a Gaussian or a mixture of Gaussian processes. However, in some change-point problems  such as the one addressed in \citet{LBM09}, a compound Poisson process appears in the limit. In such situations, Theorem \ref{th:tight}   and Lemma \ref{lm:fidiconvgen} do not apply as they address tightness and related weak convergence issues with respect  to the uniform metric and not the Skorokhod metric. In light of the conditioning arguments that we apply in this paper, we expect analogous results in Skorokhod topology to follow readily. Note, however, that the rate of convergence of the second stage estimate deduced in \citet{LBM09} can be derived from Theorem \ref{th:ratethmc6}.

\noindent
{\it Negative examples and possible solutions.} In this paper, we considered examples where multistage procedures typically accentuated the efficiency of M-estimates by accelerating the rate of convergence. As seen in Section \ref{sec:mode}, this is not always the case. In regular parametric problems, for example, where the estimates exhibit a $\sqrt{n}$-rate of convergence, acceleration to a faster rate is typically not possible. Acceleration happens when the parameter of interest has a local interpretation. Consider, for example the change-point problem. Here, the change-point is a local feature of the regression curve: not all regions of the domain contain the same amount of information about $d_0$. Regions to the far right or left of $d_0$ do not contain any information as the signal there is flat and observations in such regions can be essentially ignored.  Intensive sampling in a neighborhood of $d_0$ is a more sensible strategy as the signal here changes from one level to another, thereby suggesting a zoomed-in approach. In regular parametric models, the parameters typically capture `global' features of the curve and focusing on specific regions of the covariate space is not helpful. 
\newline
Moreover, acceleration in the rate, even for a local parameter, also depends on how the subtleties of the model interact with the method of estimation employed. Indeed, the result in Theorem \ref{th:twostagenf}, serves as a cautionary tale in this regard, illustrating that a fully non-parametric two-stage procedure that provides acceleration gains in one setting ($|\tilde{m}'(0)| > 0$) fails to do so in another ($|\tilde{m}'(0)| = 0$).  On the other hand, it is clear from the results of \cite{BGZ13} that a hybrid method that uses the `shorth' type estimate at stage one and a quadratic approximation at stage two will accelerate the rate of convergence. The potential downside of such hybrid methods, as demonstrated in \cite{TANG13} in the inverse isotonic problem, is that they may not perform well for modest budgets for which the degree of localization obtained from the first stage is typically not good enough for a parametric approximation in the second. We note here that fitting a polynomial curve at the second stage is better dealt using first principles as the $M$-estimate is then available in a sufficiently closed form. Our more abstract approach, which does not leverage on this added convenience available, may not be well suited for such situations.

\noindent 
{\it Pooling data across stages.} In certain models, it is preferred, at least from the perspective of more precise inference in the presence of 
fairly limited sample budgets, to pool the data across stages to obtain the final estimates. For example, in change-point models where the regression function is linear on either side of the threshold, e.g., $m(x) = (\alpha_0 + \alpha_1 x) 1(x\leq d_0) + (\beta_0 + \beta_1 x) 1(x>d_0)$, $\alpha_i \neq \beta_i, i =1,2$, it is recommended to estimate at least the slope parameters using the pooled data. This is due to the fact that slopes are better estimated when the design points are far apart.
The technicalities in this situation are expected to become significantly more complicated due to the more convoluted nature of the dependence. Specifically, conditional on the first stage estimate, the second stage one can no longer be viewed as a functional of i.i.d. observations. 
However, we conjecture that for parameters that are local features of the model, the second stage estimates from pooled data should exhibit the same asymptotic behavior as our current second stage estimates, since the proportion of first stage points in the shrinking sampling interval for stage two goes to zero.

\noindent
{\it Other Applications.} The approach and the results of this paper apply to a variety of other problems. For example, consider the extension of the change-point model to multiple dimensions where the regression function exhibits different functional forms in sub-regions of Euclidean space which are separated by smooth parametric boundaries, for example, hyperplanes. Determination of these separating hyperplanes could be achieved by multistage procedures: an initial fraction of the budget would be used to elicit initial estimates of these hyperplanes via least squares methods and more intensive sampling could then be carried out in a neighborhood of the hyperplanes, and the estimates updated via least squares again. This falls completely within the purview of our approach. Once again, the multistage procedure would provide gains in terms of convergence rates over one-stage methods that use the same budget. For an example of models of this type, see the problem studied in  \cite{Wei13}. Another problem involves mode estimation for a regression with higher-dimensional covariates $X$ in Section \ref{sec:mode} under an isotropic signal.  An approach similar to the one-dimensional setting can be adopted here as well with the sampling neighborhood at stage two chosen to be a ball around the initial estimate. In the presence of cusp-like signals, acceleration of the convergence rate over a competing one stage procedure would be observed. 

\noindent
{\it More than two stages:} The results of this paper can be extended to multiple ($ > 2$ but fixed) stages but caution needs to be exercised since the asymptotics will not be reliable unless the sample size invested 
at each stage is ample, which then necessitates the total sample size being large. By increasing the number of stages, the rate of convergence can be accelerated, in theory, but the gains from the theory will only become apparent for substantially large budgets. From a different perspective, one could of course consider how such multistage procedures behave if the total number of sampling stages grows like $n^{\gamma}$ ($\gamma < 1$) with order $n^{1-\gamma}$ points invested at each stage (as opposed to a fixed proportion of points that we currently consider), but again, such a framework will not be useful for realistic budgets. Our set-up is not amenable to sequential procedures where the number of stages can increase with sample size, but it should be noted that our work does not aim to develop a sequential paradigm. Rather, our results serve to illustrate that non-sequential multistage sampling (which is typically easier to implement than fully sequential procedures), used adequately, can lead to substantial gains in a variety of statistical problems.  
\appendix

\section{Proofs}\label{proofs}

\subsection{Proof of Theorem \ref{th:ratethmc6}} \label{pf:ratethm}
Note that if $\kappa_n r_n = O(1)$, i.e., there exists $C>0$, such that  $\kappa_n r_n \leq C$ for all $n$, then
\begin{eqnarray*}
P\left(r_n \rho_n(\hat{d}_n, d_n) \geq C\right) & = & P\left(r_n \kappa_n \rho_n(\hat{d}_n, d_n) \geq C \kappa_n \right) \\
& \leq & P\left(\rho_n(\hat{d}_n, d_n) \geq  \kappa_n \right), 
\end{eqnarray*} which converges to zero. Therefore, the conclusion of the theorem is immediate when $\kappa_n r_n = O(1)$. Hence, we only need to address the situation where $\kappa_n r_n \rightarrow \infty$.

For a fixed realization of $\hat{\theta} = \theta$, we use $\hat{d}_n(\theta)$ to denote our estimate, so that $\hat{d}_n = \hat{d}_n(\hat{\theta}_n)$.  For any $L>0$, 
\begin{eqnarray}
{P\left(r_n \rho_n(\hat{d}_n(\hat{\theta}_n), d_n)  \geq 2^L \right) }&\leq& { P\left( r_n \kappa_n > r_n \rho_n(\hat{d}_n(\hat{\theta}_n), d_n)  \geq 2^L , \hat{\theta}_n \in \Theta_n^\tau \right)} \nonumber \\ 
 & & + P\left(\rho_n(\hat{d}_n(\hat{\theta}_n), d_n) \geq \kappa_n \right) + \tau. 
\label{eq:splittty}
\end{eqnarray}
The second term on the right side goes to zero. 
Further,
\begin{eqnarray}
\lefteqn{P\left(r_n \kappa_n > r_n \rho_n(\hat{d}_n(\hat{\theta}_n), d_n) \geq 2^L, \hat{\theta}_n \in \Theta_n^\tau\right)} \hspace{1in}\nonumber \\ 
&=&  E \left[P\left(r_n \kappa_n > r_n \rho_n(\hat{d}_n(\hat{\theta}_n), d_n) \geq 2^L \mid \hat{\theta}_n \right) 1\left[\hat{\theta}_n \in \Theta_n^\tau\right] \right] \nonumber \\
& \leq & \sup_{\theta \in \Theta_n^\tau} P\left(r_n \kappa_n > r_n \rho_n(\hat{d}_n({\theta}), d_n) \geq 2^L \right). 
\label{eq:decomp}
\end{eqnarray}
Let $S_{j,n} = \left\{d: 2^{j} \leq r_n \rho_n(d, d_n) <  \min(2^{j+1}, \kappa_n r_n)\right\}$ for $j \in \Z$.   
If $r_n \rho_n(\hat{d}_n({\theta}), d_n)$ is larger than $2^L$ for a given positive integer $L$ (and smaller than $\kappa_n r_n$), then $\hat{d}_n(\hat{\theta}_n)$ is in one of the shells $S_{j,n}$'s for $j \geq L$. By definition of $\hat{d}_n({\theta})$, the infimum of the map $d \mapsto \M_n(d, {\theta}) - \M_n(d_n, {\theta})$ over the shell  containing $\hat{d}_n({\theta})$ (intersected with $\mathcal{D}_{{\theta}}$) is not positive. For $\theta \in \Theta_n^\tau$, 
\begin{eqnarray*}
\lefteqn{P\left(r_n \kappa_n > r_n \rho_n(\hat{d}_n({\theta}), d_n) \geq 2^L \right) } \\
& \leq & \sum_{j\geq L, 2^{j} \leq \kappa_n r_n} P^*\left( \inf_{d \in S_{j,n} \cap \mathcal{D}_{{\theta}}}\M_n(d, {\theta}) - \M_n(d_n, {\theta}) \leq 0\right). \nonumber 
\end{eqnarray*}
For every $j$ involved in the sum, $n > N_{\tau}$ and any $\theta \in \Theta_n^\tau$, \eqref{eq:cond00} gives
\be
\inf_{2^{j}/r_n \leq \rho_n(d, d_n) < \min(2^{j+1}, \kappa_n r_n)/r_n, d \in \mathcal{D}_\theta} M_n(d, \theta) - M_n(d_n, \theta) \geq c_\tau\ \frac{ 2^{2j}}{r^2_n}.
\label{eq:distrrr}
\ee
Also, for such a $j$,  $n > N_{\tau}$ and $\theta \in \Theta_n^\tau$,
\begin{eqnarray*}
\lefteqn{P^*\left( \inf_{d \in S_{j,n} \cap \mathcal{D}_\theta}\M_n(d, {\theta}) - \M_n(d_n, \theta) \leq 0\right) } \nonumber\\ 
&\leq& P^*\left( \inf_{d \in S_{j,n} \cap \mathcal{D}_\theta}[(\M_n(d, \theta) - M_n(d, \theta))- (\M_n(d_n, \theta) - M_n(d_n, \theta))]  \right. \\
&&\left. \leq - \inf_{d \in S_{j,n} \cap \mathcal{D}_\theta} M_n(d, \theta) - M_n(d_n, \theta)\right) \\
&\leq& P^*\left( \inf_{d \in S_{j,n} \cap \mathcal{D}_\theta}[(\M_n(d, \theta) - M_n(d, \theta))- (\M_n(d_n, \theta) - M_n(d_n, \theta))]  \leq - c_\tau\frac{ 2^{2j}}{r^2_n}\right) \\
&\leq& P^*\left( \sup_{d \in S_{j,n} \cap \mathcal{D}_\theta}\left|(\M_n(d, \theta) - M_n(d, \theta))- (\M_n(d_n, \theta) - M_n(d_n, \theta))\right|  \geq  c_\tau\frac{ 2^{2j}}{r^2_n}\right).
\end{eqnarray*}
 For $n > N_{\tau}$, by Markov inequality and \eqref{eq:cond01}, we get 
\begin{eqnarray}
\lefteqn{\sup_{\theta \in \Theta_{n}^\tau}\sum_{j\geq L, 2^{j} \leq \kappa_n r_n} P^*\left( \inf_{d \in S_{j,n} \cap \mathcal{D}_\theta}\M_n(d, \theta ) - \M_n(d_n, \theta) \leq 0\right) } \hspace{1in} \nonumber\\
 &\leq& C_\tau \sum_{j\geq L, 2^{j} \leq \kappa_n r_n} \frac{\phi_n(\min(2^{j+1}, r_n \kappa_n)/r_n)r^2_n}{c_\tau \sqrt{n}2^{2j}}.
\label{eq:consbrk}
\end{eqnarray}
Note that $\phi_n(c \delta) \leq c^\alpha \phi_n(\delta)$ for every $c > 1$. As $\kappa_n r_n \rightarrow \infty$, there exists $\bar{N} \in \mathbb{N}$, such that $\kappa_n r_n  >1$. Hence, for $L > 0$ and $n > \max(\bar{N}, N_{\tau})$, the above display is bounded by
$$ \frac{C_\tau}{ c_\tau} \sum_{j\geq L, 2^{j} \leq \kappa_n r_n} (\min(2^{j+1}, r_n \kappa_n))^{\alpha}\, 2^{-2j} \leq \tilde{K}\,\frac{C_\tau}{ c_\tau} \sum_{j\geq L, 2^{j} \leq \kappa_n r_n} 2^{(j+1) \alpha - 2j}, $$
for some universal constant $\tilde{K}$, by the definition of $r_n$. For any fixed $\eta>0$, take $\tau = \eta/3$ and choose $L_{\eta}>0$ such that the sum on the right side is less than $\eta/3$. Also, there exists $\tilde{N}_{\eta} \in \mathbb{N}$ such that for all $n > \tilde{N}_{ \eta} \in \mathbb{N}$,
$$ P\left( \rho_n(\hat{d}_n(\hat{\theta}_n), d_n) \geq \kappa_n \right) < \eta/3.$$
 Hence, for $n > \max( \bar{N}, N_{\eta/3}, \tilde{N}_{ \eta} )$,
$$ P\left(r_n \rho_n(\hat{d}_n(\hat{\theta}_n), d_n) > 2^{L_\eta} \right) < \eta,$$
 by \eqref{eq:splittty} and \eqref{eq:consbrk}. Thus, we get the result when conditions \eqref{eq:cond00} and \eqref{eq:cond01} hold for some sequence $\kappa_n >0$.  

Further, note that if the conditions in part (b) of the theorem hold for all sequences $\kappa_n >0$, following the arguments in \eqref{eq:splittty} and  \eqref{eq:decomp}, we have 
\begin{eqnarray}
{P\left(r_n \rho_n(\hat{d}_n(\hat{\theta}_n), d_n) > 2^L \right) }&\leq&  \sup_{\theta \in \Theta_n^\tau} P\left(r_n \rho_n(\hat{d}_n({\theta}), d_n) > 2^L \right) + \tau.  \nonumber
\end{eqnarray}
We can now use the shelling argument for $j \geq L$ letting $j$ go all the way to $\infty$ where our shell $S_{j,n}$ is now simply $\{d : 2^j \leq r_n\,\rho_n(d,d_n) < 2^{j+1}\}$. By our assumption, the bounds in \eqref{eq:distrrr} and \eqref{eq:consbrk} hold for every such shell, when $n > N_{\tau}$ and we arrive at the result by similar arguments as above  without 
needing to address the event $P\left(\rho_n(\hat{d}_n(\hat{\theta}_n), d_n) \geq \kappa_n \right)$ in \eqref{eq:splittty}  separately.
\qed

\subsection{Proof of Theorem \ref{th:tight}} \label{pf:tight}
As the sum of tight processes is tight, it suffices to show tightness of $\zeta_n(\cdot, \hat{\theta}_n)$ and $\G_n f_{n, \cdot, \hat{\theta}_n}$ separately. As $\mathcal{H}$ is totally bounded under $\tilde{\rho}$, tightness of the process $\zeta_n$ can be shown by justifying that 
\be
P^*\left[\sup_{\tilde{\rho}(h_1, h_2) < \delta_n} \left| \zeta_n(h_1, \hat{\theta}_n) - \zeta_n(h_2, \hat{\theta}_n) \right| > t \right] \rightarrow 0, \nonumber
\ee
for $\delta_n \downarrow 0$ and $t >0$. The right side of the above display is bounded by 
\begin{eqnarray*}
\lefteqn{P^*\left[\sup_{\tilde{\rho}(h_1, h_2) < \delta_n} \left| \zeta_n(h_1, \hat{\theta}_n) - \zeta_n(h_2, \hat{\theta}_n) \right| > t , \hat{\theta}_n \in \Theta_n^\tau \right] + P[\hat{\theta}_n \notin \Theta_n^\tau]} \hspace{1in}\\
&\leq& 1\left[\sup_{\substack{\theta \in \Theta_n^\tau \\ \tilde{\rho} (h_1,h_2) < \delta_n }} | \zeta_n(h_1,\theta) - \zeta_n(h_2,\theta)|  > t \right] + \tau.
\end{eqnarray*}
By \eqref{eq:cond4}, the above can be made arbitrarily small for large $n$ and hence, the process  $\zeta_n(\cdot, \hat{\theta}_n)$ is asymptotically tight.

We justify tightness of the process $\{ \G_n f_{n,h, \hat{\theta}}: h \in \mathcal{H} \}$ when \eqref{eq:cond5} holds. The proof under the condition on bracketing numbers follows along similar lines.
 As was the case with $\zeta_n$, we  consider the expression 
\be
P^*\left[\sup_{\tilde{\rho}(h_1, h_2) < \delta_n} \left| \G_n (f_{n,h_1,\hat{\theta}_n} - f_{n,h_2,\hat{\theta}_n}) \right| > t \right], \nonumber
\ee
for $\delta_n \downarrow 0$ and $t >0$. 
Let $e_i, i \geq 1$ denote Rademacher random variables independent of $V$'s and $\hat{\theta}$. By arguments similar to those at the beginning of the proof of Theorem 2.11.1 of \cite{VW96}, which use a symmetrization lemma for probabilities (Lemma 2.3.7 of the same book),  for sufficiently large $n$,
 the above display can be bounded  by 
\be
4 P^*\left[\sup_{\tilde{\rho}(h_1, h_2) < \delta_n} \left| \frac{1}{\sqrt{n}} \sum_{i=1}^n e_i (f_{n,h_1,\hat{\theta}}(V_i) - f_{n,h_2,\hat{\theta}}(V_i)) \right| > \frac{t}{4} \right]
\label{eq:outerprob}
\ee
The only difference from the proof of the cited lemma is that the arguments are to be carried out for fixed realizations of $V_i$'s and $\hat{\theta}$ (instead of fixed realizations of the $V_i$'s alone), and then outer expectations are taken. Further, from the measurability assumption, the map
$$ (V_1, V_2, \ldots, V_n, \hat{\theta}, e_1,\ldots,e_n) \mapsto \sup_{\tilde{\rho}(h_1, h_2) < \delta_n} \left| \frac{1}{\sqrt{n}} \sum_{i=1}^n e_i (f_{n,h_1,\hat{\theta}}(V_i) - f_{n,h_2,\hat{\theta}}(V_i)) \right| $$ 
is jointly measurable. Hence, the expression in \eqref{eq:outerprob} is a probability.  Let $Q_n$ denote the marginal distribution of $\hat{\theta}_n$. Then, for any $\tau > 0$, 
\begin{eqnarray*}
\lefteqn{4 P\left[\sup_{\tilde{\rho}(h_1, h_2) < \delta_n} \left| \frac{1}{\sqrt{n}} \sum_{i=1}^n e_i (f_{n,h_1,\hat{\theta}}(V_i) - f_{n,h_2,\hat{\theta}}(V_i)) \right| > \frac{t}{4} \right]}\\
& = & 4 \int  P\left[\sup_{\tilde{\rho}(h_1, h_2) < \delta_n} \left| \frac{1}{\sqrt{n}} \sum_{i=1}^n e_i (f_{n,h_1,{\theta}}(V_i) - f_{n,h_2,{\theta}}(V_i)) \right| > \frac{t}{4} \right] Q_n (d \theta)  \\
& \leq & 4\sup_{\theta \in \Theta_n^\tau} P\left[\sup_{\tilde{\rho}(h_1, h_2) < \delta_n} \left| \frac{1}{\sqrt{n}} \sum_{i=1}^n e_i (f_{n,h_1,{\theta}}(V_i) - f_{n,h_2,{\theta}}(V_i)) \right| > \frac{t}{4} \right]   + \tau \nonumber
\end{eqnarray*}
For a fixed $\theta \in \Theta_n^\tau$, let $\mathcal{F}_{n,\theta,\delta_n} = \{ f_{n,h_1,{\theta}} - f_{n,h_2,{\theta}} : \tilde{\rho}(h_1, h_2) < \delta_n \} $. For $g \in \mathcal{F}_{n,\theta,\delta_n}$, the process 
$g \mapsto (1/\sqrt{n})\,\sum_{i=1}^n e_i g (V_i)$ (given $V_i$s) is sub-Gaussian with respect to the $L_2(\P_n)$ semi-metric and hence, by Markov's inequality and chaining, Corollary 2.2.8 of \cite{VW96}, the above display can be bounded, up to a universal constant, by
\be \label{eq:one_bnd}
\frac{16}{t} \sup_{\theta \in \Theta_n^\tau} E \int_0^{\xi_n(\theta)} \sqrt{\log N\left(u, \mathcal{F}_{n, \theta, \delta_n}, L_2(\P_n)\right)} du,
\ee
with $$\xi_n^2 (\theta) = \sup_{g \in \mathcal{F}_{n, \theta, \delta_n}} \| g\|^2_{L_2(\P_n)} = \sup_{g \in \mathcal{F}_{n, \theta, \delta_n}} \left[\frac{1}{n} \sum_{i=1}^n g^2(V_i) \right].$$ It suffices to show that for all sufficiently large $n$, $\sup_{\theta \in \Theta_n^\tau} E \int_0^{\xi_n(\theta)} \sqrt{\log N\left(u, \mathcal{F}_{n, \theta, \delta_n}, L_2(\P_n)\right)} du$ can be made as small as wished. We assume, without loss of generality, that 
each $F_{n,\theta} \geq 1/2$ if necessary by adding $1/2$ to each of the original ones. (Note that this does not disturb any of the assumptions of Theorem 2.) Since, $N(u,\mathcal{F}_{n, \theta, \delta_n}, L_2(\P_n)) \leq N^2(u/2,\mathcal{F}_{n, \theta}, L_2(\P_n))$, we have:
\begin{eqnarray*}
\lefteqn{\sup_{\theta \in \Theta_n^\tau} E \int_0^{\xi_n(\theta)} \sqrt{\log N\left(u, \mathcal{F}_{n, \theta, \delta_n}, L_2(\P_n)\right)} du} \\ 
& \lesssim & \sup_{\theta \in \Theta_n^\tau} E \int_0^{\xi_n(\theta)} \sqrt{\log N\left(u/2, \mathcal{F}_{n, \theta}, L_2(\P_n)\right)} du\\
&\lesssim & \sup_{\theta \in \Theta_n^\tau} E \left[\int_0^{\xi_n(\theta)/(2\,\|F_{n,\theta}\|_n)} \sqrt{\log N\left(u\,\|F_{n,\theta}\|_n, \mathcal{F}_{n, \theta}, L_2(\P_n)\right)} du\;\|F_{n,\theta}\|_n \right] \\
& \lesssim &\sup_{\theta \in \Theta_n^\tau}\,E\,\left[\|F_{n,\theta}\|_n\,\;\int_{0}^{\xi_n(\theta)} \sup_{Q \in \mathcal{Q}} \,\sqrt{\log N\left(u\,\|F_{n,\theta}\|_{Q,2}, \mathcal{F}_{n, \theta}, L_2(Q)\right)} \,du\, \right]\,.
\end{eqnarray*}
By Cauchy-Schwarz, the above is bounded by: 
\[ \sup_{\theta \in \Theta_n^\tau}\,\left[\sqrt{\frac{1}{n}\,\sum_{i=1}^n\,E\,(F_{n,\theta}^2(V_i))}\right]\,\sqrt{E\,(h_{n,\theta}^2(\xi_n(\theta))} \,,\]
where 
\[ h_{n,\theta}(x) = \int_0^{x}\,\sup_{Q \in \mathcal{Q}}\,\sqrt{\log N\left(u\,\|F_{n,\theta}\|_{Q,2}, \mathcal{F}_{n, \theta}, L_2(Q)\right)} \,du \,.\] 
This, in turn, is bounded by: 
\[ \sup_{\theta \in \Theta_n^\tau}\,(PF_{n,\theta}^2)^{1/2} \times \sqrt{\sup_{\theta \in \Theta_n^\tau}\,E\,(h_{n,\theta}^2(\xi_n(\theta))} \,.\] 
The first term above is bounded as $n \rightarrow \infty$ by \eqref{eq:cond1}. To show that the second term can be made small for sufficiently large $n$, we {\bf claim} that it suffices to show that 
$ \sup_{\theta \in \Theta_n^\tau} E^* \xi_n(\theta)^2$ converges to zero. For the moment, assume the claim. It follows that for any $\lambda >0$, 
\[ \sup_{\theta \in \Theta_n^\tau} \,P(\xi_n(\theta) > \lambda) \rightarrow 0 \,.\] 
Next, note that $\sup_{\theta \in \Theta_n^{\tau}}\,h_{n,\theta}(\xi_n(\theta)) \leq \sup_{\theta \in \Theta_n^{\tau}}\,h_{n,\theta}(\infty) < \infty$ by \eqref{eq:cond5}. Now, for any $\lambda > 0$, 
\begin{eqnarray*}
 E(h_{n,\theta}^2(\xi_n(\theta))) &=& E(h_{n,\theta}^2(\xi_n(\theta))\,1(\xi_n(\theta) \leq \lambda)) + E(h_{n,\theta}^2(\xi_n(\theta))\,1(\xi_n(\theta) \leq \lambda)) \\
 & \leq & \lambda^2 + h_{n,\theta}^2(\infty)P(\xi_n(\theta) > \lambda) \,,
 \end{eqnarray*} 
so that 
\[ \sup_{\theta \in \Theta_n^\tau}\,E(h_{n,\theta}^2(\xi_n(\theta))) \leq \lambda^2 + \sup_{\theta \in \Theta_n^{\tau}}\,h_{n,\theta}^2(\infty)\;\sup_{\theta \in \Theta_n^\tau} \,P(\xi_n(\theta) > \lambda) \,,\]
which can be made as small as we please by first choosing $\lambda$ small enough and then letting $n \rightarrow \infty$. It remains to prove the claim. 
Note that 
\begin{eqnarray*}
E^* \xi_n(\theta)^2 \leq E^* \sup_{g \in \mathcal{F}_{n, \theta, \delta_n}} | (\P_n - P) g^2| + \sup_{g \in \mathcal{F}_{n, \theta, \delta_n}} |P g^2|
\end{eqnarray*}
By \eqref{eq:cond3}, the second term on the right side goes to zero uniformly in $\theta \in \Theta_n^\tau$. By the symmetrization lemma for expectations, Lemma  2.3.1 of \cite{VW96}, the first term on the right side is bounded by
\begin{eqnarray*}
2E^* \sup_{g \in \mathcal{F}^2_{n, \theta, \delta_n}} \left|\frac{1}{n}\sum_{i=1}^n e_i g(V_i) \right| \leq 2E^* \sup_{g \in \mathcal{F}^2_{n, \theta, \infty}} \left|\frac{1}{n}\sum_{i=1}^n e_i g(V_i) \right|
\end{eqnarray*}
Note that $G_{n,\theta}= (2 F_{n, \theta})^2$ is an envelope for the class $\mathcal{F}^2_{n,\theta,\infty}$. By condition \eqref{eq:cond2}, there exists a sequence of numbers $\eta_n \downarrow 0$ (slowly enough) such that $\sup_{\theta \in \Theta_n^\tau} P F^2_{n, \theta} 1\left[F_{n, \theta} > \eta_n \sqrt{n}\right] $  converges to zero. Let $\F^2_{n,\theta,\infty, \eta_n} = \left\{g 1[G_{n,\theta} \leq n \eta_n^2] :\ g \in \mathcal{F}^2_{n, \theta, \infty} \right\} $. Then, the above display is bounded by:
\begin{eqnarray*}
2E^*\ \sup_{g \in \mathcal{F}^2_{n, \theta, \infty, \eta_n}} \left|\frac{1}{n}\sum_{i=1}^n e_i g(V_i)\right| + 2 P^* G_{n,\theta} 1\left[G_{n,\theta} > n \eta_n^2\right]
\end{eqnarray*}
The second term in the above display goes to zero (uniformly in $\theta$) by \eqref{eq:cond2} and it remains to show the convergence of the first term (to 0) uniformly in $\theta$. By the $P$-measurability of the class $\mathcal{F}^2_{n, \theta, \infty, \eta_n}$, the first term in the above display is an expectation. For $u>0$, let $\mathcal{G}_{u,n}$ be a minimal $u R_n$-net in $L_1(\P_n)$ over $\F^2_{n,\theta,\infty, \eta_n}$, where $R_n = 4\|F_{n,\theta}\|_n^2$. Note that the cardinality of $\mathcal{G}_{u,n}$ is $N(u R_n, \mathcal{F}^2_{n, \theta, \infty, \eta_n}, L_1(\P_n))$ and that 
\begin{equation}
\label{turns-out-imp} 
2E^*\ \sup_{g \in \mathcal{F}^2_{n, \theta, \infty, \eta_n}} \left|\frac{1}{n}\sum_{i=1}^n e_i g(V_i)\right|  \leq 2E\ \sup_{g \in \mathcal{G}_{u,n}} \left|\frac{1}{n}\sum_{i=1}^n e_i g(V_i)\right| + u E(R_n)\,.
\end{equation}
Note that $\sup_{\theta \in \Theta_n^{\tau}}\,u E(R_n) = 4u \sup_{\theta \in \Theta_n^{\tau}}\,u \,PF_{n,\theta}^2 \lesssim u$, 
by \eqref{eq:cond1}. Using the fact that the $L_1$ norm is bounded up to a (universal) constant by the $\psi_2$ Orlicz norm and letting $\psi_2|V$ denote the conditional Orlicz norm given fixed realizations of the $V_i$'s, we obtain the following bound on the first term of the above display: 
\begin{eqnarray*} 
\frac{2}{n}E_V\,E_e\,\left[\sup_{g \in \mathcal{G}_{u,n}} \left|\sum_{i=1}^n e_i g(V_i)\right|\right] &\lesssim& \frac{2}{n}E_V\,\left\|\sup_{g \in \mathcal{G}_{u,n}} \left|\sum_{i=1}^n e_i g(V_i)\right|\right \|_{\psi_2 \mid V} \\
&\lesssim& 
\frac{2}{n}\,E_V\,\left[\sqrt{1 + \log N(u R_n, \mathcal{F}^2_{n, \theta, \infty, \eta_n}, L_1(\P_n)) } \right.\\
& & \left. \qquad \qquad  \times \mbox{max}_{g \in \mathcal{G}_{u,n}}\left\|\sum_{i=1}^n e_i g(V_i)\right \|_{\psi_2 \mid V} \right]\,, 
\end{eqnarray*}  
where the last inequality follows by an application of a maximal inequality for Orlicz norms (Lemma 2.2.2. of \cite{VW96}). 
By Hoeffding's inequality, for each $g \in \mathcal{G}_u$, $\left\|\sum_{i=1}^n e_i g(V_i)\right \|_{\psi_2 \mid V} \leq [\sum_i g^2(V_i) ]^{1/2}$ which is at most $\left[\sum_i n \eta^2_n G_{n,\theta}(V_i) \right]^{1/2}$. We 
conclude that the first term on the right side of \ref{turns-out-imp} is bounded, up to a universal constant, by: 
\[E \left[\frac{\left[\sum_i n \eta^2_n G_{n,\theta}(V_i) \right]^{1/2}}{n} \sqrt{1 + \log N(u\, 4\|F_{n,\theta}\|_n^2, \mathcal{F}^2_{n, \theta, \infty, \eta_n}, L_1(\P_n)) }\right] \,.\]
Next, 
\begin{eqnarray*} 
\log\,N(u\, 4\|F_{n,\theta}\|_n^2, \mathcal{F}^2_{n, \theta, \infty, \eta_n}, L_1(\P_n)) &\leq & 
\log\,N(u\, 4\|F_{n,\theta}\|_n^2, \mathcal{F}^2_{n, \theta,\infty}, L_1(\P_n)) \\
& \leq & \log N(u\, \|F_{n,\theta}\|_n, \mathcal{F}_{n, \theta, \infty}, L_2(\P_n)) \\
& \leq & \log N^2((u/2)\,\|F_{n,\theta}\|_n, \mathcal{F}_{n, \theta}, L_2(\P_n)) \\
& \leq & 2\,\sup_{Q}\,\log N((u/2)\,\|F_{n,\theta}\|_{Q,2}, \mathcal{F}_{n, \theta}, L_2(Q)) \,. 
\end{eqnarray*} 
Conclude that the expectation preceding the above display is bounded by: 
\begin{eqnarray*}
\lefteqn{\frac{\eta_n}{\sqrt{n}} E \left[\sum_{i=1}^n G_{n,\theta}(V_i) \right]^{1/2} \sqrt{1 + 2\sup_Q \log N((u/2)\,\|F_{n,\theta}\|_{Q,2}, \mathcal{F}_{n, \theta}, L_2(Q)) }} \\
& \leq & \frac{\eta_n}{\sqrt{n}} \left[E \left[\sum_{i=1}^n  G_{n,\theta}(V_i) \right]\right]^{1/2} \sqrt{1 + 2\sup_Q \log N((u/2)\,\|F_{n,\theta}\|_{Q,2}, \mathcal{F}_{n, \theta}, L_2(Q)) } \\
& \leq & 4 \eta_n  \left[P F^2_{n,\theta}\right] \sqrt{1 + 2\sup_Q \log N((u/2)\,\|F_{n,\theta}\|_{Q,2}, \mathcal{F}_{n, \theta}, L_2(Q)) }. 
\end{eqnarray*}
Now, note that $u$ is arbitrary (and can therefore be as small as wished), $\sup_{\theta \in \Theta_n^{\tau}}\,P F^2_{n,\theta}$ is $O(1)$ from\eqref{eq:cond1}, and, 
\[ \sup_{\theta \in \Theta_n^{\tau}}\,\sqrt{1 + 2\sup_Q \log N((u/2)\,\|F_{n,\theta}\|_{Q,2}, \mathcal{F}_{n, \theta}, L_2(Q)) } = O(1) \,,\]
since, 
\[\sup_{\theta \in \Theta_n^{\tau}}\,h_{n,\theta}(u/2) \geq \sup_{\theta \in \Theta_n^{\tau}}(u/2)\,\sup_Q \sqrt{\log N((u/2)\,\|F_{n,\theta}\|_{Q,2}, \mathcal{F}_{n, \theta}, L_2(Q))} \,,\]
showing that
\[ \sup_{\theta \in \Theta_n^{\tau}}\,\sup_Q \sqrt{\log N((u/2)\,\|F_{n,\theta}\|_{Q,2}, \mathcal{F}_{n, \theta}, L_2(Q))} \leq (2/u)\,\sup_{\theta \in \Theta_n^{\tau}}\,h_{n,\theta}(u/2) \,,\]
and from \eqref{eq:cond5}, $\sup_{\theta \in \Theta_n^{\tau}}\,h_{n,\theta}(u/2)$ is $O(1)$. Hence, by choosing $u$ small enough and then letting $n \rightarrow \infty$, the first term on the right side of \ref{turns-out-imp} can be made as small as wished, uniformly over $\theta \in \Theta_n^{\tau}$, for $n$ sufficiently large, since $\eta_n \rightarrow 0$. \qed

\subsection{Proof of  Theorem \ref{th:ratecp}}\label{pf:ratecp}
 As $n_1$, $n_2$ and $n$ are of the same order, we deduce bounds in terms of $n$ only. For notational ease, we first consider the situation where $d \geq d_0$. Recall that $\theta = (\alpha, \beta, \mu)$. 
Also, let \be \label{eq:thetaset}
\begin{split}
\Theta_{n_1}^\tau =& \left[\alpha_n - \frac{K_\tau}{ \sqrt{n_1}}, \alpha_n + \frac{K_\tau}{ \sqrt{n_1}}\right] \times \left[\beta_n - \frac{K_\tau}{ \sqrt{n_1}}, \beta_n + \frac{K_\tau}{ \sqrt{n_1}}\right] \times \\
& \left[d_0 - \frac{K_\tau}{ {n_1}^\nu}, d_0 + \frac{K_\tau}{ {n_1}^\nu}\right],
\end{split}
	\ee
 where $K_\tau$ is chosen such that $P\left(\hat{\theta}_{n_1} \in \Theta_{n_1}^\tau  \right) > 1 - \tau$. For $\theta \in \Theta_{n_1}^\tau$, $\beta - \alpha \geq c_0 n^{-\xi} - 2 K_\tau/ \sqrt{n_1}$. As $\xi < 1/2$, $\mbox{sgn}(\beta - \alpha) =1$ for  $n > N^{(1)}_{\tau} := (2K_\tau/(\sqrt{p} c_0))^{2/(2-\xi)}$.  Also, for $x >d_0$, $m_n(x) = \beta_n$ and thus,     
\begin{eqnarray*}
\M_{n_2}(d, {\theta}) & = &  \P_{n_2} \left[g_{n_2,d, \theta} (V)\right],
\end{eqnarray*}
where for $V = (U, \epsilon)$, $U \sim \mbox{Uniform}[-1,1]$,
\begin{eqnarray*}
g_{n_2,d,\theta}(V) & = &\left(\beta_n + \epsilon - \frac{\beta +\alpha}{2}  \right) 1\left[\mu + K n_1^{-\gamma} U \in (d_0, d]\right]\\
& = &\left(\beta_n + \epsilon - \frac{\beta +\alpha}{2}  \right) 1\left[U \in \left(\frac{d_0- \mu}{Kn_1^{-\gamma}}, \frac{d- \mu}{Kn_1^{-\gamma}}\right]\right].
\end{eqnarray*}
Consequently, for  $n > N^{(1)}_{\tau}$,
\begin{eqnarray*}
M_{n_2}(d, \theta) & = &\frac{1}{2}\left(\beta_n - \frac{\beta +\alpha}{2}  \right) \lambda\left([-1,1] \cap \left(\frac{d_0- \mu}{Kn_1^{-\gamma}}, \frac{d- \mu}{Kn_1^{-\gamma}}\right] \right).
\label{eq:mn2cp}
\end{eqnarray*}
As $\gamma < \nu$,  $d_0 \in \mathcal{D}_\theta$ for all $\theta \in \Theta_{n_1}^\tau$, for $n > N^{(2)}_\tau := (1/p) (K_\tau/K)^{1/(\nu-\gamma)}$ 
 the intervals  $$\left\{ \left((d_0-\mu)/(Kn_1^{-\gamma}), (d -\mu)/(Kn_1^{-\gamma})\right]: d>d_0,  d \in \mathcal{D}_\theta, \theta \in \Theta_{n_1}^\tau\right\}$$ are all contained in $[-1,1]$. Therefore, for $n > N^{(3)}_{\tau} := \max(2N^{(1)}_{\tau}, N^{(2)}_{\tau}  )$,
\begin{eqnarray*}
M_{n_2}(d, \theta) & =  &  \frac{1}{2}\left(\beta_n - \frac{\beta +\alpha}{2}  \right) \frac{d - d_0}{K n_1^{-\gamma}}. 
\end{eqnarray*}
Note that $M_{n_2}(d_0, \theta) = 0$ for all $\theta \in \R^3$. Further, let $\rho_n^2(d, d_0) = n^{\gamma-\xi} |d - d_0|$. Then, for $n > N^{(3)}_{\tau}$,
\begin{eqnarray}
M_{n_2}(d, \theta) - M_{n_2}(d_0, \theta) & \geq & \left(\beta_n - \frac{\beta_n +  \alpha_n}{2}  - \frac{K_\tau}{\sqrt{n_1}} \right) \frac{d - d_0}{2K n_1^{-\gamma}} \nonumber \\
& =& \left( \frac{\beta_n - \alpha_n}{2}  - \frac{K_\tau}{\sqrt{n_1}} \right) \frac{d - d_0}{2K n_1^{-\gamma}} \nonumber \\
& = & \left(\frac{c_0 n^{-\xi}}{2} - \frac{K_\tau}{\sqrt{n_1}}  \right) \frac{d - d_0}{2K n_1^{-\gamma}} \nonumber \\
& \geq& c_\tau \rho_n^2(d, d_0),
\label{eq:distcp}
\end{eqnarray}
for some $c_\tau>0$ (depending on $\tau$ through $K_\tau$). The last step follows from the fact that $\xi < 1/2$. Also, the above lower bound can be shown to hold for the case $d > d_0$ as well.
Further, to apply Theorem \ref{th:ratethmc6}, we need to  bound 
\be \label{eq:modcont}
\sup_{\theta \in \Theta_{n_1}^\tau} E^* \sup_{\substack{|d- d_0| < n^{\xi-\gamma}\delta^2, \\d \in \mathcal{D}_\theta }} \sqrt{n_2}\left| (\M_{n_2} (d, \theta) -M_{n_2}(d, \theta))  - (\M_{n_2}(d_0, \theta)  -M_{n_2}(d_0, \theta) ) \right|.  
\ee
Note that for $d> d_0$, the expression in $| \cdot |$ equals $(1/\sqrt{n_2})  \G_{n_2} g_{n_2, d,\theta}$. 
The class of functions $\mathcal{F}_{\delta,\theta}= \{g_{n_2, d,\theta}: 0 \leq  d-d_0 < n^{\xi-\gamma}\delta^2, d \in \mathcal{D}_\theta \}$ is VC with index at most 3 (for every $(\delta, \theta)$) and is enveloped by
\be
M_{\delta, \theta}(V) = \left(|\epsilon| + \frac{\beta_n -\alpha_n}{2} + \frac{K_\tau}{\sqrt{n_1}}  \right) 1\left[U  \in \left[\frac{d_0-\mu}{K {n_1}^{-\gamma}}, \frac{d_0 - \mu +\delta^2 n^{\xi-\gamma}}{K {n_1}^{-\gamma}}\right]\right]. \nonumber
\ee
Note that 
\begin{eqnarray*}
\lefteqn{E\left[M_{\delta, \theta}(V)\right]^2}\\
 & =&  \frac{1}{2}E\left[ \left(|\epsilon| + \frac{\beta_n -\alpha_n}{2} + \frac{K_\tau}{\sqrt{n_1}}  \right)^2\right] \lambda \left[[-1,1] \cap \left[\frac{d_0-\mu}{K {n_1}^{-\gamma}}, \frac{d_0 - \mu +\delta^2 n^{\xi-\gamma}}{K {n_1}^{-\gamma}}\right]\right]\\
& \leq&  \frac{1}{2}E\left[ \left(|\epsilon| + \frac{\beta_n -\alpha_n}{2} + \frac{K_\tau}{\sqrt{n_1}}  \right)^2\right] \lambda \left[\frac{d_0-\mu}{K {n_1}^{-\gamma}}, \frac{d_0 - \mu +\delta^2 n^{\xi-\gamma}}{K {n_1}^{-\gamma}}\right]\\
& \leq&  C^2_\tau \frac{n^{\xi -\gamma}\delta^2}{n^{-\gamma}}  = C^2_\tau n^{\xi} \delta^2,
\end{eqnarray*}
where $C_\tau $ is positive constant (it depends on $\tau$ through $K_\tau$). Further, the uniform entropy integral for $\mathcal{F}_{\delta,\theta}$ is bounded by a constant which only depends upon its VC-index (which, as noted above, is uniformly bounded in $(\delta, \theta)$), i.e., the quantity
$$ J(1, \mathcal{F}_{\delta,\theta}) = \sup_Q \int_0^1 \sqrt{1 + \log N(u \|M_{\delta, \theta}\|_{Q,2}, \mathcal{F}_{\delta,\theta}, L_2(Q)) } d u $$
is uniformly bounded in $(\delta, \theta)$; see Theorems 9.3  and 9.15 of \citet{K08} for more details. Using Theorem 2.14.1 of \citet{VW96}, 
\be \label{eq:bndI}
E^* \sup_{\substack{0 \leq  d- d_0 < n^{\xi-\gamma}\delta^2 \\ d \in \mathcal{D}_\theta }} \left|\G_{n_2} g_{n_2, d,\theta}\right|  \leq J(1, \mathcal{F}_{\delta,\tau}) \|M_{\delta, \theta}\|_2 \leq  C_\tau n^{\xi/2} \delta.
\ee
Note that this bound does not depend on $\theta$ and can be shown to hold  for the case $d \leq  d_0$ as well. 
Hence, we  get the bound $\phi_n(\delta) = n^{\xi/2} \delta  $ on the modulus of continuity. 
 Further, for $n > N^{(3)}_{\tau}$, \eqref{eq:distcp} holds for all $d \in \mathcal{D}_\theta$,  and \eqref{eq:bndI} is valid for all $\delta>0$. Hence, we do not need to justify a condition of the type $P\left(\rho_n(\hat{d}_n, d_n)\geq  \kappa_n \right) \rightarrow 0$ to apply Theorem \ref{th:ratethmc6}. 
For $r_n= n^{1/2 - \xi/2} $, the relation $r^2_n \phi_n(1/r_n) \leq \sqrt{n}$ is satisfied. Consequently, $r^2_n (n^{\gamma - \xi } (\hat{d}_n -d_0)) =  n^\eta(\hat{d}_n -d_0) = O_p(1)$.
\qed
\subsection{Proof of  Theorem \ref{th:limproccp}}\label{pf:limproccp}
 For any $L>0$, we start by justifying the conditions of Theorem \ref{th:tight} to prove tightness of the process $Z_{n_2}(h, \hat{\theta}_{n_1})$, for $h \in [-L, L]$. For sufficiently large $n$, the set $\{ {h: d_0 + h/n^{\eta} \in \mathcal{D}_{{\theta}}} \}$ contains $[-L, L]$ for all $\theta \in \Theta_{n_1}^\tau$ and hence, it is not necessary to extend $Z_{n_2}$ (equivalently, $f_{n_2, h, \theta}$) as done in \eqref{zextn}.  Further, for a fixed $\theta \in \Theta_{n_1}^\tau$ (defined in \eqref{eq:thetaset}), an envelope for the class of functions $\{ f_{n_2,h,{\theta}}: |h| \leq L \}$ is given by
\begin{eqnarray*}
F_{n_2, \theta} (V)  & =& n_2^{1/2- \xi} \left(\frac{\beta_n - \alpha_n}{2}+ \frac{K_\tau}{\sqrt{n_1}} + |\epsilon| \right) \times \\
& & 1\left[\mu + U K n_1^{-\gamma} \in  [ d_0 - L n^{-\eta}, d_0 + L n^{-\eta}] \right].   \nonumber
\end{eqnarray*}
Note that 
\be
P F^2_{n_2, \theta} \lesssim n^{1 - 2\xi} \left(\left(\frac{\beta_n - \alpha_n}{2} + \frac{K_\tau}{\sqrt{n_1}} \right)^2 + \sigma^2\right) \frac{2L n^{-\eta}}{2K n_1^{-\gamma}} \nonumber
\ee
As  $\eta = 1 + \gamma - 2 \xi$, the right side (which does not depend on $\theta$) is $O(1)$. Moreover, the bound is uniform in $\theta$, $\theta \in \Theta_{n_1}^\tau$. 
Let $K_0$ be a constant (depending on $\tau$) such that $K_0 \geq {(\beta_n - \alpha_n)}/{2} + {K_\tau}/{\sqrt{n_1}}$. Then, for $t>0$, $P F^2_{n_2, \theta} 1[F_{n_2, \theta} > \sqrt{n_2} t] $ is bounded by
\begin{eqnarray*}
& & n^{1 - 2\xi} P\left( (K_0 + |\epsilon|)^2 1\left[\mu + U K n_1^{-\gamma} \in  [ d_0 - L n^{-\eta}, d_0 + L n^{-\eta}] \right] \times \right. \\
& &   \left.  1\left[ n^{1/2 - \xi}(K_0 + |\epsilon|) > \sqrt{n_2} t  \right]\right).
\end{eqnarray*} 
As $\epsilon$ and $U$ are independent, the above is bounded up to a constant by 
$$ P (K_0 + |\epsilon|)^2 1\left[ (K_0 + |\epsilon|) > \sqrt{p}{n^\xi} t  \right] $$
which goes to zero. This justifies condition \eqref{eq:cond1} and \eqref{eq:cond2} of Theorem \ref{th:tight}. Let $\tilde{\rho} (h_1, h_2) = |h_1 - h_2|$. For any $L>0$, the space $[-L,L]$ is totally bounded with respect to $\tilde{\rho}$.
For $h_1, h_2 \in [-L,L]$ and $\theta \in \Theta_{n_1}^\tau$, we have
\begin{eqnarray*}
P(f_{n_2, h_1, \theta} - f_{n_2, h_2, \theta})^2 & \lesssim &  n^{1 - 2\xi} \frac{|h_1 - h_2|n^{-\eta}}{2K n_1^{-\gamma}}  E\left[K_0 + |\epsilon|\right]^2.
\end{eqnarray*} 
The right side is bounded (up to a constant multiple depending on $\tau$) by $|h_1 - h_2|$ for all choices of $\theta$, $\theta \in \Theta_{n_1}^\tau$. Hence, condition \eqref{eq:cond3} is satisfied as well. Condition \eqref{eq:cond4} can be justified in a manner mentioned later. 
Further, the class of functions $\{ f_{n_2,h,{\theta}}: |h| \leq L \}$ is VC of index at most 3 with envelope $F_{n_2,\theta}$. Hence, it has a bounded entropy integral with the bound only depending on the VC index of the class (see Theorems 9.3  and 9.15 of \citet{K08}) and hence, condition \eqref{eq:cond5} is also satisfied. Also, the measurability condition \eqref{eq:condmes} can be shown to hold by approximating $\mathcal{F}_{n_2,\delta} = \{ f_{n_2, h_1, \theta} - f_{n_2, h_2, \theta}: |h_1 -h_2| < \delta \} $ (defined in Theorem \ref{th:tight})  by the countable class involving only rational choices of $h_1$ and $h_2$. Note that the  supremum over this countable class is measurable and it agrees with supremum over $\mathcal{F}_{n_2,\delta}$.  
Thus $\G_{n_2} f_{n_2,h,\hat{\theta}}$ is tight in $l^\infty([-L, L])$. 

Next, we apply Corollary \ref{lm:fidiconv} to deduce the limit process. Note that for $\theta \in \Theta_{n_1}^\tau$ and $|h| \leq L$,
\begin{eqnarray*}
\zeta_{n_2}(h, \theta) & = &  n_2^{1 - \xi } \left(\alpha_n 1(h \leq 0) + \beta_n 1(h > 0) - \frac{\alpha + \beta}{2}\right) \frac{h n^{-\eta}}{2K n_1^{-\gamma}}\\
& = &  (1-p)^{1 - \xi } \left(\alpha_n 1(h \leq 0) + \beta_n 1(h > 0) - \frac{\alpha + \beta}{2}\right) \frac{h n^\xi}{2K p^{-\gamma}} \\
& = &  \frac{(1-p)^{1 - \xi} p^{\gamma} n^\xi}{2K} h \left(\alpha_n 1(h \leq 0) - \beta_n 1(h > 0) - \frac{\alpha_n  + \beta_n }{2}\right)+ R_n. 
\end{eqnarray*}
The remainder term $R_n$ in the last step accounts for replacing $\alpha + \beta$ by $\alpha_n + \beta_n$ in the expression for $\zeta_{n_2}$ and  is bounded (uniformly in $\theta \in \Theta_{n_1}^\tau$) up to a constant by 
$$ n^\xi L \left(|\alpha_n - \alpha| + |\beta_n - \beta| \right) = O(n^{\xi-1/2}) .$$ As $\xi < 1/2$, $\sqrt{n_2} P f_{n_2, h, \theta}$ converges uniformly to $|h| \left({(1-p)^{1 - \xi} p^{\gamma} c_0}\right)/({4K}) $. Condition \eqref{eq:cond4} can be justified by calculations parallel to the above.  
Further, $P f_{n_2, h, \theta} = \zeta_{n_2}(h,\theta)/\sqrt{n_2}$ converges to zero (uniformly over $\theta \in \Theta_n^{\tau}$) and hence, the covariance function of the limiting Gaussian process (for $h_1, h_2 >0$) is given by
\begin{eqnarray*}
\lefteqn{\lim_{n \rightarrow \infty} P f_{n_2,h_1,{\theta}} f_{n_2,h_1,{\theta}} } \\
& = & \lim_{n \rightarrow \infty} n_2^{1- 2 \xi} \left[\left(\alpha_n 1(h \leq 0) + \beta_n 1(h > 0) - \frac{\alpha + \beta}{2}\right)^2 + \sigma^2\right] \frac{h_1 \wedge h_2  n^{-\eta}}{2K n_1^{-\gamma}}\\
& = & \frac{(1-p)^{1 - 2\xi} p^{\gamma} \sigma^2 }{2K} (h_1 \wedge h_2). 
\end{eqnarray*}
Analogous results can be established for other choices of $(h_1, h_2)  \in [-L, L ]^2$. Also, the above convergence can be shown to be uniform in $\theta \in \Theta_n^{\tau}$ by a calculation similar to that  done for $\zeta_{n_2}$. 
This justifies the form of the limit $Z$. Hence, we get the result.
\qed
\subsection{Proof of  Theorem \ref{limitcp}}\label{pf:limitcp}
 As Var($Z(t) - Z(s)) \neq 0$, uniqueness of the argmin follows immediately from Lemma 2.6 of \cite{KP90}. Also, $Z(h) \rightarrow \infty$ as $|h| \rightarrow \infty$ almost surely. This is true as $$Z(h) = |h| \left[ \sqrt{\frac{(1-p)^{1-2\xi} p^\gamma}{2K}}\sigma \frac{B(h)}{|h|} + \frac{(1-p)^{1-\xi} p^\gamma}{2K} \frac{c_0}{2} \right]$$ with $B(h)/|h|$ converging to zero almost surely as $|h| \rightarrow \infty$. Consequently, the unique argmin of $Z$ is tight and $Z \in C_{min}(\mathbb{R})$ with probability one. An application of argmin continuous mapping theorem \cite[Theorem 2.7]{KP90} then gives us distributional convergence. 
By dropping a constant multiple, it can be seen that 
$$ \argmin_h Z(h) = \argmin_h \left[ \sigma B(h) + \sqrt{\frac{(1-p)p^\gamma}{2K}} \frac{c_0}{2} |h|\right]. $$
As $\sigma \sqrt{\lambda_0 }  = \sqrt{({(1-p)p^\gamma})/({2K})} ({c_0}\lambda_0 )/{2} $, by the rescaling property of Brownian motion,
{\allowdisplaybreaks \begin{eqnarray*}
\lefteqn{\argmin_h \left[ \sigma B(h) + \sqrt{\frac{(1-p)p^\gamma}{2K}} \frac{c_0}{2} |h|\right]}\hspace{1in}\\
 & = & \lambda_0\,  \argmin_{v} \left[ \sigma B(\lambda_0 v) + \sqrt{\frac{(1-p)p^\gamma}{2K}} \frac{c_0}{2}|\lambda_0| |v|\right] \\   
& \stackrel{d}{=} &  \lambda_0\, \argmin_{v} \left[ \sigma \sqrt{\lambda_0 } B(v) + \sqrt{\frac{(1-p)p^\gamma}{2K}} \frac{c_0}{2} \lambda_0 |v|
\right] \\   
& =  &  \lambda_0\, \argmin_{v} \left[ B(v) + |v| \right]. 
\end{eqnarray*}}
The result follows.
 \qed
\bibliography{References}

\newpage
\section{Supplementary Material} 
\subsection{Proof of Lemma \ref{lm:const}} \label{pf:const}
Note that $\M_n (\hat{d}_n(\hat{\theta}_n), \hat{\theta}_n) - \M_n ({d}_n, \hat{\theta}_n)$ is not positive by definition of $\hat{d}_n(\hat{\theta}_n)$. Hence, 
\begin{eqnarray*}
\lefteqn{P\left[\rho_n(\hat{d}_n(\hat{\theta}_n), d_n) \geq  \kappa_n, \hat{\theta}_n \in \Theta_n^\tau \right] }\\
& \leq & E \left[ P\left[\rho_n(\hat{d}_n(\hat{\theta}_n), d_n) \geq  \kappa_n \mid \hat{\theta}_n\right] 1 \left[\hat{\theta}_n \in \Theta_n^\tau \right] \right]\\
& \leq & \sup_{\theta \in \Theta_n^\tau} P\left[2\rho_n(\hat{d}_n({\theta}), d_n) \geq  \kappa_n \right] \\
& \leq & \sup_{\theta \in \Theta_n^\tau} P\left[ M_n(\hat{d}_n({\theta}), \theta) - M_n(d_n, \theta) \geq c^\tau_n(\kappa_n) \right]  \\
& \leq & \sup_{\theta \in \Theta_n^\tau} P\left[ M_n(\hat{d}_n({\theta}), \theta) - M_n(d_n, \theta) - \left(\M_n (\hat{d}_n({\theta}), {\theta}) - \M_n ({d}_n, {\theta}) \right)\geq c^\tau_n(\kappa_n) \right] \\
& \leq & \sup_{\theta \in \Theta_n^\tau} P\left[ 2 \sup_{\substack{d \in \mathcal{D}_\theta}} \left| \M_n({d}, \theta) - M_n(d, \theta) \right| \geq c^\tau_n(\kappa_n)  \right]. 
\end{eqnarray*}
As the probability in right side converges to zero and $\tau >0$ is arbitrary, we get the result. \qed

\subsection{Proof of Lemma \ref{lm:fidiconvgen}}\label{pf:fidiconv}
In light of Theorem \ref{th:tight}, we only need to establish the finite dimensional convergence.  Given the independence of vectors $V_i$s with $\hat{\theta}_n$, the drift process $\zeta_n(\cdot,\hat{\theta}_n)$ is independent of the centered process $(Z_n - \zeta_n)(\cdot,\hat{\theta}_n)$ given $\hat{\theta}_n$. Hence, it suffices to show the finite dimensional convergence of these two processes separately.   
On the set $\hat{\theta} \in \Theta_n^\tau$, 
\begin{eqnarray*}
{|\zeta_n(h, \theta_n + n^{-\nu} \Delta_{\hat{\theta}_n}) -  \zeta(h, \xi)|} &\leq & \sup_{\theta \in \Theta_n^\tau} |\zeta_n(h, \theta_n + n^{-\nu} \Delta_{{\theta}}) -  \zeta(h,  \Delta_{{\theta}}) | \\
& & + |\zeta(h, \Delta_{\hat{\theta}_n}) -  \zeta(h, \xi) |.
\end{eqnarray*}
In light of conditions 3 and 4, an application of Skorokhod representation theorem then ensures the convergence of finite dimensional  marginals of $\zeta_n(\cdot, \theta_n + n^{-\nu} \Delta_{\hat{\theta}_n})$ to that of the process $\zeta(\cdot, \xi)$. To establish the finite dimensional convergence of the centered process $Z_n - \zeta_n$, we require the following result that arises from a careful examination of the proof of the Central Limit Theorem for sums of independent zero mean random variables \cite[pp. 359 - 361]{BL66}. 
\begin{thm}\label{th:lberg}
For $n \geq 1$, let $\{ X_{i,n} \}_{i=1}^n$ be independent and identically distributed random variables with 
mean zero and variance $\sigma^2_n >0$. Let $S_n = (1/ \sqrt{n} ) \sum_{i\leq n} X_{i,n}$, $F_n$ be the distribution function of $S_n$ and for $\kappa >0$, 
\be
	L_{n}(\kappa) = E\left[X_{1, n}^2 1\left[ |X_{1,n}| > \kappa \sqrt{n} \right]  \right] \nonumber
\ee 
Then, for any $t \in \R$ with $|\sigma_n t| \leq \sqrt{2 n}$, we have
\be
|\hat{F}_n(t) - \hat{\Phi}(\sigma_n t)| \leq \kappa \sigma^2_n |t|^3 + t^2 L_n (\kappa) + \frac{\sigma^4_n t^4 \exp(\sigma_n^2 t^2)}{n} 
\ee
Here $\hat{\ }$  denotes characteristic function, so that $\hat{\Phi}(t) = \int_{\R} e^{\imath tx}\Phi\{dx\}$.
\end{thm}
We now prove Lemma \ref{lm:fidiconvgen}. Let $k \geq 1$, $c = (c_1, \ldots c_k) \in \R^k$, $h = (h_1, \ldots, h_k) \in \R^k$ and for $\Delta_\theta = n^\nu(\theta - \theta_n)$,
\begin{eqnarray*}
T_n (\Delta_\theta) = T_n (h, c, \Delta_\theta)  & = & \sum_{j \leq k} c_j \G_n f_{n,h_j, \theta_n + n^{-\nu} {\Delta}_\theta}.
\end{eqnarray*}
Note that 
\begin{eqnarray*}
\pi^2_n(\Delta_\theta) = \mbox{Var} (T_n (\Delta_\theta)) = \mbox{Var}\left( \sum_{j \leq k} c_j f_{n,h_j, \theta_n + n^{-\nu} {\Delta}_\theta} \right).
\end{eqnarray*}
converges uniformly in $\Delta_\theta$, $\theta \in \Theta_n^\tau$ to  $$\pi_0^2(\Delta_\theta) : = \sum_{j_1,j_2}c_{j_1}c_{j_2}C(h_{j_1},h_{j_2}, \Delta_\theta).$$ 
By L\'{e}vy continuity theorem, it suffices to show that the characteristic function
$$(c_1, \ldots c_k) \mapsto E \exp \left[\imath T_n({\Delta}_{\hat{\theta}_n}) \right]$$
converges to $E \exp \left[\imath \pi_0(\xi) Z \right]$, where $Z$ is a standard normal random variable independent of $\xi$ and ${\Delta}_{\hat{\theta}_n}$.  Note that
\begin{eqnarray*}
\lefteqn{\left|E \exp \left[\imath T_n({\Delta}_{\hat{\theta}_n}) \right] - E \exp \left[\imath \pi_0(\xi) Z \right] \right|} \\
& \leq & \left|E \exp \left[\imath T_n({\Delta}_{\hat{\theta}_n}) \right] - E \exp \left[\imath \pi_n({\Delta}_{\hat{\theta}_n}) Z \right] \right|\\
& & + \left| E \exp \left[\imath \pi_n({\Delta}_{\hat{\theta}_n}) Z \right] - E \exp \left[\imath \pi_0(\xi) Z \right] \right|.
\end{eqnarray*}
The right side is further bounded (up to $4 \epsilon$) by
{\be
\begin{split}
\lefteqn{\sup_{\theta \in \Theta_n^\tau} \left|E \exp \left[\imath T_n({\Delta}_{{\theta}}) \right] - E \exp \left[\imath \pi_n({\Delta}_{{\theta}}) Z \right] \right|}\\
 & +   \sup_{\theta \in \Theta_n^\tau} \left| E \exp \left[\imath \pi_n({\Delta}_{{\theta}}) Z \right] - E \exp \left[\imath \pi_0({\Delta}_{{\theta}}) Z \right] \right|  \\
&  + \left| E \exp \left[\imath \pi_0({\Delta}_{\hat{\theta}_n}) Z \right] - E \exp \left[\imath \pi_0(\xi) Z \right] \right|.
\end{split}
\label{eq:fidibnd}
\ee
}
The second term in the above display is precisely $\sup_{\theta \in \Theta_n^\tau} |\exp(-\pi_n^2(\Delta_\theta)/2) - \exp(-\pi_0^2(\Delta_\theta)/2) |$ which converges to zero. The third term converges to zero by continuous mapping theorem. 
To control the first term, we apply Theorem \ref{th:lberg}. Let 
\begin{eqnarray*}
L_n(\kappa, \Delta_\theta) & = & P\left[\left[\sum_{j \leq k} c_j (f_{n,h_j, \theta_n + n^{-\nu} {\Delta}_\theta} - Pf_{n,h_j, \theta_n + n^{-\nu} {\Delta}_\theta}) \right]^2 \times \right.\\
& & \left. 1\left[\left|\sum_{j \leq k} c_j (f_{n,h_j, \theta_n + n^{-\nu} {\Delta}_\theta} - P f_{n,h_j, \theta_n + n^{-\nu} {\Delta}_\theta})\right| > \sqrt{n} \kappa \right]\right].
\end{eqnarray*}
Then, by Theorem \ref{th:lberg}, the first term in \eqref{eq:fidibnd} is bounded by
\begin{eqnarray*}
\sup_{\theta \in \Theta_n^\tau}\left[\kappa \pi^2_n(\Delta_\theta) + L_n (\kappa, \Delta_\theta) + \frac{\pi^4_n(\Delta_\theta) \exp(\pi^2_n(\Delta_\theta) )}{n} \right]
\end{eqnarray*}
whenever $\sup_{\theta \in \Theta_n^\tau} |\pi_n(\Delta_\theta)|  \leq 2 \sqrt{n}$, which happens eventually as the right side is $O(1)$. To see this, note that 
\begin{eqnarray}
\left|\sum_{j \leq k} c_j f_{n,h_j, \theta_n + n^{-\nu} {\Delta}_\theta} \right| \leq 2k \max_j (|c_j |\vee 1) F_{n,\theta}.
\label{eq:boundvar}
\end{eqnarray}
Then, by \eqref{eq:cond1},  $\sup_{\theta \in \Theta_n^\tau} |\pi_n(\Delta_\theta)| \leq 2k \max_j (|c_j |\vee 1) \sup_{\theta \in \Theta_n^\tau} P F^2_{n, \theta} = O(1)$.
Further, using \eqref{eq:boundvar}, 
\begin{eqnarray*}
L_n(\kappa, \Delta_\theta) & \leq&  \left(2k \max_j (|c_j |\vee 1)\right)^2 \times \\
& & P \left[\left[F^2_{n,\theta} + P F^2_{n,\theta}\right] 1\left[F > \frac{\sqrt{n} \kappa}{\max_j (|c_j |\vee 1)} -  P F_{n,\theta} \right]\right], 
\end{eqnarray*}  
which converges to zero uniformly in $\theta \in \Theta_n^\tau$ due to conditions \eqref{eq:cond1} and \eqref{eq:cond2}. Hence,
\begin{eqnarray*}
{\lim \sup_{n \rightarrow \infty } \sup_{\theta \in \Theta_n^\tau} \left|E \exp \left[\imath T_n({\Delta}_{{\theta}}) \right] - E \exp \left[\imath \pi_n({\Delta}_{{\theta}}) Z \right] \right|} & \leq & \kappa \lim \sup_{n \rightarrow \infty }\sup_{\theta \in \Theta_n^\tau} \pi^2_n(\Delta_\theta). \nonumber
\end{eqnarray*}
As $\sup_{\theta \in \Theta_n^\tau} \pi^2_n(\Delta_\theta) = O(1)$ and $\kappa>0$ is arbitrary, we get the result.
\qed
\subsection{Proof of Proposition \ref{lm:vwcont}}\label{pf:vwcont}
We  show that the result holds for $h>0$. The case $h<0$ can be shown analogously. In what follows, the dependence on $h$ is suppressed in the notations for convenience.

To start with, note that $\xi_n = n^\nu(\hat{d}_1 - d_0)$ is $O_p(1)$ and it converges in distribution to a tight random variable $\xi$ with a continuous bounded density on $\R$. In particular, $P\left[ |\xi_n| < \delta, |\xi_n| > K_{\delta/2}  \right]$ converges to $P\left[ |\xi| < \delta , |\xi| > K_{\delta/2}  \right] \leq C \delta$, for some $C>0$.  

For $u \in \R $, let $F_{n_2}^{u}$ denote the distribution function of $T_{n_2}(u)$, where $$ T_{n_2}(u) = Z_{n_2}(h, \alpha_n, \beta_n, d_0 + u n^{-\nu}) - Z_{n_2}(h,{\alpha}_n, {\beta}_n, {d}_0).$$
Also, let $\pi_{n_2}^2  := \pi_{n_2}^2 ( u) = \mbox{Var}[T_{n_2}( u)].$ Conditional on $\xi_n = u$, $T_{n_2}$ is distributed as $T_{n_2}(u)$. Also, let $\hat{\ }$  denote characteristic function, so that $\hat{\Phi}(t) = \int_{\R} e^{\imath tx}\Phi \{dx\}$. By L\'{e}vy continuity theorem, it suffices to show that  for any $t \in \R$,
$$ E \left[ \exp\left(\imath t T_{n_2} \right)\right] - \hat{\Phi} ( t \pi_0) $$
converges to zero.
Note that
\begin{eqnarray}
\lefteqn{\left|E \left[ \exp\left(\imath t T_{n_2} \right)\right] - \hat{\Phi} ( t \pi_0)\right|} \nonumber\\
& = & \left|E\left[E \left[ \left. \exp\left(\imath t T_{n_2} \right) - \hat{\Phi} ( t \pi_0) \right| \xi_n \right]\right]\right| \nonumber\\
& = &  \sup_{\delta \leq |u| \leq K_{\delta/2}} \left|\hat{F}_{n_2}^{u}(t) - \hat{\Phi} ( t \pi_0)\right| + 2 P\left[ |\xi| < \delta , |\xi| > K_{\delta/2}  \right]  \nonumber\\
& = &  \sup_{\delta \leq |u| \leq K_{\delta/2}} \left|\hat{F}_{n_2}^{u}(t) - \hat{\Phi} (t \pi_{n_2}(u))\right| \nonumber\\
& & + \sup_{\delta \leq |u| \leq K_{\delta/2}}\left|\hat{\Phi} (t \pi_{n_2}(u)) - \hat{\Phi} ( t \pi_0)\right| + C \delta
\label{eq:termtocon}
\end{eqnarray}
We first show that $\pi_{n_2}(u)$ converges to $\pi_0$ uniformly over $u$, $ \delta \leq |u| \leq K_{\delta/2}$ which will ensure that the second term on the right side of the above display converges to zero. To show this, note that
\begin{eqnarray}
\lefteqn{T_{n_2}(u) } \nonumber \\
 &= & \frac{1}{n_2^\xi} \sum_{i=1}^{n_2} \left(\frac{\beta_n- \alpha_n }{2} + \epsilon_i \right)\left[ 1\left[ U_i K {n_1}^{-\gamma} \in (-u n^{-\nu}, -u n^{-\nu} + hn^{-\eta}]\right]\right. \nonumber \\
  & & \left.- 1\left[ U_i K {n_1}^{-\gamma} \in (0, hn^{-\eta}]\right]\right]\nonumber \\
& = & \frac{1}{n_2^\xi} \sum_{i=1}^{n_2} \left(\frac{\beta_n- \alpha_n }{2} + \epsilon_i \right)\left[ 1\left[ U_i K p^{-\gamma}  \in (-u n^{-\nu+ \gamma}, -u n^{-\nu + \gamma} + hn^{-\nu}]\right]\right. \nonumber \\
&  & \left.- 1\left[ U_i K p^{-\gamma} \in (0, h n^{-\nu}]\right]\right].\nonumber
\end{eqnarray}
Hence, $\pi_{n_2}$ can be simplified as 
\begin{eqnarray*}
\lefteqn{\pi_{n_2}^2(u) = \mbox{Var}[T_{n_2}(u)]  }\\
 & =& \frac{n_2}{n_{2}^{2\xi}} E \left[ \left({(\beta_n- \alpha_n )}/{2} - \epsilon \right)\left[ 1\left[ U K p^{-\gamma}  \in (-u n^{-\nu+ \gamma}, -u n^{-\nu + \gamma} + h n^{-\nu}]\right] \right.\right.\\
& & \left. \left.- 1\left[ U K p^{-\gamma} \in (0, h n^{-\nu}]\right] \right]
 \right]^2 \\
& = & \frac{n_2}{n_{2}^{2\xi}} E \left[ \left({(\beta_n- \alpha_n )^2}/{4} + \sigma^2 \right) \right.\times \\
&  & \left. 1\left[ U K p^{-\gamma}  \in (-u n^{-\nu+ \gamma}, -u n^{-\nu + \gamma} + h n^{-\nu} ] \triangle (0, h n^{-\nu}]\right]\right].
\end{eqnarray*}
For $n > N_1 = (h/|\delta|)^{1/\nu} $, the sets $(-u n^{-\nu+ \gamma}, -u n^{-\nu + \gamma} + h n^{-\nu}]$ and $(0, h n^{-\nu}]$  are disjoint  and hence,
\begin{eqnarray}
\pi_{n_2}^2(u) & = & \frac{n_2}{n_{2}^{2\xi}} \left(\frac{c_0^2}{4}n^{-2\xi} + \sigma^2 \right)\left[ \frac{2 h n^{-\nu }}{2 K p^{-\gamma}}\right] = \pi_0^2 + \tilde{C} n^{-2\xi},
\label{eq:taunexp}
\end{eqnarray}
where $\tilde{C} = c_0^2(1-p)^{1- 2\xi} h /(4K)$. Consequently, $\pi_{n_2}^2( u)$ converges to $\pi_0^2$ uniformly over $u$.

Next, we apply Theorem \ref{th:lberg} to show that the first term in \eqref{eq:termtocon} converges to zero.
Write $T_{n_2}(h)$ as $(1/\sqrt{n_2})\sum_{i\leq n_2} R_{i, n_2}(u)$, where 
\begin{eqnarray*}
\lefteqn{R_{i, n_2}(u)}\\
 & = & n_2^{1/2-\xi } \left(\frac{\beta_n- \alpha_n }{2} + \epsilon_i \right)\left[ 1\left[ U_i K p^{-\gamma}  \in (-u n^{-\nu+ \gamma}, -u n^{-\nu + \gamma} + hn^{-\nu}]\right] \right.\\
& & \left.- 1\left[ U_i K p^{-\gamma} \in (0, h n^{-\nu}]\right]\right]. 
\end{eqnarray*}
As $\gamma < \nu$, the intervals $(-u n^{-\nu+ \gamma}, -u n^{-\nu + \gamma} + h n^{-\nu}]$ and $(0, h n^{-\nu}]$ are both contained in $[- K p^{-\gamma}, K p^{-\gamma}]$ for $n > N_2 = \max\left\{(K_{\delta/2}/ Kp^{-\gamma})^{1/(\nu-\gamma)}, (h/ Kp^{-\gamma})^{1/\nu}\right\}$ and have the same Lebesgue measure $h n^{-\nu}$. Hence, 
$E[T_{n_2}(u)] = E[ R_{i, n_2}(u)]= 0$ for $n > N_1$. Thus $T_{n_2}(u)$ is a normalized sum of mean zero random variables.
Let
\be
	L_{n_2}(\kappa, u) = E\left[R_{i, n_2}(u)^2 1\left[ |R_{i, n_2}(u)| > \sqrt{n_2}\kappa \right]  \right].
\ee 
Using Theorem \ref{th:lberg},  for any $\kappa > 0$, $n_2 > \max(N_1, N_2)$ and $|\pi_{n_2}(u) t| \leq \sqrt{2 n_2}$ (which holds eventually) we have
\be
|\hat{F}_{n_2}^{u}(t) - \hat{\Phi}(\pi_{n_2}(u) t)| \leq \kappa \pi^2_{n_2}(u)|t|^3 + t^2 L_{n_2} (\kappa, u) + \frac{\pi^4_{n_2}(u) t^4 \exp(\pi^2_{n_2}(u) t^2)}{n_2} 
\ee
As $ \sup_{\delta \leq |u| \leq K_{\delta/2}} \pi_{n_2}(u) =O(1)$ and $\kappa$ is arbitrary, it suffices to show that 
$$ \sup_{\delta \leq |u| \leq K_{\delta/2}} 	L_{n_2}(\kappa, u) $$ converges to zero.
Using the expression for $\pi_{n_2}$ in \eqref{eq:taunexp}, we have
\begin{eqnarray*}
\lefteqn{	L_{n_2}(\kappa, u) } \\
& \leq & \frac{n_2}{n_{2}^{2\xi}} E\left[ \epsilon^2 \left[ 1\left[ U K p^{-\gamma}  \in (-u n^{-\nu+ \gamma}, -u n^{-\nu + \gamma} + h n^{-\nu}] \triangle (0, h n^{-\nu}]\right]\right] \times \right.\\
&& \left. 1\left[n_2^{1/2-\xi}|\epsilon| > \sqrt{n_2} \kappa \right]\right] \nonumber \\
& & + \tilde{C} n^{-2\xi} \\
& \lesssim & n^{-2\xi}  + E\, \epsilon^2 1\left[|\epsilon| > \kappa n_2 ^\xi \right], 
\end{eqnarray*}
 which converges to zero uniformly in $u$. Hence, the first term in right side of \eqref{eq:termtocon} converges to zero. 
As $\delta >0$ is arbitrary, we get the result.
\qed

\subsection{Proof of Theorem \ref{th:rateu}} \label{pf:rateu}
We derive bounds in terms of $n$ ($n_1$, $n_2$ and $n$ have the same order). Firstly, note that $0 \in \mathcal{D}_\theta$, for all $\theta \in \Theta_{n_1}^\tau$, whenever $n > N^{(1)}_\tau := (1/p) (K_\tau/ K)^{3/(1- 3\gamma)}$.  Further, as $r'(d_0) >0$ and $r$ is continuously differentiable, there exists $\delta_0>0$ such that $|r'(x) - r'(d_0)| < r'(d_0)/2$ (equivalently, $  r'(d_0)/2 < r'(x) < 3 r'(d_0)/2$) for $x \in [d_0 - \delta_0, d_0 + \delta_0]$. As $ u \in \mathcal{D}_\theta$ and $\theta \in \Theta_{n_1}^\tau$, $|d_0 + u n_2^{-\gamma}| <K_\tau n_1^{-1/3} + K n_1^{-\gamma} <\delta_0$ for $n > N^{(2)}_{\tau, \delta_0} :=(1/p) ((K_\tau + K)/\delta_0)^{1/\gamma}$. Hence, for $n > N^{(3)}_{\tau, \delta_0} : = \max(N^{(1)}_\tau, N^{(2)}_{\tau, \delta_0})$, by a change of variable,
\begin{eqnarray*}
M_{n_2}(u, \theta) & =& n_2^\gamma \left[\int_{d_0}^{d_0 + u n_2^{-\gamma}} \left(r(t) - r(d_0)\right) \frac{n_1^\gamma}{2K} dt\right] \\
& \geq &  n_2^\gamma \left[\int_{d_0}^{d_0 + u n_2^{-\gamma}} \frac{r'(d_0)}{2} \left(t - d_0\right) \frac{n_1^\gamma}{2K} dt\right] \gtrsim u^2 =: \rho^2_{n_2}(u, 0). \nonumber
\end{eqnarray*}
Using Theorem \ref{th:ratethmc6}, we need to  bound 
\begin{equation}\label{eq:modcontiso}
\sup_{\theta \in \Theta_n^\tau}E^* \sup_{\substack{|u| \leq \delta, u \in \mathcal{D}_\theta}} \left|\left(\M_{n_2}(u,\theta) - M_{n_2}(u, \theta)\right) - \left(\M_{n_2}(0,\theta) - M_{n_2}(0, \theta)\right)\right|
\end{equation}
Recall that $\M_{n_2}(0,\theta) = M_{n_2}(0,\theta) = 0$. Also, 
\begin{eqnarray*}
\sqrt{n} |\M_{n_2}(u,\theta) - M_{n_2}(u, \theta)| & = & \left|\G_{n_2} {g}_{n_2,u,\theta}\right| 
\end{eqnarray*}
The class of functions $\mathcal{F}_{\delta, \theta} = \{ {g}_{n_2,u,\theta} : |u| \leq \delta, u \in \mathcal{D}_\theta \}$ is a VC class of index at most 3, with a measurable envelope (for $n > N^{(3)}_{\tau, \delta_0}$)
\begin{eqnarray*}
 M_{\delta, \theta}  & =& n_2^{\gamma}(2\|r\|_\infty+ |\epsilon| ) \times \\
 & & 1\left[U K n_1^{-\gamma} \in \left[d_0-\theta- \delta n_2^{-\gamma}, d_0-\theta +\delta n_2^{-\gamma}\right]\right].
\end{eqnarray*} 
Note that $$ E\left[M_{\delta, \theta}  \right]^2 \lesssim n_2^{\gamma} P\left[U K n_1^{-\gamma} \in \left[d_0-\theta- \delta n_2^{-\gamma}, d_0-\theta +\delta n_2^{-\gamma}\right]\right] \lesssim \delta.$$  
Further, the uniform entropy integral for $\mathcal{F}_{\delta,\theta}$ is bounded by a constant which only depends upon the VC-indices, i.e., the quantity
$$ J(1, \mathcal{F}_{\delta,\theta}) = \sup_Q \int_0^1 \sqrt{1 + \log N(u \|M_{\delta, \theta}\|_{Q,2}, \mathcal{F}_{\delta,\theta}, L_2(Q)) } d u $$
is bounded. 
 Using Theorem 2.14.1 of \citet{VW96}, we have
\be 
E^* \sup_{\substack{|u| \leq \delta u \in \mathcal{D}_\theta }} n_2^{\gamma} \left|\G_{n_2} {g}_{n_2,u,\theta}\right|  \lesssim J(1, \mathcal{F}_{\delta,\theta}) \|M_{\delta,\theta} \|_2 \lesssim  \delta^{1/2}.
\nonumber 
\ee
Note that this bound is uniform in $\theta \in \Theta_n^\tau$.
Hence, a candidate for $\phi_n(\cdot)$ to apply Theorem \ref{th:ratethmc6}  is $\phi_n (\delta) = \delta^{1/2}$. The sequence $r_n = n^{(1-2\gamma)/3}$ satisfies the conditions $
r_n^2 \phi_n(1/r_n ) \leq \sqrt{n_2}$. As a consequence,
 $r_n\hat{u} = O_p(1).$ \qed
\subsection{Proof of Theorem \ref{th:limprociso}}\label{pf:limprociso} 
We outline the main steps of the proof below.
 Note that
\begin{eqnarray*}
f_{n_2,w,{\theta}} & = & n_2^{1/6- \gamma/3} (r(\theta + U K n_1^{-\gamma} ) + \epsilon - r(d_0)) \times \\ 
& & \left(1\left[\theta + U Kn_1^{-\gamma} \leq d_0 + w n_2^{-(\alpha + \gamma) }\right] - 1\left[\theta + U Kn_1^{- \gamma} \leq d_0 \right]\right). \nonumber
\end{eqnarray*}
For any $L>0$, we use Theorem \ref{th:tight} to justify the tightness of $Z_{n_2} (w, \hat{\theta}_{n_1})$ for $w \in [-L, L]$. For sufficiently large $n$, the set $\{ {w: w/n_2^{\alpha} \in \mathcal{D}_{{\theta}}} \}$ contains $[-L, L]$ for all $\theta \in \Theta_{n_1}^\tau$ and hence, it is not necessary to extend $Z_{n_2}$ (equivalently, $f_{n_2, w, \theta}$) as done in \eqref{zextn}. 
 For a fixed $\theta \in \Theta_{n_1}^\tau$ and an envelope for $\{ f_{n_2,w,\theta}: w \in [-L,L] \}$ is given by $F_{n_2, \theta} (V)  $ which equals
\be 
n_2^{1/6- \gamma/3} (2 \|r\|_\infty + |\epsilon| ) 1\left[\theta + U Kn_1^{-\gamma} \in [ d_0 -L n_2^{-(\alpha + \gamma)}, d_0 + L n_2^{-(\alpha + \gamma)}]\right] . \nonumber
\ee
Further, $P F^2_{n, \theta}  \lesssim n^{1/3 - 2\gamma/3} n^{-\alpha} = O(1)$. Also,
\begin{eqnarray*}
P \left[F^2_{n_2, \theta} 1[F_{n_2,\theta} > \sqrt{n_2}t]\right] & \lesssim  &  E \epsilon^2 1\left[ 2 \|r\|_\infty + |\epsilon| >\sqrt{n_2} n^{-1/6+ \gamma/3} t  \right],
\end{eqnarray*}
which goes to zero (uniformly in $\theta$) as $E \left[\epsilon^2\right] < \infty$. Hence, conditions \eqref{eq:cond1} and \eqref{eq:cond2} of Theorem \ref{th:tight} are verified. With $\tilde{\rho}(w_1, w_2) = |w_1 - w_2|$,  conditions \eqref{eq:cond3} and \eqref{eq:cond4} can be justified by elementary calculations. We justify \eqref{eq:cond4} below. For $-L \leq w_2 \leq w_1 \leq L$ and sufficiently large $n$ (such that $(K_\tau n_1^{-1/3}+ L n_2^{-(1+\gamma)/3})< \min(K n_1^{-\gamma}, \delta_0)$ with $\delta_0$ as defined in the proof of Theorem \ref{th:rateu}), a change of variable and boundedness of $r'$ in a $\delta_0$-neighborhood of $d_0$ yields
\begin{eqnarray*}
|\zeta_{n_2} (w_1, \theta) - \zeta_{n_2} (w_2, \theta) | & \leq & {n_2^{2/3-\gamma/3} \int_{d_0+ w_2 n_2^{-(1+\gamma)/3}}^{d_0 + w_1 n_2^{-(1+\gamma)/3}} (r(s) - r(d_0) ) \frac{n_1^\gamma}{2K} ds }\\
& = & {n_2^{1/3-2\gamma/3} \int_{w_2 }^{w_1 } (r(d_0 + t n_2^{-(1+\gamma)/3}) - r(d_0) ) \frac{n_1^\gamma}{2K} ds }\\
& \lesssim &  \frac{3 r'(d_0)}{4} (w_1 - w_2)^2 .
\end{eqnarray*}
The above bound does not involve $\theta$ and converges to zero when $|w_1 - w_2|$ goes to zero. Hence, condition \eqref{eq:cond4} holds.

Further, for a fixed $\theta$, the class $\{f_{n_2,w,\theta}: w \in [-L,L] \}$ is VC of index at most 3 with envelope $F_{n, \theta}$. Hence, the entropy condition in \eqref{eq:cond5} is satisfied. The measurability condition \eqref{eq:condmes} can be readily justified  as well. Hence, the processes $ Z_{n_2}$ are asymptotically tight for $w$ in any fixed compact set.

For a fixed $\theta \in \Theta_n^\tau$, $w \in [0, L]$ and sufficiently large $n$, $\zeta_{n_2}(w,\theta)$ equals 
\begin{eqnarray*}
\lefteqn{n_2^{2/3-\gamma/3} \int_{d_0}^{d_0 + w n_2^{-(1+\gamma)/3}} (r(s) - r(d_0) ) \frac{n_1^\gamma}{2K} ds }\\
& = & \frac{(1-p)^{2/3 - \gamma/3}p^\gamma n^{2/3+2\gamma/3}}{2K} \int_{d_0}^{d_0 + w n_2^{-(1+\gamma)/3}} (r(s) - r(d_0) ) ds \\
& = & \frac{(1-p)^{2/3 - \gamma/3}p^\gamma n^{1/3+\gamma/3}}{2K (1-p)^{(1+\gamma)/3}}  \int_{0}^{w} (r(d_0 + t n_2^{-(1+\gamma)/3} ) - r(d_0) ) dt \\
& =& \frac{(1-p)^{-\gamma}p^\gamma }{2K }\frac{r'(d_0)}{2}w^2 +o(1).
\end{eqnarray*}
This convergence is uniform in $\theta$ by arguments paralleling those for justifying condition \eqref{eq:cond4}. 

Note that $P f_{n_2, w , \theta} = \zeta_{n_2}(w, \theta)/\sqrt{n_2}$ converges to zero. Hence, for a fixed $\theta \in \Theta_n^\tau$ and $w_1, w_2 \in [0, L], L>0$, the covariance function  of $Z_{n_2}$ eventually equals (up to an $o(1)$ term which does not depend on $\theta$ due to a change of variable)
\begin{eqnarray*}
\lefteqn{P\left[f_{n_2,w_1, {\theta}}  f_{n_2,w_2,{\theta}} \right]}\\
& = & n_2^{1/3 - 2\gamma/3} \int_0^{(w_1 \wedge w_2)n_2^{-(1+ \gamma)/3}} \left[\sigma^2 + (r(d_0 + s) - r(d_0) )^2\right] \frac{n_1^\gamma}{2K} ds\\
& = & \frac{ p^\gamma n^{1/3 + \gamma/3}}{2K(1-p)^{-1/3+2\gamma/3}} \times \\
& & \int_0^{(w_1 \wedge w_2)n_2^{-(1+ \gamma)/3}} \left[\sigma^2 + (r(d_0 + s) - r(d_0) )^2\right] ds\\
& = & \frac{ p^\gamma }{2K (1-p)^{\gamma}} \int_0^{(w_1 \wedge w_2)} \left[\sigma^2 + (r(d_0 + t n_2^{-(1+ \gamma)/3}) - r(d_0) )^2\right]  ds\\
& =&  \frac{p^\gamma}{2K (1-p)^{\gamma}} (w_1 \wedge w_2)\sigma^2 + o(1).
\end{eqnarray*}
This justifies the form of the limit process $Z$. Note that the process $Z \in C_{min}(\R)$ (using argmin versions of Lemmas 2.5 and 2.6 of \citet{KP90}) and it possesses a unique argmin almost surely which is tight (the Chernoff random variable). An application of argmin continuous mapping theorem \cite[Theorem 2.7]{KP90} along with \eqref{eq:switchrel} yields
\begin{eqnarray*}
 n_2^{\alpha + \gamma} (\hat{d}_2 - d_0) &\stackrel{d}{\rightarrow}& \argmin_{w}\left\{\sigma \sqrt{\frac{p^\gamma}{2K (1-p)^{\gamma}}} + \frac{(1-p)^{-\gamma}p^\gamma }{2K }\frac{r'(d_0)}{2}w^2 \right\}.
\end{eqnarray*}
Consequently,
\begin{eqnarray*}
\lefteqn{n^{(1+ \gamma)/3} (\hat{d}_2 - d_0)}\\
 &\stackrel{d}{\rightarrow}& (1-p)^{-(1+\gamma)/3}\argmin_{w}\left\{\sigma \sqrt{\frac{p^\gamma}{2K (1-p)^{\gamma}}} B(w) + \frac{(1-p)^{-\gamma}p^\gamma }{2K }\frac{r'(d_0)}{2}w^2 \right\}.
\end{eqnarray*}
Letting $\tilde{\lambda} = \left({8 \sigma^2 K(1-p)^\gamma}/({(r'(d_0))^2 p^\gamma })\right)^{1/3} $ so that $\sigma \sqrt{{\tilde{\lambda}p^\gamma}/({2K (1-p)^{\gamma}})} = {(1-p)^{-\gamma}p^\gamma r'(d_0)\tilde{\lambda}^2}/{(4K) }$, the rescaling property of Brownian motion gives
{\allowdisplaybreaks \begin{eqnarray*}
\lefteqn{(1-p)^{-(1+\gamma)/3} \argmin_w \left\{\sigma \sqrt{\frac{p^\gamma}{2K (1-p)^{\gamma}}} B(w) + \frac{(1-p)^{-\gamma}p^\gamma }{2K }\frac{r'(d_0)}{2}w^2 \right\}}\\
 & = & (1-p)^{-(1+\gamma)/3} \tilde{\lambda}\,  \argmin_{v} \left\{\sigma \sqrt{\frac{p^\gamma}{2K (1-p)^{\gamma}}} B(\tilde{\lambda} v)  + \frac{(1-p)^{-\gamma}p^\gamma }{2K }\frac{r'(d_0)}{2} (\tilde{\lambda}v)^2 \right\} \\   
& \stackrel{d}{=} &  (1-p)^{-(1+\gamma)/3} \tilde{\lambda}\, \argmin_{v} \left\{\sigma \sqrt{\frac{\tilde{\lambda}p^\gamma}{2K (1-p)^{\gamma}}} B(v)  + \frac{(1-p)^{-\gamma}p^\gamma }{2K }\frac{r'(d_0)}{2} (\tilde{\lambda}v)^2 \right\}\\ 
& =  &  (1-p)^{-(1+\gamma)/3} \tilde{\lambda}\, \argmin_{v} \left\{ B(v) + v^2 \right\}\\ 
& =  &  \left(\frac{8 \sigma^2 K}{(r'(d_0))^2 p^\gamma (1-p)}\right)^{1/3}\, \argmin_{v} \left\{ B(v) + v^2 \right\}.
\end{eqnarray*}}
The result follows.
\qed

\subsection{Proof of Theorem \ref{th:risk}} \label{pf:risk}
Note that for $f(x) = 1\left[x \geq a\right]$ 
\begin{eqnarray*}
\mathcal{R}(f) = \int_0^a r(x) dx + \int_a^1 (1- r(x)) dx = \int_0^1 (1- r(x)) dx + \int_{0}^a (2r(x) -1) dx.  
\end{eqnarray*}
For notational ease, we use $\int_c^d$ to denote $-\int_d^c$ whenever $c>d$. 
Then, by a change of variable, 
\begin{eqnarray*}
\lefteqn{n^{2(1+\gamma)/3}(\mathcal{R}(\hat{f}) - \mathcal{R}(f^*)) }\\
& = & n^{(1+\gamma)/3}\int_{d_0}^{\hat{d}_2} 2(r(x) - 1/2) dx \\
& = & n^{(1+\gamma)/3}\int_0^{\left(n^{(1+\gamma)/3}(\hat{d}_2 - d_0)\right)} 2(r(d_0 + h n^{-(1+\gamma)/3}) - r(d_0)) dh.
\end{eqnarray*}
By Skorokhod's representation theorem, a version of $n^{(1+\gamma)/3}(\hat{d}_2 - d_0)$, say $\xi_n(\omega)$, converges almost surely to a tight random variable $\xi(\omega)$ which has the same distribution as the random variable on right side of \eqref{limiisoh}. As $r$ is continuously differentiable in a neighborhood of $d_0 =r^{-1}(1/2)$, there exists $\delta_0>0$, such that $|r'(x)| < 2 r'(d_0)$, whenever $|x- d_0| < \delta_0$. Hence, for a $\tau >0$ and a fixed $\omega$, there exist $N_{\omega, \tau, \delta_0} \in \mathbb{N}$, such that $|\xi_n(\omega) - \xi(\omega) | < \tau$ and $(|\xi(\omega)| + \tau)n^{-(1+\gamma)/3} <\delta_0$ whenever $n > N_{\omega, \tau, \delta_0}$. Hence, for $n > N_{\omega, \tau, \delta_0}$,  
\begin{eqnarray*}
\lefteqn{n^{(1+\gamma)/3}\int_0^{\xi_n(\omega)} 2(r(d_0 + h n^{-(1+\gamma)/3}) - r(d_0)) dh } \\
& = & n^{(1+\gamma)/3} \int_0^{\xi_n(\omega)} 2(r(d_0 + h n^{-(1+\gamma)/3}) - r(d_0)) 1\left[|h| \leq |\xi(\omega)| + \tau\right] dh\\
& = & \int_0^{\xi_n(\omega)} 2r'(d^\star_h)h 1\left[|h| \leq |\xi(\omega)| + \tau\right] dh,\\
\end{eqnarray*}
where $d^\star_h$ is an intermediate point between $d_0$ and $d_0 + h n^{-(1+\gamma)/3}$. Note that $r'(d^\star_h)$ converges (pointwise in $h$) to $r'(d_0)$.  As the integrand is bounded by $4 r'(d_0) h 1\left[|h| \leq |\xi(\omega)| + \tau\right] $ which is integrable, by the dominated convergence theorem, the above display then converges to $r'(d_0) \xi^2(\omega)$. Consequently,
$$ P\left({n^{(1+\gamma)/3}\int_0^{\xi_n} 2(r(d_0 + h n^{-(1+\gamma)/3}) - r(d_0)) dh {\not\rightarrow} r'(d_0)\xi^2 } 
 \right) \leq P\left(\xi_n {\not\rightarrow} \xi\right) =0.$$
Thus, we establish the result. \qed

\subsection{Proof of Theorem \ref{th:fstage}}\label{pf:fstge}
Let $M(d) = P \left[Y^{(1)} 1\left[|X^{(1)} - d| < b\right]\right]$. For $F(t) = \int_0^t m(x+ d_0) dx$, we have
$$ M(d) = F(d- d_0+b) - F(d- d_0 -b).$$
Note that $M'(d) = 0$ implies $m(d  + b) = m(d  - b)$ which holds for $d = d_0$. Hence, $d_0$ maximizes $M(\cdot)$. Also, note that $M''(d_0) = m'(d_0+b) - m'(d_0-b) = 2 m'(d_0+b) < 0$.  For $d$ in a small neighborhood of $d_0$ (such that $d+b> d_0$ and $2m'(d+b) \leq m'(d_0+b)$), we get
$$ M(d) - M (d_0) \leq -|m'(d_0+b)| (d -d_0)^2. $$ 
Note that we derived an upper bound here as our estimator is an argmax (instead of an argmin) of the criterion $\M_{n_1}$. 
Hence, the distance for applying Theorem 3.2.5 of \cite{VW96} can be taken to be $\rho(d, d_0) = |d - d_0|$.
The consistency of $\hat{d}_1$ with respect to $\rho$ can be deduced through standard Glivenko-Cantelli arguments and an application of argmax continuous mapping theorem \citep[Corollary 3.2.3]{VW96}. 
For sufficiently small $\delta >0$, consider the modulus of continuity
\begin{eqnarray*}
\lefteqn{E^* \sup_{|d - d_0| < \delta} \sqrt{n_1}| (\M_{n_1} - M)(d) - (\M_{n_1} - M)(d_0) | }\\
 &=& E^* \sup_{|d - d_0| < \delta} \left| \G_{n_1} Y^{(1)}\left\{1\left[|X^{(1)} - d| \leq b\right] - 1\left[|X^{(1)} - d_0| \leq b\right]\right\} \right|
\end{eqnarray*}
An envelope for the class of functions $\mathcal{F}_{\delta} = \{ g_d(x,y) = y\left\{1\left[|x - d| \leq b\right] - 1\left[|x - d_0| \leq b\right]\right\}: |d - d_0| < \delta \}$  is given by
$$F_{\delta}(X^{(1)}, \epsilon) = (\|m\|_{\infty}  + |\epsilon| ) 1\left[|X^{(1)} - d_0| \in [b- \delta, b+  \delta] \right].$$
Note that $\|F_{\delta}\|_2 \lesssim \delta^{1/2}$. 
Further, the uniform entropy integral for $\mathcal{F}_{\delta}  $ is bounded by a constant which only depends upon the VC-indices, i.e., the quantity
$$ J(1, \mathcal{F}_{\delta}) = \sup_Q \int_0^1 \sqrt{1 + \log N(u \|F_{\delta}\|_{Q,2}, \mathcal{F}_{\delta}, L_2(Q)) } d u $$
is bounded. 
 Using Theorem 2.14.1 of \citet{VW96}, we have
\be 
E^* \sup_{|d - d_0| < \delta} \sqrt{n_1}| (\M_{n_1} - M)(d) - (\M_{n_1} - M)(d_0) | \lesssim J(1, \mathcal{F}_{\delta}) \|F_{\delta} \|_2 \lesssim  \delta^{1/2}. \nonumber
\ee
Hence, a candidate for $\phi_n(\delta)$ in Theorem 3.2.5 of \cite{VW96} is $\phi_n(\delta) = \delta^{1/2}$. 
This  yields $n_1^{1/3}(\hat{d}_1 - d_0) = O_p(1)$.
Next, consider the local process, 
$$ Z_{n_1}(h) = n_1^{2/3} \P_{n_1} Y^{(1)}\left[1\left[|X^{(1)} - (d_0 + h n_1^{-1/3})| <  b\right] - 1\left[|X^{(1)} - d_0 | <  b\right]\right]. $$
Note that 
\begin{eqnarray*}
E\left[Z_{n_1}(h)\right] & = & n_1^{2/3} \left\{M(d_0 + h n_1^{-1/3}) - M(d_0)\right\} \\
& =  & \frac{M''({d_0})+ o(1)}{2} (h n_1^{-1/3})^2 n_1^{2/3} \\
& =  & m'(d_0+b) h + o(1)  =   -c h + o(1).
\end{eqnarray*}
Let $G(t) = \int_0^t m^2(d_0+x) dx$. Then,
{\allowdisplaybreaks \begin{eqnarray*}
\lefteqn{\mbox{Var}( Z_{n_1} (h) )}\\
 & = & \frac{n_1^{4/3}}{n_1^2} \mbox { Var} \left[Y^{(1)}\left[1\left[|X^{(1)} - (d_0 + h n_1^{-1/3})| <  b\right] - 1\left[|X^{(1)} - d_0 | <  b\right]\right] \right]\\
& = & n_1^{1/3} E \left[ (Y^{(1)})^2 \left[1\left[|X^{(1)} - (d_0 + h n_1^{-1/3})| <  b\right] - 1\left[|X^{(1)} - d_0 | <  b\right]\right]^2\right]\\
& &  + o(1)\\
& = & n_1^{1/3} \left[G(b + h n_1^{-1/3})- G(b) + G(-b + h n_1^{-1/3})- G(-b) + 2 \sigma^2 h n_1^{-1/3}\right]\\
& = & (m^2(d_0+b) + m^2(d_0-b)  + 2 \sigma^2)h  + o(1) \\
& = & 2(m^2(d_0+b) + \sigma^2)h  + o(1) =  a^2 h  + o(1). 
\end{eqnarray*} }
The limiting covariance function can be derived in an analogous manner and the tightness of the process follows from an application of Theorem 2.11.22 of \cite{VW96} involving routine justifications. An application of argmax continuous mapping theorem \citep[Theorem 3.2.2]{VW96} gives
\be
n_1^{1/3} (\hat{d}_1 - d_0) \stackrel{d}{\rightarrow} \argmax \left\{a B(h) - c h^2\right\}. \nonumber
\ee
 By rescaling arguments, we get the result. \qed

\subsection{Proof of Theorem \ref{th:twostage}}\label{pf:twostage}
{\it Rate of convergence.} 
Choose $K_{\tau} >0$, such that for $\Theta_{n_1}^\tau = [\theta_0 - K_\tau n_1^{-1/3}, \theta_0 + K_\tau n_1^{-1/3} ]$, $P\left[\hat{d}_1 \notin \Theta_{n_1}^\tau \right] < \tau$.
As $\gamma < 1/3$, for all $\theta \in \Theta_{n_1}^\tau $, $d_0 \in \mathcal{D}_\theta$,  whenever $n > N^{(1)}_\tau := (1/p)(K_\tau/ (K -b))^{3/(1- 3\gamma)}$. For $d \in \mathcal{D}_\theta$, the set $\{ u:  |\theta + u K n_1^{-\gamma} - d| \leq b n_1^{-\gamma} \} \subset [-1,1]$. 
Hence, by a change of variable, 
\begin{eqnarray}
M_{n_2} (d, \theta) & := & E \left[\M_{n_2} (d, \theta)\right] \nonumber\\
& = & \frac{1}{2}\int_{-1}^{1} m(\theta + u K n_1^{-\gamma}) 1\left[|\theta + u K n_1^{-\gamma} - d| \leq b n_1^{-\gamma} \right] du\nonumber \\
& =  & \frac{1}{2}\int_{\R} m(\theta + u K n_1^{-\gamma}) 1\left[|\theta + u K n_1^{-\gamma} - d| \leq b n_1^{-\gamma} \right] du \nonumber\\
& =  &  \frac{n_1^{\gamma}}{2K} \int_{\R} m(x) 1\left[|x - d| \leq b n_1^{-\gamma} \right] dx \nonumber\\
& =  &  \frac{n_1^{\gamma}}{2K} \int_{d- bn_1^{-\gamma}}^{d+ bn_1^{-\gamma}} m(x)  dx.
\label{eq:pop_second}
\end{eqnarray}
Let  
$$F_{n}(d) = \int_{d- bn_1^{-\gamma}}^{d+ bn_1^{-\gamma}} m(x)  dx.$$
Note that $F_n'(d) = m(d+ bn_1^{-\gamma}) - m(d- bn_1^{-\gamma})$. Also,
\begin{eqnarray*}
F_n''(d) & =&  m'(d + b n_1^{-\gamma}) -  m'(d- bn_1^{-\gamma})\\
& =&  m'(d + b n_1^{-\gamma}) + m'(2d_0 -d + b n_1^{-\gamma}),
\end{eqnarray*}
whenever $d \neq d_0 \pm bn_1^{-\gamma}$. Here, the last step follows from the anti-symmetry of $m'$ around $d_0$ (but not at $d_0$). 
Further, as $-m'(d_0+) >0$ and $\tilde{m}$ is continuously differentiable in a neighborhood of 0, there exists $\delta_0>0$ such that $|m'(x) - m'(d_0+)| < -m'(d_0+)/2$ (equivalently, $ 3m'(d_0+)/2 < m'(x) <  m'(d_0+)/2$) for $x \in (d_0, d_0 + \delta_0]$. For $ d \in \mathcal{D}_\theta$  and $\theta \in \Theta_{n_1}^\tau$, $|d \pm b n_1^{-\gamma} - d_0| <K_\tau n_1^{-1/3} + K n_1^{-\gamma} <\delta_0$ for $n > N^{(2)}_{\tau, \delta_0} :=(1/p) ((K_\tau + K)/\delta_0)^{1/\gamma}$. Let $\rho^2_n(d, d_0) = n_1^{\gamma} (d- d_0)^2.$ For $n > N^{(3)}_{\tau, \delta_0} : = \max(N^{(1)}_\tau, N^{(2)}_{\tau, \delta_0})$ and $\rho_n(d, d_0) < \kappa_n:= b n_1^{-\gamma/2}$ (so that $d_0 \in [d- bn_1^{-\gamma}, d+ bn_1^{-\gamma}]$), 
\begin{eqnarray*}
F_n''(d) & =&  m'(d + b n_1^{-\gamma}) + m'(2d_0 -d + b n_1^{-\gamma})\\
& \leq & 2  (-m'(d_0+)/2) = m'(d_0+) = -|{m}'(d_0+)|.
\end{eqnarray*}
%
%
%
%
Consequently, by a second order Taylor expansion,
\begin{eqnarray}\label{eq:disttaylor}
{M_{n_2} (d, \theta)  - M_{n_2}(d_0,\theta)} & = & \frac{n_1^{\gamma}}{2K}\left[F_n(d) - F_n(d_0)\right]\\
& \leq & -\frac{n_1^{\gamma}}{2K} \frac{|{m}'(d_0+)|}{2} (d- d_0)^2 \nonumber\\
& \lesssim & - n_1^{\gamma} (d- d_0)^2 =  (-1)\rho^2_n(d, d_0).\nonumber
\end{eqnarray}
Again, an upper bound is deduced here as we are working with an argmax estimator.

{\bf Claim A.} We claim that $P\left[\rho_n(\hat{d}_n,d_0) \geq  \kappa_n\right]$ converges to zero.
We first use the claim to prove the rate of convergence.
To apply Theorem \ref{th:ratethmc6}, we need to  bound 
\be \label{eq:modcontshrt}
 \sup_{\theta \in \Theta_{n_1}^\tau }E^* \sup_{\substack{|d- d_0| < n_1^{-\gamma/2}\delta \\ d \in \mathcal{D}_\theta}} \sqrt{n_2}\left| (\M_{n_2} (d, \theta) -M_{n_2}(d, \theta))  - (\M_{n_2}(d_0, \theta)  -M_{n}(d_0, \theta) ) \right|.  
\ee
%
Note that 
\begin{eqnarray*}
\sqrt{n_2}\left( (\M_{n_2} (d, \theta) -M_{n_2}(d, \theta))  - (\M_{n_2}(d_0, \theta)  -M_{n}(d_0, \theta) ) \right) = \G_{n_2} g_{n_2, d, \theta} (V),
\end{eqnarray*}
where
\begin{eqnarray}
g_{n_2, d,\theta}(V)  &=& \left[m(\theta + U K n_1^{-\gamma}) + \epsilon\right] \times \nonumber\\
&  & \left[1\left[|\theta + U K n_1^{-\gamma} - d| < b n_1^{-\gamma} \right] - 1\left[|\theta + U K n_1^{-\gamma} - d_0| < b n_1^{-\gamma} \right]\right]. \nonumber
\end{eqnarray}
The class of functions $\mathcal{F}_{\delta,\theta}= \{g_{n_2, d,\theta}: |d-d_0| < n_1^{-\gamma/2}\delta, d \in \mathcal{D}_\theta \}$ is VC with index at most 3 and has a measurable envelope
\begin{eqnarray*}
\lefteqn{M_{\delta, \theta}(V) }\\
& = &\left(\|m\|_\infty + |\epsilon| \right) \times \\
& & \left[1\left[bn_1^{-\gamma}-(d_0 + n_1^{-\gamma/2}\delta) < \theta_0 + U K n_1^{-\gamma}  < bn_1^{-\gamma}-(d_0 - n_1^{-\gamma/2}\delta)\right] \right.\\
& & \left.+ 1\left[-bn_1^{-\gamma}-(d_0+ n_1^{-\gamma/2}\delta) < \theta_0 + U K n_1^{-\gamma}  < -bn_1^{-\gamma}-(d_0 - n_1^{-\gamma/2}\delta)\right]\right].
\end{eqnarray*}
Note that $E\left[M_{\delta, \theta}(V) \right]^2 \lesssim n^{-\gamma/2}\delta$.
Hence, the uniform entropy integral for $\mathcal{F}_{\delta,\theta}$ is bounded by a constant which only depends upon the VC-indices, i.e., the quantity
$$ J(1, \mathcal{F}_{\delta,\theta}) = \sup_Q \int_0^1 \sqrt{1 + \log N(u \|M_{\delta, \theta}\|_{Q,2}, \mathcal{F}_{\delta,\theta}, L_2(Q)) } d u $$
is bounded.  
 Using Theorem 2.14.1 of \citet{VW96}, we have
\be \label{eq:secondq}
E^* \sup_{\substack{| d- d_0| < n_1^{-\gamma/2}\delta \\ d \in \mathcal{D}_\theta}} \left|\G_{n_2} g_{n_2, d,\theta}\right|  \leq J(1, \mathcal{F}_{\delta,\theta}) \|M_{\delta, \theta} \|_2 \lesssim  n^{\gamma/4}\delta^{1/2}. 
\ee
 The above bound is uniform in $\theta \in \Theta_{n_1}^\tau$. Hence, a candidate for $\phi_n$ to apply Theorem \ref{th:ratethmc6} is 
$\phi_{n_2}(\delta) = n^{\gamma/4} \delta^{1/2} $. This yields $n^{(1+\gamma)/3} (\hat{d}_2 - d_0) = O_p(1)$.

{\it Proof of} {\bf Claim A}.  Note that $\rho_n(d, d_0) \geq  \kappa_n \Leftrightarrow |d - d_0| \geq b n_1^{-\gamma}$. Also, for such $d \in \mathcal{D}_\theta$, the bin $(d - bn_1^{-\gamma}, d + bn_1^{-\gamma})$ does not contain $d_0$ and is either completely to the right of $d_0$ or to the left (regions where $m$ is continuously differentiable). In particular, for such $d$'s with  $d> d_0$ and $n > N^{(3)}_{\tau, \delta_0} $,
\begin{eqnarray*}
F_n'(d) & =&  m(d + b n_1^{-\gamma}) -  m(d- bn_1^{-\gamma}) \leq  - (|m'(d_0+)|/2) (2b n_1^{-\gamma}) = -|m'(d_0+)|b n_1^{-\gamma}.
\end{eqnarray*}
As a consequence, 
\be
 M_{n_2}(d, \theta) - M_{n_2}(d_0 + bn_1^{-\gamma}, \theta) \leq (n_1^\gamma/2K) (- (|m'(d_0+)|b n_1^{-\gamma})  |d -(d_0+ bn_1^{-\gamma})|) \leq 0, 
\label{eq:crd1}
\ee
for $ d > d_0 + bn_1^{-\gamma}$. Also, for $n > N^{(3)}_{\tau, \delta_0} $,
{\allowdisplaybreaks \be
\begin{split}
\lefteqn{M_{n_2}(d_0 + bn_1^{-\gamma}, \theta)  - M_{n_2}(d_0, \theta)}\\
  &= \frac{n_1^\gamma}{2K}\left[\int_{d_0 }^{d_0 + 2 bn_1^{-\gamma}} m(x) dx - 2 \int_{d_0 }^{d_0 +  bn_1^{-\gamma}} m(x) dx \right] \\
&= \frac{n_1^\gamma}{2K}\left[\int_{d_0 +b n_1^{-\gamma}}^{d_0 + 2bn_1^{-\gamma}} m(x) dx - \int_{d_0 }^{d_0 +  bn_1^{-\gamma}} m(x) dx \right]\\
&= \frac{n_1^\gamma}{2K}\int_{d_0 }^{d_0 + bn_1^{-\gamma}} (m(x+b n_1^{-\gamma}) - m(x)) dx \\
&\leq \frac{n_1^\gamma}{2K}\int_{d_0 }^{d_0 + bn_1^{-\gamma}} (m'(d_0)/2) b n_1^{-\gamma})  dx \leq \frac{-|m'(d_0)|b^2}{4K} n_1^{-\gamma}.
\label{eq:crd2}
\end{split}
\ee}
Using \eqref{eq:crd1} and \eqref{eq:crd2},
\begin{eqnarray*}
{c^\tau_n(\kappa_n) }&=& \sup_{\theta \in \Theta_n^\tau}\sup_{\substack{\rho_n(d, d_n) \geq  \kappa_n, d > d_0 \\ d \in \mathcal{D}_\theta}} \left\{M_{n_2}(d, \theta) - M_{n_2}(d_0, \theta) \right\}\\
&\leq& \sup_{\theta \in \Theta_n^\tau}\sup_{\substack{\rho_n(d, d_n) \geq  \kappa_n, d > d_0 \\ d \in \mathcal{D}_\theta}} \left\{M_{n_2}(d, \theta) -M_{n_2}(d_0 + bn_1^{-\gamma}, \theta)\right\}\\
& &  + \sup_{\theta \in \Theta_n^\tau}\sup_{\substack{\rho_n(d, d_n) \geq  \kappa_n, d > d_0 \\ d \in \mathcal{D}_\theta}} \left\{M_{n_2}(d_0 + bn_1^{-\gamma}, \theta) - M_{n_2}(d_0, \theta) \right\}\\
& \lesssim & - n^{-\gamma}.
\end{eqnarray*}
Note that an upper bound is derived as we are working with argmax type estimators instead of argmins. The same upper bound can deduced for the situation $d< d_0$. Further, $\M_{n_2}(d, \theta) - M_{n_2}(d, \theta)  = (\P_{n_2} - P)\tilde{g}_{n_2,d, \theta}$, where
\begin{eqnarray*}
\tilde{g}_{n_2, d,\theta}(V)  &=& \left[m(\theta + U K n_1^{-\gamma}) + \epsilon\right]  1\left[|\theta + U K n_1^{-\gamma} - d| < b n_1^{-\gamma} \right]. \nonumber
\end{eqnarray*}
 The class of functions 
$\mathcal{G}_{n_2, \theta} = \{\tilde{g}_{n_2,d, \theta}:  d \in \mathcal{D}_\theta\} $ is VC of index at most 3 and is enveloped by the function 
$$G_{n_2}(V) = \left(\|m\|_{\infty} + |\epsilon| \right) $$
with $\|G_{n_2}\|_{L_2(P)} = O(1)$.
Further, the uniform entropy integral for $\mathcal{G}_{n_2,\theta}$ is bounded by a constant which only depends upon the VC-indices, i.e., the quantity
$$ J(1, \mathcal{G}_{n_2,\theta}) = \sup_Q \int_0^1 \sqrt{1 + \log N(u \|G_{n_2}\|_{Q,2}, \mathcal{G}_{n_2,\theta}, L_2(Q)) } d u $$
is bounded. 
 Using Theorem 2.14.1 of \citet{VW96}, 
\be 
E^* \sup_{\mathcal{G}_{n_2, \theta} } \left|\G_{n_2} \tilde{g}_{n_2,d, \theta}\right|  \lesssim J(1, \mathcal{G}_{n_2,\theta}) \|G_{n_2} \|_2 = O(1),
\ee
where the $O(1)$ term does not depend on $\theta$ (as the envelope $G_{n_2}$ does not depend on $\theta$). Consequently, by Markov inequality,
\begin{eqnarray*}
\sup_{\theta \in \Theta_{n_1}^\tau} P\left[ 2 \sup_{\substack{d \in \mathcal{D}_\theta}} \left| \M_n({d}, \theta) - M_n(d, \theta) \right| > -c^\tau_n(\kappa_n)  \right]& \leq & \frac{O(1)}{\sqrt{n} n^{-\gamma}}.
\end{eqnarray*}
As $\gamma < 1/3 < 1/2$, the right side converges to zero. Hence, {\bf Claim A} holds.
 %
%

{\it Limit distribution.} For deriving the limit distribution, let 
\begin{eqnarray*}
Z_{n_2} (h, \theta)  = \G_{n_2} f_{n_2,h, \theta}(V) + \zeta_{n_2}(h,\theta),
\end{eqnarray*}
where $\zeta_{n_2}(h,\theta) = \sqrt{n_2 } P \left[f_{n_2,h, \theta}(V)\right]$ and 
\begin{eqnarray*}
f_{n_2,h, \theta}(V) & = & n_2^{1/6 - \gamma/3} (g_{n_2, d_0 + h n_2^{(1+\gamma)/3}, \theta} (V) -  g_{n_2, d_0, \theta} (V)). 
\end{eqnarray*}
Further, the asymptotic tightness of processes of the type 
\be
\sqrt{n_2} \G_{n_2} (m(\theta+ UKn_1^{-\gamma})+ \epsilon) 1\left[d_0 - b n_1^{-\gamma}< \theta+ UKn_1^{-\gamma} \leq  d_0 + h n_2^{-(1+\gamma)/3}+ b n_1^{-\gamma}\right]
\label{eq:splitproc}
\ee
can be established by arguments analogous to those in the proof of Theorem \ref{th:limprociso}.
As indicators with absolute values can be split as $$1\left[ |a_1 - a_2| \leq a_3\right] = 1\left[ a_1 - a_2 \leq a_3\right] - 1\left[ a_3 < a_1 - a_2| \right],$$
 the process $Z_{n_2}$ can be broken into process of the form \eqref{eq:splitproc}. As the sum of tight processes is tight, we get tightness for the process $Z_{n_2}$.
Further, 
\begin{eqnarray*}
\zeta_{n_2}(h,\theta) = n_2^{1/2 + 1/6 - \gamma/3} \left[M_{n_2}(d_0 +  h n_2^{-(1+\gamma)/3}, \theta) - M_{n_2}(d_0, \theta)  \right].
\end{eqnarray*}
Fix $L>0$. For $h \in [-L,L]$ and $\theta \in \Theta_{n_1}^\tau$, both $d_0 +  h n_2^{(1+\gamma)/3}$ and $d_0$ lie in the set $\mathcal{D}_\theta$ and hence,
\begin{eqnarray*}
\zeta_{n_2}(h,\theta) & = & n_2^{2/3 - \gamma/3} \frac{n_1^\gamma}{2K}\left[F_n(d_0 + h n_2^{-(1+\gamma)/3}) - F_n(d_0) \right].
\end{eqnarray*}
Note that $$F_n''(d_0 + h n_2^{-(1+\gamma)/3}) =m'(d_0 + h n_2^{-(1+\gamma)/3} + b n_1^{-\gamma}) -  m'(d_0 + h n_2^{-(1+\gamma)/3}- bn_1^{-\gamma}). $$
For any $h \in [-L,L]$, $d_0 \in [d_0 + h n_2^{-(1+\gamma)/3} - b n_1^{-\gamma}, d_0 + h n_2^{-(1+\gamma)/3} - b n_1^{-\gamma}]$ eventually and hence, $F_n''(d_0 + h n_2^{-(1+\gamma)/3}) = 2m'(d_0+) + o(1)$. 
Consequently,
\begin{eqnarray*}
\zeta_{n_2}(h,\theta) 
& = & \frac{p^\gamma n_2^{2/3 + 2\gamma/3}}{2K(1-p)^\gamma} \frac{F_n''(d_0 + o(1))}{2} h^2 n_2^{-2(1+\gamma)/3} \\
& = & -\frac{p^\gamma}{(1-p)^\gamma} \frac{|{m}'(d_0+)|}{2K} h^2 + o(1).
\end{eqnarray*}
Note that the above convergence is uniform in $\theta \in \Theta_{n_1}^\tau$ (due to a change of variable allowed for large $n$).
Next, we justify the form of the limiting variance function for simplicity. The covariance function can be deduced along to same lines in a notationally tedious manner.  
As $P \left[f_{n_2,h, \theta}(V)\right]  = \zeta_{n_2}(h,\theta)/\sqrt{n}$ converges to zero, for $\theta \in \Theta_{n_1}^\tau$ and $h \in [0, L]$,  the variance of $Z_{n_2}(h)$ eventually equals (up to an $o(1)$ term)
{\small\begin{eqnarray}
\lefteqn{P\left[f^2_{n_2,h_1, {\theta}}   \right] =} \nonumber\\
& & \frac{n_2^{1/3 - 2\gamma/3} }{{2K}{n_1^{-\gamma}} } \int_{\R} \left(\sigma^2 + m^2(x)\right) 
 \left[1\left[|x - d_0 + h n_2^{-(1+\gamma)/3}| \leq b n_1^{-\gamma}\right] - 1\left[|x - d_0 | \leq b n_1^{-\gamma}\right]\right]^2 dx.\nonumber
\end{eqnarray}}
Note that 
\begin{eqnarray*}
\lefteqn{\left[1\left[|x - (d_0 + h n_2^{-(1+\gamma)/3})| \leq b n_1^{-\gamma}\right] - 1\left[|x - d_0 | \leq b n_1^{-\gamma}\right]\right]^2} \hspace{1.5in}\\
& = &  1\left[ d_0 + b n_1^{-\gamma}< x \leq d_0 + h n_2^{-(1+\gamma)/3} + b n_1^{-\gamma} \right] \\
&  &  + 1\left[ d_0 - b n_1^{-\gamma}< x \leq d_0 + h n_2^{-(1+\gamma)/3} - b n_1^{-\gamma} \right].
\end{eqnarray*}
Further,
\begin{eqnarray*}
\lefteqn{\frac{n_2^{1/3 - 2\gamma/3} {n_1^\gamma}}{{2K} }\int_{\R}  \left(\sigma^2 + m^2(x)\right)1\left[ d_0 + b n_1^{-\gamma}< x \leq d_0 + h n_2^{-(1+\gamma)/3} + b n_1^{-\gamma} \right]dx} \hspace{1.5in}\\
& = &  \frac{p^\gamma n_2^{1/3 +\gamma/3} }{{2K} (1-p)^\gamma } (\sigma^2 + m^2(d_0) + o(1)) h n_2^{-(1+\gamma)/3} \\
& = &  \frac{p^\gamma  }{{2K} (1-p)^\gamma } (\sigma^2 + m^2(d_0)) h  + o(1).  
\end{eqnarray*}
Hence, the process $Z_{n_2}$ converges weakly to the process
$$ Z(h) = \sqrt{\frac{p^\gamma  }{{K} (1-p)^\gamma } (m^2(d_0)+\sigma^2)} B(h)-\frac{p^\gamma}{(1-p)^\gamma} \frac{|{m}'(d_0+)|}{2K} h^2.$$
By usual rescaling arguments we get the result.
\qed
\begin{rmk}\label{nonuniftech}
If a non-flat design centered at $\hat{d}_1$ is used instead of a uniform design at the second stage, i.e., if the second stage design points are sampled as $X_i^{(2)} = \hat{d}_1 + V_i K n_1^{-\gamma}$, where $V_i$'s are i.i.d. realizations from a distribution with a non-flat density $\psi$ supported on $[-1,1]$, then the second stage population criterion
function $M_{n_2} (d,\theta) = E\left[\M_{n_2}(d,\theta)\right]$ need not be at its maximum at $d_0$. To see this, consider the situation where $m(x) = \exp(-|x- d_0|)$ and $\psi(x) = C \exp(-|x|) 1\left[|x| \leq 1\right]$ for some constant $C>0$. From calculations parallel to those in \eqref{eq:pop_second} (a change of variable), it can be deduced that 
\begin{eqnarray*}
M_{n_2}(d, \theta) & =& \frac{n_1^{\gamma}}{K}\int_{d - b n_1^{-\gamma}}^{d + b n_1^{-\gamma}} m(x) \psi \left(\frac{n_1^\gamma}{K} (x - \theta) \right) dx\nonumber
\\
& =&  \frac{Cn_1^{\gamma}}{K}\int_{d - b n_1^{-\gamma}}^{d + b n_1^{-\gamma}} \exp\left(-|x-d_0|-\frac{n_1^\gamma}{K} |x - \theta| \right) dx.\nonumber
\end{eqnarray*}
 It can be shown that $M_{n_2}(d, \hat{d}_1)$ is maximized at $d^\star = (d_0 + (n_1^\gamma/K) \hat{d}_1)/(1 + n_1^\gamma/K)$ with probability converging to 1. Using Theorem \ref{th:fstage}, $(d^\star - d_0) = O_p(n_1^{-1/3})$. As $\hat{d}_2$ is a guess for $d^\star$, it is not expected to converge to $d_0$ at a rate faster than $n_1^{1/3}$. Moreover, a simpler analysis along these lines shows that $\hat{d}_1$ is not guaranteed to be consistent if a non-flat design is used to generate the covariates $X_i^{(1)}$s at the first stage.
 \end{rmk}
\begin{rmk}\label{derzero}
For the situation where $m'(d_0)=0$ but $m''(d_0) <0$, note that $F'_n(d_0) = m'(d_0 + bn_1^{-\gamma}) - m'(d_0 - b n_1^{-\gamma) } \leq - m''(d_0) b n_1^{-\gamma}$, for sufficiently large $n$. Consequently, from derivations similar to those in \eqref{eq:disttaylor},
$M_{n_2}(d, \theta) - M_{n_2}(d_0, \theta) \lesssim - (d- d_0)^2,$ and hence, a choice for the distance is $\rho_n(d, d_0) = |d- d_0|$. Paralleling the steps in the above proof, it can be shown that the modulus of continuity is bounded by $n^{\gamma/4} (n^{\gamma/2} \delta)^{1/2} = n^{\gamma/2} \delta^{1/2}$ ($\delta$ in  \eqref{eq:secondq} gets replaced by $n^{\gamma/2} \delta$).   This yields $n^{(1- \gamma)/3} (\hat{d}_2 - d_0) = O_p(1)$.
\end{rmk}

\subsection{Proof of Theorem \ref{th:twostagenf}}\label{pf:twostagenf}

{\it Rate of convergence. } 
We provide an outline of the proof below. Let $\theta_0 = d_0$ and 
\be
M_{n_2} (d, \theta) = P\ m(\theta + W K n_1^{-\gamma} ) 1\left[|\theta + W K n_1^{-\gamma} - d| \leq b n_1^{-\gamma} \right].
\ee
We take our population criterion function to be $M_{n_2} (d) := M_{n_2} (d, \theta_0)$. Let $\tilde{F}_{n}(t) = \int_0^t m(\theta_0 + w Kn_1^{-\gamma})  g(w) dw$. Then
\begin{eqnarray*}
M_{n_2} (d) &= & P\ m(\theta_0 + W K n_1^{-\gamma}) 1\left[|\theta_0 + W K n_1^{-\gamma} - d| \leq b n_1^{-\gamma} \right]\\
&= & P\ m(\theta_0 + W K n_1^{-\gamma}) 1\left[n_1^{\gamma}(d - \theta_0) -b  \leq W K \leq n_1^{\gamma}(d - \theta_0) +b  \right]\\
&= & \tilde{F}_{n} \left(\frac{n_1^{\gamma}(d - \theta_0) + b}{K}\right) - \tilde{F}_{n} \left(\frac{n_1^{\gamma}(d - \theta_0) - b}{K}\right). \nonumber
\end{eqnarray*}
By symmetry of $m$ around $\theta_0$ and that of $g$ around zero,
\begin{eqnarray*}
\frac{\partial M_{n_2}}{\partial d} (d_0, \theta_0) & = & m(\theta_0 + b n_1^{-\gamma}) g\left(\frac{b}{K}\right) - m(\theta_0 -b n_1^{-\gamma}) g\left(\frac{-b}{K}\right) = 0 \mbox{ and }\\
\frac{\partial^2 M_{n_2}}{\partial d^2} (d_0, \theta_0) & = & \frac{2 n^{2\gamma}}{K^2} \tilde{F}_{n}''\left(\frac{b}{K}\right).
\end{eqnarray*}
Note that $$\tilde{F}_{n}''(t) = K n_1^{-\gamma} m'(\theta_0 + t K n_1^{-\gamma} ) g(t) + m(\theta_0 + t K n_1^{-\gamma} ) g'(t),$$
 $m'(\theta_0 + b n_1^{-\gamma}) = 0 + o(1) $ and $m(\theta_0 + b n_1^{-\gamma}) = m(\theta_0) +  o(1)$. Therefore, 
\begin{eqnarray}
\frac{\partial^2 M_{n_2}}{\partial d^2} (d_0, \theta_0) & = &  \frac{2 n_1^{2\gamma}}{K^2} (m(\theta_0)+ o(1)) g'\left(\frac{b}{K}\right) +\frac{2 n_1^{\gamma} (0+o(1)) }{K} g\left(\frac{b}{K}\right) \nonumber \\
& = & \frac{2 n_1^{2\gamma}}{K^2} \left[m(\theta_0) g'\left(\frac{b}{K}\right)  + o(1) \right].
\label{eq:secndder}
\end{eqnarray}
The leading term in the above display is of the order $n^{2 \gamma}$. Let $\rho_n(d, d_0) = n_1^{ \gamma}|d - d_0|$. Following the arguments in the proof of Theorem \ref{th:twostage}, it can be shown that for sufficiently large $n$ and $d$ such that $|d- d_0| < b n_1^{-\gamma}$ (equivalently, $\rho_n(d, d_0) < \kappa_n = b n_1^{-2\gamma}$),
\be
M_{n_2} (d, \theta_0) - M_{n_2} (d_0, \theta_0) \lesssim - \rho_n^2(d, d_0).  \nonumber
\ee
The condition $P\left[\rho_n(\hat{d}_2, d_0) \geq \kappa_n \right] = P\left[|d- d_0| \geq b n_1^{-\gamma}\right]$ converging to zero can be established through analogous arguments. Further, to use Theorem \ref{th:ratethmc6}, we need to bound 
\be \label{eq:modcontshrth}
\sup_{ \theta \in \Theta_{n_1}^\tau} E^* \sup_{\substack{|d- d_0| < n^{-\gamma}\delta, \\ d \in \mathcal{D}_\theta }} \sqrt{n_2}\left| (\M_{n_2} (d, \theta) -M_{n_2}(d))  - (\M_{n_2}(d_0, \theta)  -M_{n_2}(d_0) ) \right|.  
\ee
Here $\Theta_{n_1}^\tau$ (and  $K_\tau$) is same as in the proof of Theorem \ref{th:twostage}.
Split the expression in $| \cdot |$ in\eqref{eq:modcontshrth} as $I + II$, where 
\be
I = \left(\M_{n_2}(d,\theta) - \M_{n_2}(d_0, \theta)\right) - \left(M_{n_2}(d,\theta) - M_{n_2}(d_0, \theta)\right) \mbox{ and } \nonumber
\ee
\be
II = \left(M_{n_2}(d,\theta) - M_{n_2}(d_0, \theta)\right) - \left(M_{n_2}(d,\theta_0) - M_{n_2}(d_0, \theta_0)\right).\nonumber
\ee
We first resolve I. 
Note that $\sqrt{n_2}I  =  \G_{n_2} \tilde{g}_{n_2, d,\theta}$ with 
\begin{eqnarray*}
\lefteqn{\tilde{g}_{n_2, d,\theta}(\epsilon, W)  = \left[m(\theta + W K n_1^{-\gamma}) + \epsilon\right] \times }\\
& &\left[1\left[|\theta + W K n_1^{-\gamma} - d| < b n_1^{-\gamma} \right] - 1\left[|\theta + W K n_1^{-\gamma} - d_0| < b n_1^{-\gamma} \right]\right]. \nonumber
\end{eqnarray*}
The class of functions $\mathcal{F}_{\delta,\theta}= \{\tilde{g}_{n_2, d,\theta}: 0< |d-d_0| < n^{-\gamma}\delta\}$ is VC with index at most 3 with a measurable envelope
\be
\begin{split}
\lefteqn{M_\delta(\epsilon, W) = \left(\|m\|_\infty + |\epsilon| \right) \times}\\
& 1\left[bn_1^{-\gamma} - 2\delta n_1^{-\gamma} - 2K_{\tau} n^{-1/3} < |\theta_0 + W K n_1^{-\gamma} - d| < bn_1^{-\gamma} + 2\delta n_1^{-\gamma} +  2K_{\tau} n^{-1/3}\right].
\end{split}
\ee
Note that the envelope does not depend on $\theta$. Further,  the uniform entropy integral for $\mathcal{F}_{\delta,\theta}$ is bounded by a constant which only depends upon the VC-indices, i.e., the quantity
$$ J(1, \mathcal{F}_{\delta,\theta}) = \sup_Q \int_0^1 \sqrt{1 + \log N(u \|M_\delta\|_{Q,2}, \mathcal{F}_{\delta,\theta}, L_2(Q)) } d u $$
is bounded. Using Theorem 2.14.1 of \citet{VW96}, we have
\be \label{eq:bndIshrt}
E^* \sup_{\substack{0< d- d_0 < n^{-\gamma}\delta \\ d \in \mathcal{D}_\theta}} \left|\G_{n_2} g_{n_2, d,\theta}\right|  \leq J(1, \mathcal{F}_{\delta,\theta}) \|M_\delta \|_2 \lesssim  C_\tau (\delta + n^{-1/3 + \gamma})^{1/2}, 
\ee
for some $C_\tau >0$ (depending on $\tau$ through $K_\tau$). Note that the above bound does not depend on $\theta$.
For simplifying $II$, let $\Delta_\theta = n_1^{1/3}(\theta - \theta_0)$ and  $\Delta_d = n_1^{\gamma}(d - d_0)$ and  
\begin{eqnarray*}
\lefteqn{\tilde{M}_{n_2}(\Delta_d, \Delta_\theta, b ) }\\
& =& P\ m(\theta_0 +n^{-1/3}\Delta_\theta + W K n^{-\gamma}) 1\left[{n_1^{-1/3}\Delta_\theta}+ W K n_1^{-\gamma}- \Delta_d n_1^{-\gamma} \leq b n_1^{-\gamma}\right] \\
& = & P\ m(\theta_0 +n^{-1/3}\Delta_\theta + W K n^{-\gamma}) 1\left[{n_1^{-1/3}\Delta_\theta}+ W K n_1^{-\gamma}-b n_1^{-\gamma} \leq \Delta_d n_1^{-\gamma} \right] 
\end{eqnarray*}
Note that $ M_{n_2}(d, \theta)= \tilde{M}_{n_2}(\Delta_d, \Delta_\theta, b) - \tilde{M}_{n_2}(\Delta_d, \Delta_\theta, -b)$. 
Also, by a change of variable ($un_1^{-\gamma} =  {n_1^{-1/3}\Delta_\theta}+ w K n_1^{-\gamma} - b n_1^{-\gamma}$),
\begin{eqnarray*}
\lefteqn{\bar{M}_{n_2}(\Delta_d, \Delta_\theta, b) }\\
&:=& (\tilde{M}_{n_2}(\Delta_d, \Delta_\theta, b)  - \tilde{M}_{n_2}(0, \Delta_\theta, b)) - (\tilde{M}_{n_2}(\Delta_d, 0, b)  - \tilde{M}_{n_2}(0, 0, b)) \\
& = & \frac{1}{K}\int_{0}^{\Delta_d} \Bigg[ m(\theta_0 +  (u + b) n_1^{-\gamma}) \times \\
& & \left\{g\left(\frac{(u+b)n_1^{-\gamma} + n_1^{-1/3} \Delta_\theta}{K n_1^{-\gamma}}\right) - g\left(\frac{(u+b)n_1^{-\gamma} }{K n_1^{-\gamma}}\right)\right\}\Bigg] du.
\end{eqnarray*}
A similar expression can be obtained for $\bar{M}_{n_2}(\Delta_d, \Delta_\theta, -b)$. As  $g$ is Lipschitz of order 1, we have 
\begin{eqnarray*}
\lefteqn{\sup_{\substack{|\Delta_d| < \delta, \\|\Delta_\theta| < K_{\tau}}} \sqrt{n_2}\left|(\tilde{M}_{n_2}(\Delta_d, \Delta_\theta, b)  - \tilde{M}_{n_2}(0, \Delta_\theta, b)) - (\tilde{M}_{n_2}(\Delta_d, 0, b)  - \tilde{M}_{n_2}(0, 0, b))\right|} \hspace{3in}\\ 
&\lesssim & \sqrt{n_2} \delta \frac{n_1^{-1/3}}{n_1^{-\gamma}} \lesssim \tilde{C}_\tau n_2^{1/6 + \gamma} \delta,
\end{eqnarray*}
for some $\tilde{C}_\tau >0$ (depending on $\tau$ through $K_\tau$).
As $II= \bar{M}_{n_2}(\Delta_d, \Delta_\theta, b)- \bar{M}_{n_2}(\Delta_d, \Delta_\theta, -b)$, a bound on the modulus of continuity is
$\phi_{n_2}(\delta) = (\delta + n^{-1/3 + \gamma})^{1/2} + n_2^{1/6 + \gamma} \delta$. This yields $n_2^{1/3} (\hat{d}_2 - d_0) = O_p(1)$.

{\it Limit Distribution. } Here, we outline the steps for deriving the form of the limit process.
Let  
\be
\begin{split}
\lefteqn{\tilde{f}_{n_2, h, \theta}(\epsilon, W) = n_2^{1/6 - 2\gamma} (m(\theta + W K n^{-\gamma}) + \epsilon) \times } \\
&  \left[1\left[|\theta - d_0 + W K n_1^{-\gamma}- h n_2^{-1/3}| \leq b n_1^{-\gamma}\right] - 1\left[|\theta - d_0 + W K n_1^{-\gamma}| \leq b n_1^{-\gamma}\right] \right],  
\end{split}
\ee
and \begin{eqnarray*}
Z_{n_2} (h, \theta)  = \G_{n_2} \tilde{f}_{n_2,h, \theta}(\epsilon, W) + \zeta_{n_2}(h,\theta),
\end{eqnarray*}
where $\zeta_{n_2}(h,\theta) = \sqrt{n_2 } P \left[\tilde{f}_{n_2,h, \theta}(\epsilon, W)\right]$. %
For $\Delta_\theta = n_1^{1/3}(\theta - \theta_0)$, note that 
\begin{eqnarray*}
\lefteqn{\zeta_{n_2}(h,\theta_0 + n_1^{-1/3}\Delta_\theta) }\\
 & = & n_2^{2/3 - 2\gamma} \left[M_{n_2}(d_0 + h n_2^{-1/3}, \theta_0 + \Delta_{\theta} n_1^{-1/3}) - M_{n_2}(d_0, \theta_0 + \Delta_{\theta} n_1^{-1/3}) \right] \\
& = & n_2^{2/3 - 2\gamma} \left[M_{n_2}(d_0 + h n_2^{-1/3}, \theta_0 ) - M_{n_2}(d_0, \theta_0) \right]\\
& & {{}}+  n_2^{2/3 - 2\gamma} \left[\bar{M}_{n_2}(h n_2^{-1/3}/n_1^{-\gamma}, \Delta_\theta, b)- \bar{M}_{n_2}(h n_2^{-1/3}/n_1^{-\gamma}, \Delta_\theta, -b)\right].
\end{eqnarray*}
Using the expression for partial derivatives of $M_{n_2}$ at $d_0$, we have
\begin{eqnarray*}
\lefteqn{n_2^{2/3 - 2\gamma} \left[M_{n_2}(d_0 + h n_2^{-1/3}, \theta_0 ) - M_{n_2}(d_0, \theta_0 ) \right] }\\
& = & \frac{m(\theta_0)}{K^2}g'\left(\frac{b}{K}\right)n_1^{2\gamma} (hn_2^{-1/3})^2 n_2^{2/3 - 2\gamma}  + o(1) \\
& = & \left(\frac{p}{1-p}\right)^{2\gamma} \frac{m(\theta_0)}{K^2}g'\left(\frac{b}{K}\right)h^2   + o(1).
\end{eqnarray*} 
Further,
\begin{eqnarray*}
\lefteqn{n_2^{2/3 - 2\gamma} \bar{M}_{n_2}(h n_2^{-1/3}/n_1^{-\gamma}, \Delta_\theta, b)}\\
& = & n_2^{2/3 - 2\gamma} \frac{1}{K}\int_{0}^{h n_2^{-1/3}/n_1^{-\gamma}} \Bigg[ m(\theta_0 +  (u + b) n_1^{-\gamma}) \times\\
& &  \left\{g\left(\frac{(u+b)n_1^{-\gamma} + n_1^{-1/3} \Delta_\theta}{K n_1^{-\gamma}}\right) - g\left(\frac{(u+b)n_1^{-\gamma} }{K n_1^{-\gamma}}\right)\right\} \Bigg] du \\
& = & n_2^{2/3 - 2\gamma} \frac{n_2^{-1/3}}{n_1^{-\gamma}}\frac{1}{K}\int_{0}^{h} \Bigg[ m(\theta_0 +  w ((1-p)/p)^\gamma n_2^{-1/3} + b n_1^{-\gamma}) \times \\
& & {{}} \left\{g\left(\frac{w  n_2^{-1/3} + b n_1^{-\gamma} }{K n_1^{-\gamma}} + \frac{\Delta_\theta n_1^{-1/3}}{Kn_1^{-\gamma}}\right) - g\left(\frac{w n_2^{-1/3} +bn_1^{-\gamma} }{K n_1^{-\gamma}}\right)\right\} \Bigg]du \\
& =& n_2^{2/3 - 2\gamma} \frac{n_2^{-1/3}}{n_1^{-\gamma}} h \frac{m(\theta_0 )}{K} g'\left(\frac{b}{K}\right)\frac{\Delta_\theta n_1^{-1/3}}{Kn_1^{-\gamma}} + o(1) \\
& =& h \frac{m(\theta_0 )}{K} g'\left(\frac{b}{K}\right)\frac{\Delta_\theta}{K}\left(\frac{1-p}{p}\right)^{1/3- 2\gamma} + o(1).
\end{eqnarray*}
As $g'(x) = - g'(-x)$, we have 
\begin{eqnarray}
  \lefteqn{n_2^{2/3 - 2\gamma} \left[\bar{M}_{n_2}(hn^{-1/3 + \gamma}, \Delta_\theta, b)- \bar{M}_{n_2}(hn^{-1/3 + \gamma}, \Delta_\theta, -b)\right] } \nonumber\\
	&= & \left(\frac{1-p}{p}\right)^{1/3- \gamma} \frac{2m(\theta_0)}{K^2} g'\left(\frac{b}{K}\right)\Delta_\theta h+ o(1).
\label{eq:abbb}
\end{eqnarray} 
We next show that  Var$(Z_{n_2}(h, \Delta_\theta))$ converges to zero. Let
\begin{eqnarray*}
\lefteqn{f_{n, h, \Delta_\theta} ( \epsilon, W) = (m(n_1^{-1/3}\Delta_\theta + W K n^{-\gamma}) + \epsilon) \times } \\
& & \left[1\left[|{n_1^{-1/3}\Delta_\theta}+ W K n_1^{-\gamma}- h n_2^{-1/3}| \leq b n_1^{-\gamma}\right] - 1\left[|{n_1^{-1/3}\Delta_\theta}+ W K n_1^{-\gamma}| \leq b n_1^{-\gamma}\right] \right].  
\end{eqnarray*}
Consequently, $P \tilde{f}_{n_2, h, \theta_0 + n_1^{-1/3} \Delta_\theta} = \zeta_{n_2}(h,\theta_0 + n_1^{-1/3} \Delta_\theta)/\sqrt{n_2}$ converges to zero. Thus
\begin{eqnarray*}
\mbox{Var$\left(\tilde{f}_{n_2, h, \theta_0 + n_1^{-1/3} \Delta_\theta}\right)$} & = &  E \left[\tilde{f}^2_{n_2, h, \theta_0 + n_1^{-1/3} \Delta_\theta}\right] + o(1) \\
& \lesssim & \frac{n_2^{4/3 - 4\gamma}}{n_2}  (\|m\|^2_\infty + \sigma^2) h n_2^{-1/3 + \gamma} + o(1) =o(1).
\end{eqnarray*}
Using \eqref{eq:abbb} and the above, it can be shown by applying Theorem \ref{th:tight} and Lemma \ref{lm:fidiconvgen} that
\begin{eqnarray*}
\lefteqn{n_2^{1/3} (\hat{d}_2 - d_0) \stackrel{d}{\rightarrow} }\\
 & & \argmax_h \left\{\left(\frac{p}{1-p}\right)^{2\gamma} \frac{m(\theta_0)}{K^2}g'\left(\frac{b}{K}\right)h^2  +  \left(\frac{1-p}{p}\right)^{1/3- 2\gamma} \frac{2m(\theta_0)}{K^2} g'\left(\frac{b}{K}\right) \mathcal{Z} h \right\} \\
& = &  -\left(\frac{1-p}{p}\right)^{1/3 } \mathcal{Z}. 
\end{eqnarray*}
As the Chernoff random variable $\mathcal{Z}$ is symmetric, we get the result.
\qed
\begin{rmk}\label{gexpln}
We reiterate here that when $m'(d_0) =0$, the regression function is essentially flat in the zoomed-in neighborhood which hinders estimating $d_0$ through a two stage procedure. Using a (second stage) design peaking at the first stage  estimate adds to the curvature (second derivative) of the second stage population criterion function (see \eqref{eq:secndder} and the resulting $\rho_n$ in comparison with the distance in Remark \ref{derzero}) which alleviates this problem to an extent. However, as was the case in Remark \ref{nonuniftech} with a non-smooth $m$, there is a bias introduced by the non-uniform design which does not allow an acceleration in the rate of convergence. 
\end{rmk}

\end{document}